\documentclass[10pt,journal,compsoc]{IEEEtran}	
	\usepackage{amssymb}
	\usepackage{graphicx}
	\usepackage{epstopdf}
	\usepackage{gensymb}
	\usepackage{color}
	\usepackage{amsmath}
	\usepackage{enumerate}
	\usepackage{cite}
	\usepackage{comment}
	
	\usepackage{amsthm}
	
	\theoremstyle{definition}
	\newtheorem{definition}{Definition}
	
	\usepackage{algorithmicx}
	\usepackage[ruled]{algorithm}
	\usepackage{algpseudocode}
	\alglanguage{pseudocode}
    
	\usepackage{array}
	\usepackage{booktabs}
	\usepackage{multirow}

	\usepackage{threeparttable} 
	\usepackage{hyperref}
	
	\usepackage{mathtools}
	
	\DeclarePairedDelimiter\floor{\lfloor}{\rfloor}

	\usepackage{url}
	\urldef{\mailsa}\path|{alfred.hofmann, ursula.barth, ingrid.haas, frank.holzwarth,|
		\urldef{\mailsb}\path|anna.kramer, leonie.kunz, christine.reiss, nicole.sator,|
		\urldef{\mailsc}\path|erika.siebert-cole, peter.strasser, lncs}@springer.com|

\definecolor{garrisonpink1}{rgb}{0.858, 0.188, 0.478}
\newcommand{\garrison}[1]{\textcolor{garrisonpink1}{Garrison: #1}\PackageWarning{garrison:}{#1!}}
	
\begin{document}
\title{TREVERSE: \underline{TR}ial-and-\underline{E}rror Lightweight Secure Re\underline{VERSE} Authentication with Simulatable PUFs}

\author{
Yansong~Gao,~Marten van Dijk,~Lei Xu,~Wei Yang,~Surya Nepal, and Damith C.~Ranasinghe
  
\IEEEcompsocitemizethanks{\IEEEcompsocthanksitem Yansong~Gao, Lei Xu and Wei Yang are with School of Computer Science and Engineering, NanJing University of Science and Technology (NJUST), Nanjing, China. Yansong Gao is also with Data61, CSIRO, Sydney, Australia. yansong.gao@njust.edu.au; xulei\_marcus@126.com;generalyzy@gmail.com}
\IEEEcompsocitemizethanks{\IEEEcompsocthanksitem Marten van Dijk is with Secure Computation Laboratory, Department of Electrical and Computer Engineering, University of Connecticut, USA. marten.van\_dijk@uconn.edu}
\IEEEcompsocitemizethanks{\IEEEcompsocthanksitem Surya Nepal is with Data61, CSIRO, Sydney, Australia. surya.nepal@data61.csiro.au}.
\IEEEcompsocitemizethanks{\IEEEcompsocthanksitem Damith~C.~Ranasinghe is with Auto-ID Labs, School of Computer Science, The University of Adelaide, SA 5005, Australia. damith.ranasinghe@adelaide.edu.au}.
\IEEEcompsocitemizethanks{\IEEEcompsocthanksitem Usage Permission goes to IEEE. Cite as: Yansong Gao, Marten van Dijk, Lei Xu, Wei Yang, Surya Nepal, and Damith C. Ranasinghe. "TREVERSE: Trial-and-error Secure Lightweight Reverse Authentication with Simulatable PUFs." IEEE Transactions on Dependable and Secure Computing (2020).}.
\vspace{-0.5cm}
}
 
\IEEEtitleabstractindextext{		
\begin{abstract}
A physical unclonable function (PUF) generates hardware intrinsic volatile secrets by exploiting uncontrollable manufacturing randomness. Although PUFs provide the potential for lightweight and secure authentication for increasing numbers of low-end Internet of Things devices, practical and secure mechanisms remain elusive. We aim to explore simulatable PUFs (SimPUFs) that are physically unclonable but efficiently modeled mathematically through privileged one-time PUF access to address the above problem. Given a challenge, a securely stored SimPUF in possession of a trusted server computes the corresponding response and its bit-specific reliability. Consequently, naturally noisy PUF responses generated by a resource limited prover can be immediately processed by a one-way function (OWF) and transmitted to the server, because the resourceful server can exploit the SimPUF to perform a trial-and-error search over likely error patterns to recover the noisy response to authenticate the prover. Security of trial-and-error reverse (TREVERSE) authentication under the random oracle model is guaranteed by the hardness of inverting the OWF. We formally evaluate the TREVERSE authentication capability with two SimPUFs experimentally derived from popular silicon PUFs. 
\end{abstract}
}
\maketitle

\vspace{-0.4cm}	
\section{Introduction}\label{Sec:Intro}
Physical unclonable functions (PUFs) exploit manufacturing imperfections to extract hardware instance-specific secrets on demand. The unavoidable fabrication variations of devices endows a PUF with physical unclonability. Thus, even the same manufacturer is incapable of forging two PUFs exhibiting identical behaviors. As a function, the PUF takes inputs (challenges) and react with instance-specific outputs (responses) referred  to as challenge-response pairs (CRPs). The first silicon PUF, coined the Arbiter PUF~\cite{gassend2002silicon} was created in 2002. Since then, various other microelectronic PUF types such as ring oscillator PUF (ROPUF)\cite{suh2007physical,zhang2017xor,rahman2016aging}, SRAM PUF~\cite{holcomb2007initial,cao2015low}, DRAM PUF~\cite{kim2018dram,sutar2018d,sutar2018memory}, and nanoelectronic PUFs~\cite{gao2015emerging} have emerged. Primarily, PUF primitives are a fundamentally different  solution to addressing the secure key storage problem and authentication~\cite{suh2007physical,herder2014physical,delvaux2017security,gunlu2018privacy}. In contrast to requiring of the secure non-volatile memory (NVM) to permanently store a the key in digital form, a PUF key is volatile and only present on demand.
As a consequence of the ability to derive hardware intrinsic secrets, PUF primitives provide the potential for building authentication mechanisms with inherent key protection~\cite{suh2007physical,herder2014physical,yu2017pervasive}.
	
Realizing a lightweight authentication mechanisms with PUFs is a non-trivial task in practice. 
As a comprehensive examination of lightweight authentication mechanisms by Delvaux \textit{et al.}~\cite{delvaux2015survey,delvaux2017security,delvaux2017machine} highlighted the difficulty of realizing PUF based authentication that is lightweight, secure and practical. A key hurdle is the naturally noisy nature of PUF responses. Most studied and popular silicon PUFs
yield responses susceptible to thermal noise and environmental parameter fluctuations such as supply voltage and temperature. Therefore, PUF primitives must directly deal with noise inherent to the source of entropy used for deriving keys~\cite{bhargava2013high,xu2015digital,miao2016lrr,bhargava2014efficient,suh2007physical,maes2016secure,delvaux2017security,yu2014noise,rostami2014robust,gao2016obfuscated} (as detailed in Section~\ref{Sec:Related}). Consequently: 

\vspace{1mm}
{\noindent \textit{Realizing secure, lightweight and practical authentication in the presence of noisy PUF responses remains an open problem}.}

\vspace{2mm}
We observe that previous studies have established the challenge-response specific nature of PUF response reliability~\cite{maes2009soft,maes2013accurate,herder2017trapdoor,gao2020physical}. In essence, the homogeneous application of a reliability measure such as bit error rate across the entire challenge-response space presents only a limited characterization of response unreliability and ignores the challenge-response specific nature of unreliability. 
Following this observation, we unfold a new method in cryptography and security in this paper, so-called TREVERSE---trial and error authentication method where:
\begin{enumerate}
\item The resourceful server authenticates a prover exploiting its unique ability to \textit{estimate the bit specific nature} of PUF response reliability together with its response.
\item The prover is oblivious to the noisy nature of the PUF response and treats the PUF response as a secure digital key; as in a classical crypo system without applying error correction.
\item The adversary is forced to build a method to discover \textit{both} the bit specific reliability and the response information hidden from an adversary in a computationally intractable problem---a trapdoor function. 
\end{enumerate}

\subsection{Goals and Contributions}


The aim of this study is to investigate a new methodology\footnote{We contrast our approach with existing methodologies in detailed in Section~\ref{Sec:Related}.} to achieve a secure and lightweight authentication implementation on a PUF embedded device---referred to as prover---in the presence of noisy PUF responses. Our authentication mechanism takes advantage of response-specific nature of PUF reliability and outsources the overhead of dealing with noisy nature of PUF responses from the prover to the server; here, we rely on the fact that the computational power can be flexibly configured at the server. Overall, our work makes the following contributions:
\begin{itemize}
\item We propose TREVERSE authentication. We challenge the commonly employed approach for dynamic authentication with PUFs where response bits are either: i) corrected using stored or sent helper to the prover; or ii) hashed response value generated on the prover and sent to the server is used to reconcile noisy response bits. Although counter to intuition, we directly employ \textit{noisy} PUF responses without error correction or helper data generation. 

\item To the best of our knowledge,
we are the first to develop a generic PUF based authentication mechanism, where: i) the PUF integrated device can be oblivious to the noisy nature of the PUF response; and ii) the PUF responses are directly employed for security functions, similar to a digital key stored in a secure NVM in a classical crypo system, without applying error correction.

\item In order to realize TREVERSE, we propose a server-aided trial and error authentication algorithm, where the goal of the server is to recover the noise corrupted PUF response processed by a \textsf{OWF} on the prover. The capability is only held by the server because of a securely managed \textsf{SimPUF} that estimates \textit{both} PUF responses and their bit specific reliability. 

\item We develop two sets of TREVERSE protocols: i) unilateral authentication; and ii) mutual authentication. 
We evaluate the security of these protocols, in particular, against {\it all} known modeling attacks. We show that TREVERSE authentication mechanisms are secure in the random oracle model. 

\item We validate the authentication capability and practicability of TREVERSE authentication through concrete formal analyses, and further extensive empirical experiments based on Virginia Tech's public silicon PUF dataset~\cite{maiti2010large}. 
\end{itemize}

\subsection{Paper Organization}
Followed by an overview of TREVERSE in Section~\ref{sec:treverse-overview}, we illustrate the existence of simulatable PUFs and techniques for enrolling \textsf{SimPUF}s in Section~\ref{Sec:ResponseReliability}. In Section~\ref{Sec:TREVERSE}, we elaborate on the TREVERSE instantiation on the prover and then analyze its security. Section~\ref{Sec:AuthenCap} concretely formalizes the TREVERSE authentication capability with respect to both false rejection rate and false acceptance rate. We also validate the formula based on empirical results. Based on public silicon PUF dataset, Section~\ref{Sec:AugAuthen} experimentally evaluates authentication capability of TREVERSE from two different PUF types: ROPUF and linear additive PUF (LAPUF). In Section~\ref{Sec:Comparison}, we compare TREVERSE with other existed works and further discuss TREVERSE. In Section~\ref{Sec:Related}, we present related works, followed by conclusion in Section~\ref{Sec:Conclusion}.
\vspace{-0.3cm}

\section{TREVERSE Overview}\label{sec:treverse-overview}

Here we denote binary vectors with a bold lowercase character, e.g., challenge $ \bf c $ and response $ \bf e $. All vectors are row vectors. A set is denoted with calligraphic character, e.g., challenge set $ {\mathcal C} $ and response set $ {\mathcal E} $. A procedure or function is printed in a sans-serif font, e.g. {\textsf{PUF}}(${\bf c}$). 

\begin{figure}[t]
	\centering
	\includegraphics[trim=0 0 0 0,clip,width=0.50\textwidth]{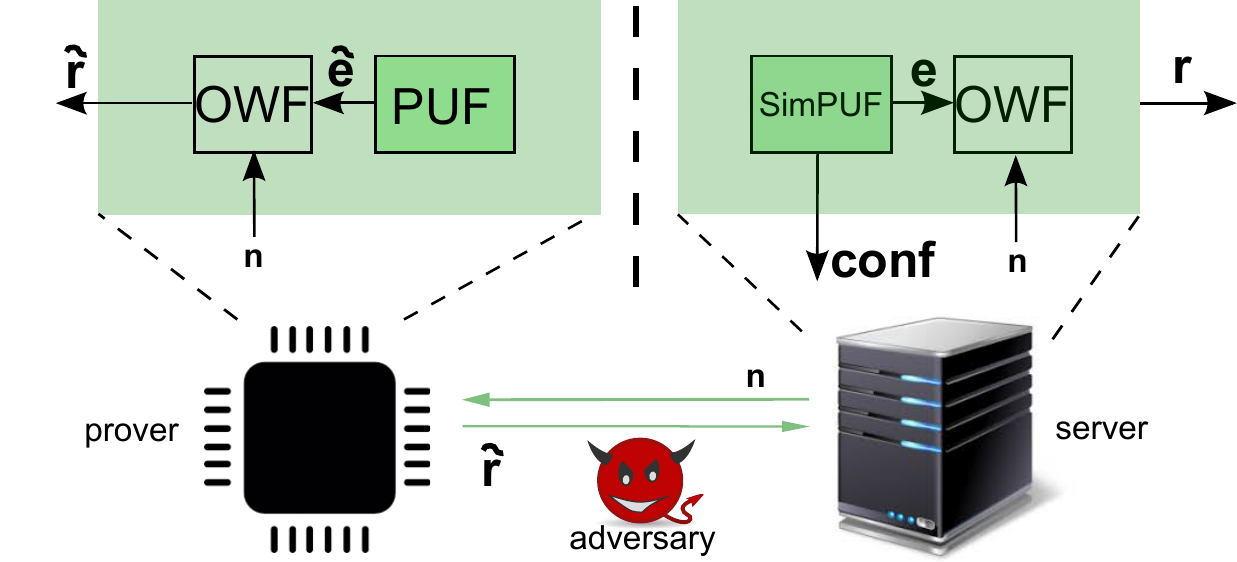}
	\caption{TREVERSE considers three parties: the server, the prover and the adversary. The server holds a SimPUF that is a parameterized model of the physical PUF to not only accurately emulate the response $\bf e$ but also its corresponding reliability confidence $\bf conf$.}
	\label{fig:overviewTREVERSE}
\end{figure}

Consider the dynamic authentication scenario realized with a key derived from a prover integrated PUF illustrated in Fig.~\ref{fig:overviewTREVERSE}. For simplicity, we start by assuming that a hard-wired or fixed PUF challenge $\bf c$ (not shown) is employed to derive a fixed response $ \tilde{\bf e} $ on the prover. 
Following~\cite{delvaux2017security}, we define a physical PUF and One-Way Function (OWF) as below.

\begin{definition}{\bf PUF:}\label{def:PUF}
For a given manufacturing process, a PUF is a manufactured building block that realizes a non-deterministic mapping from a set $ {\mathcal C}\in\{0,1\}^{\lambda} $ to a set $ {\tilde{\mathcal E}}\in\{0,1\}^{\eta} $, where the distribution of each random variable $ {\tilde {\bf e}}_{i}$, with $ i\in [1,|{\mathcal C}|] $, depends on process variations, noise, environmental variables, and aging. Therefore, two random evaluations of the response given the same challenge might slightly vary but with an upper bound {\textsf{HD}}($ {\bf e}, \tilde{{\bf e}}$)$\le {\rm th}$, with threshold ${\rm th}$ a constant.
\end{definition}

In general, assuming that the PUF is a function, given input challenge $\bf c$, it returns output response $\bf e \leftarrow$\textsf{PUF}($\bf c$).
\begin{definition}{\bf One-Way Function}
A function \textsf{OWF} is one-way if and only the function can be computed by a polynomial time algorithm, but any polynomial time randomized $\textsf{OWF}^{-}$---pseduo-inverse function of \textsf{OWF} that attempts to compute a pseduo-inverse for \textsf{OWF} succeeds with negligible probability.
\end{definition}

We refer to those PUFs that can be simulated, for example, using a mathematical model or exhaustive characterization where the characterization complexity is linear with respect to the number of challenges, as simulatable PUFs. The server in Fig. \ref{fig:overviewTREVERSE} holds a \textsf{SimPUF(.)}; a parameterized model of the simulatable physical PUF embedded within the prover. We formally define a \textsf{SimPUF} below.

\begin{definition}{\bf Simulatable PUF:}~\footnote{We are aware that the term Simulatable PUF was previously used by Ruhrmair {\it et. al} in 2013~\cite{ruhrmair2013pufs}. In general, the Simulatable PUF in~\cite{ruhrmair2013pufs} emulates only the response. This concept, while similar, is limited for our needs since the simulatable PUFs we describe predict the bit-specific reliability as well as the response.}\label{def:SimuPUF}
A PUF with a parameterized model \textsf{SimPUF} capable of computing a response $\bf e$ and its corresponding reliability confidence $\bf conf$ in polynomial time for any given challenge $\bf c$ is said to be a simulatable PUF.  Here: i) $({\bf e},{\bf conf}) \leftarrow$\textsf{SimPUF}($\bf c$) where \textsf{SimPUF} is constructed using one-time privileged access by an authorized party in a secure environment and subsequent acquisition of \textsf{SimPUF} by any party is disabled;
ii) $\bf e$ is indistinguishable from the response $\tilde{\bf e}\leftarrow$\textsf{PUF}($\bf c$), that is $ \mathbb{P}(\tilde{\bf e}={\bf e}) $ is $ \epsilon $-close to 1; and iii) the estimated $\bf conf$ is $ \epsilon $-close to the reliability confidence of $\tilde{\bf e}$.  

\end{definition}  
 

In the TREVERSE authentication scenario in Fig.~\ref{fig:overviewTREVERSE}, a \textsf{SimPUF} is held by the server. Subsequent acquisition of \textsf{SimPUF} by any party is disabled, for example, by fusing the access wire to the PUF response~\cite{yulockdown}. The TREVERSE authentication protocol can be described as follows:
\begin{enumerate}
\item A nonce $\bf n$ is issued by the server and sent to the prover\footnote{Notably, the nonce $\bf n$ may also be generated by the prover and sent to the server as in our TREVERSE-B instantiation, in Section~\ref{sec:TREVERSE-B}, to allow the use of LAPUFs and/or mutual authentication functions.}.
\item At the prover, an output $ \tilde{\bf r}\leftarrow\textsf{OWF}$($ \bf \tilde{e},\bf n)$ is generated based on PUF response $ \bf \tilde{e}\leftarrow\textsf{PUF}$() and the $\bf n$. The prover transmits $\tilde {\bf r}$ to the server.
\item The server securely manages the \textsf{SimPUF} and computes $ \bf e $ and the corresponding reliability confidence $ \bf conf $ where $({\bf e},{\bf conf}) \leftarrow$\textsf{SimPUF}().
\item Given $ \bf conf $, the server has the exclusive ability to identify bits that are least reliable in the response $ \bf e $. These bits will have a high chance to be different from that in $ \bf \tilde{e} $. The server exhaustively tries all error patterns for these unreliable bits to form a set of trial responses $ {\mathcal E}^t$. The server successfully authenticates the prover if for any trial response ${\bf e}^t\in {\mathcal E}^t$,  \textsf{OWF}(${\bf e}^t$,$\mathbf{n})=\tilde{\bf r}$, otherwise, the authenticity of the prover is rejected.
\end{enumerate}
In general, the TREVERSE authentication relies on the server's \textit{unique} ability to discover the response $ \bf \tilde{e} $ according to their securely managed knowledge of \textsf{SimPUF}---detailed in Section~\ref{sec:aut_halgorithm}. Notably, the adversary is forced to build a method to discover \textit{both} the bit specific reliability information and the response information obfuscated from an adversary in a computationally intractable trapdoor function. 
The TREVERSE inevitably raises three important questions: 

\begin{itemize}
\item Do simulatable PUFs defined in Definition \eqref{def:SimuPUF} exist in practice?
\item What is the probability of false acceptance and false rejection---we refer to as the \textit{authentication capability}---with respect to the number of unreliable response bits selected to generate the trial response set $ {\mathcal E}^t$?
\item What is the security of the proposed protocol? 
\end{itemize}
The following of this paper answers these questions.
\begin{figure}[t]
	\centering
	\includegraphics[trim=0 0 0 0,clip,width=0.50\textwidth]{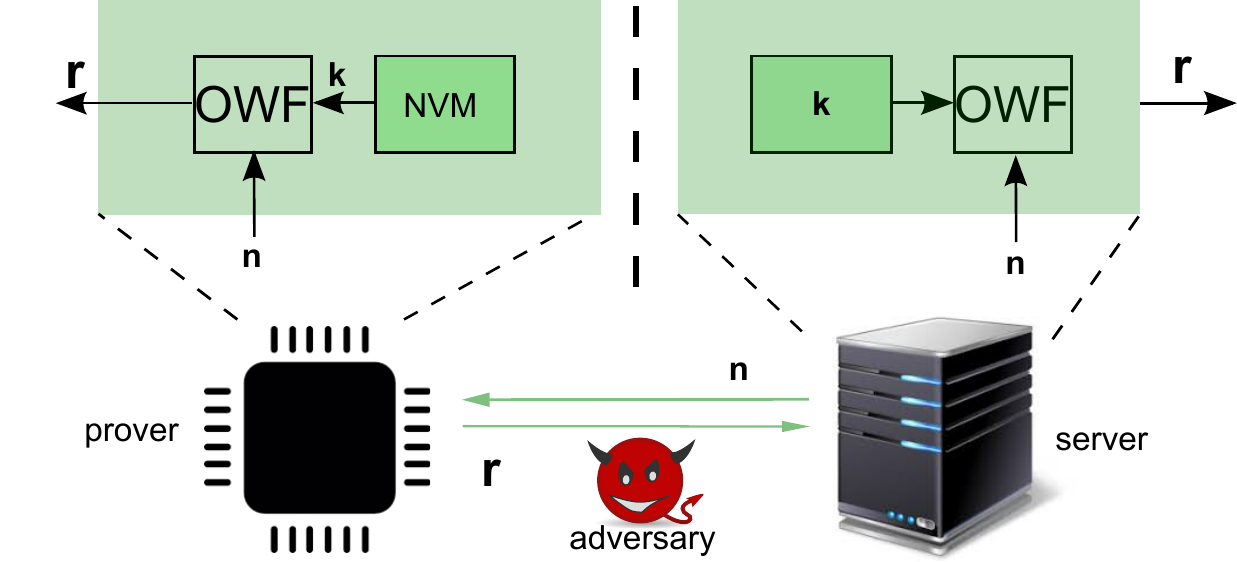}
	\caption{Classical entity/client digital key based authentication. The key is commonly stored on the secure NVM.}
	\label{fig:serverNVM}
\end{figure}

\vspace{-0.3cm}
\section{Simulatable PUFs}\label{Sec:ResponseReliability}
In practice, any PUF with the ability to exhaustively and repeatedly readout CRPs and associated bit-specific reliability in polynomial time can be a simulatable PUF. Unsurprisingly, PUFs with CRPs exponential in the number of challenge bits capable of being mathematically modeled also fall into the class of Simulatable PUFs. 
Without loss of generality, we elaborate on techniques to acquire a \textsf{SimPUF} for three different and popular silicon PUFs suitable for microelectronic devices: i) Linear Additive delay PUFs (LAPUFs); ii) ROPUFs; iii) and SRAM PUFs. The first one is a strong PUF owing to its exponential CRP space~\cite{ruhrmair2013puf} while the later two are examples of weak PUFs with limited CRP space while the 
Recall that a \textsf{SimPUF} must be capable of estimating for any given challenge: i)~the response; and ii)~the associated bit specific reliability.

\subsection{Linear Additive PUF}


A popular PUF topology is the linear additive PUF (LAPUF)~\cite{yu2011lightweight,yu2012performance,yu2014noise}. Representatives of the LAPUF are the Arbiter PUF (APUF) and the $k$-sum ROPUF~\cite{yu2011lightweight,yu2014noise}. LAPUFs 
yield a massive number of CRPs in a limited area footprint in silicon.

\vspace{2mm}
\noindent{\bf Response:} For LAPUFs, it is impractical to exhaustively readout its responses due to its very large CRP space. From a modeling perspective, the APUF and the $ k $-sum ROPUF can be reduced to the same topology~\cite{yu2014noise}. It has been widely shown that LAPUFs can be modeled ~\cite{ruhrmair2013puf,lim2004extracting,ruhrmair2010modeling,becker2015pitfalls,becker2015gap}. 
TREVERSE authentication benefits from the existing body of methods for constructing a mathematical model of an LAPUF. The server can learn LAPUF model parameters using a limited number of direct CRP measurements during the secure enrollment phase to subsequently emulate the response of any chosen challenge.

\vspace{1.5mm}
\noindent{\bf Bit specific reliability:} In fact, an LAPUF model is not only able to emulate the response to a given a random challenge, it can also be employed to accurately predict the bit-specific reliability of the response~\cite{xu2016using}. To be precise, 
given a challenge, the LAPUF model predicts a numerical value that is linear with the time difference between the top and bottom paths in the APUF, or frequency difference between the top and and bottom paths in the $k$-sum ROPUF. Such an numerical value can be utilized as the bit-specific reliability. As for the binary response, if the value is larger than zero, then the predicted response is `1', otherwise `0'.

\subsection{Ring Oscillator PUF (ROPUF)}
An ROPUF has a number, $k$, of ring oscillators (ROs); each RO has an odd number of inverters. The frequency of each RO is designed to be identical but varying in practice due to fabrication randomness. The ROPUF produces a response upon comparing frequencies of a pair of ROs where the given challenge selects the pair to be compared~\cite{suh2007physical}. 

\vspace{1.5mm}
\noindent{\bf Response:} The ROPUF also has a limited CRP space. Specifically, the number of CRPs is $\floor{\frac{k}{2}}$ when a response is produced from independent ROs and ${k \choose 2}$ when the response generated from all possible combinations of ROs. Therefore, the responses can be fully characterized by the server. 

\vspace{1.5mm}
\noindent{\bf Bit specific Reliability:} 
The response bit-specific reliability can be evaluated conveniently by subtracting the frequencies of the two ROs selected by a given challenge. The magnitude of the difference in the frequencies can then be employed to estimate the reliability of the response bit~\cite{yu2010secure}. 

\vspace{1.5mm}
\noindent\subsection{SRAM PUF}
The SRAM PUF exploits random but repeatable power-up states of SRAM cells as responses; each cell consists of two cross-coupled inverters, where the cell address is the challenge~\cite{holcomb2007initial}. 

\vspace{1.5mm}
\noindent\textbf{Response}: Given the limited CRP space of a SRAM PUF, an exhaustive readout of CPRs can be performed by the server to enroll response bits.

\vspace{1.5mm}
\noindent\textbf{Bit specific reliability:} 
To gain a bit-specific reliability model, current methods is to apply multiple physical measurements of the same SRAM PUF response. The number of measurements is in the order of 10 to 100~\cite{maes2009soft,maes2009low} that has been shown to be applicable to the soft-decision based error correction. More number of measurements performed, more accurate the bit-specific reliability \footnote{Exhaustive repeated measurement is possible but is not preferable in practice. Therefore, new methods to gain accurate bit-specific reliability of the SRAM PUF without relying on exhaustive repeated measurement as an interesting future work should be investigated.}.  

\section{TREVERSE Authentication}\label{Sec:TREVERSE}
We presented an overview of TREVERSE authentication in Section~\ref{sec:treverse-overview}. In this section, we begin with a description of the algorithm employed by a server.
Then we detail both TREVERSE unilateral and mutual authentication along with specific prover architectures and analyze the security.

\subsection{Trial and Error Authentication}\label{sec:aut_halgorithm}

\begin{figure}
	\centering
	\includegraphics[trim=0 0 0 0,clip,width=0.25\textwidth]{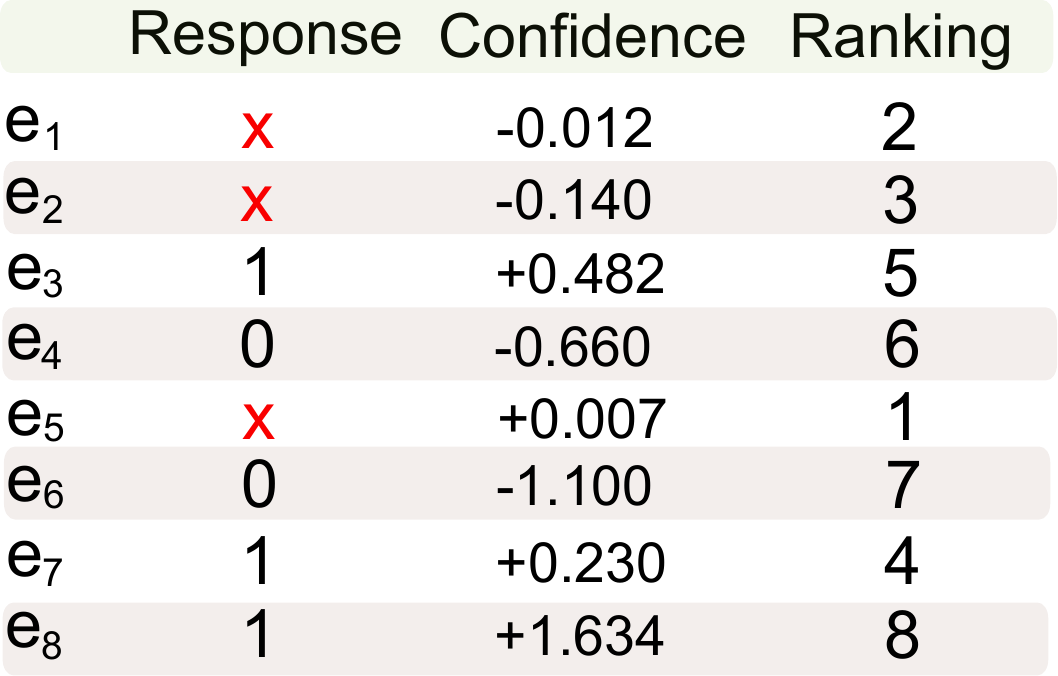}
	\caption{An example of sorting reliability confidence of responses. Lower the reliability (closer to zero), higher the ranking. The highest ranked $ m $---$m=3$ for example---response bits are marked as unknown, because their regenerations, e.g., $ {\tilde{e}_5}, {\tilde{e}_1}, {\tilde{e}_2}$, are more likely to be flipped and thus differing from $ {e}_5, {e}_1, {e}_2$.}
	\label{fig:confidence}
\end{figure}

We take a motivating example to help describe the algorithm~\ref{Algorithm:TAE} for trial and error conducted by a server. Assuming that for a chosen challenge $\bf c$, the server employs its \textsf{SimPUF} to compute the response $ {\bf e} =\{e_1,..,e_8\} $, a response bit vector of length $k=8$, with the associated response bit confidence as shown Fig.~\ref{fig:confidence}. Here, an $ \bf index $ vector ranks each response bit's reliability in descending order; lower the reliability, higher the ranking. For example, $ {\rm index}_1 $ corresponds to $ {e}_5 $ and $ {\rm index}_2 $ corresponds to $ {e}_1 $. Then for $m$ lowest confidence bits where $ m<k $, response bits $ e_{\rm index_1},...,e_{\rm index_m} $ are selected as $m$ unreliable response bits.

In order to conduct trial and error authentication, the server can exhaustively iterate over all possible error patterns for these $m$ unreliable response bits, $ e_{\rm index_1},...,e_{\rm index_m} $. The  $ k-m $ reliable response bits emulated using the \textsf{SimPUF} is kept unaltered during this trial and error phase. 

To illustrate the algorithm, suppose that the prover generated response $ \bf \tilde{e} $ is "{\it 01}10{\it 0}011". Then, consider for the  enrolled response shown in Fig.~\ref{fig:confidence}, the selected number of lowest confidence bits $m=3$. Consequently, the server can generate the possible set of error patterns  \{0,0,0; 0,0,1; 0,1,0; 0,1,1; 1,0,0; 1,0,1; 1,1,0; 1,1,1\} for $ e_5, e_1 $ and $ e_2 $. Each error pattern is injected into the unreliable response bit positions 5, 1, and 2 to form a trial response ${\bf e}^t\in\mathcal{E}^t$. Subsequently, the server iterates over  all ${\bf e}^t\in\mathcal{E}^t$ to compute the corresponding response ${\bf r}^t$
and compares with the received response $ \bf \tilde{r} $ from the prover. Suppose the server tries each error pattern sequentially, we can see that within two trials, the authenticity of the token is accepted. 
Notably, in the worst case, {\it at most}, $ 2^m=8 $ trials are performed with $m=3$. If {\it all} the trials fail, the authenticity of the token is rejected. 

\begin{algorithm}[h]
	\small
	\caption{TREVERSE authentication}
	\label{Algorithm:TAE}
	\begin{algorithmic}[1]
    \State $\mathbf{\tilde{r}}$ \Comment {received response vector from the token}
    \State $\mathbf{n}$ \Comment {nonce employed in the protocol}
    \State $[{\bf e},{\bf conf}]$ 
    \Comment {generated using the securely held \textsf{SimPUF}}
    \State [$\bf index$] = sort(abs($ \bf conf $), 'ascending') \Comment{sort() is a sorting function, abs($x$) is a function returning the absolute value of $x$ }
	\Procedure{$\mathbf{authenticate}$~} {$ \bf e $, $ \bf \tilde{r}$, $ \bf index $}
    \State $\mathit{authState}\leftarrow$ Fail
	\For{$ i=1:2^m $}
	\State generate a new trial response $ {\bf e}_i^t $ by altering response\par
        \hskip\algorithmicindent bits $e_{\rm index(1)},...,e_{\rm index(m)}$ 
	\State compute response $ {\bf r}$ using $ {\bf e}_i^t$ and $\mathbf{n}$
	\If {$\bf r$=$ \bf \tilde{r} $} 
	\State $\mathit{authState}\leftarrow$ Success
    \Comment{prover is authenticated}
	\State \Return $\mathit{authState}$
	\EndIf
	\EndFor 
	\EndProcedure		
	\Statex
\end{algorithmic}
\vspace{-0.4cm}%
\end{algorithm}

\subsection{Unilateral Authentication}
We have established a method for trial and error authentication by a server. Now, we consider the realization of unilateral authentication with a simulatable PUF on a prover. We propose two prover architectures that we refer to as: i) TREVERSE-A; and ii) TREVERSE-B. 

\begin{figure}
	\centering
	\includegraphics[trim=0 0 0 0,clip,width=0.25\textwidth]{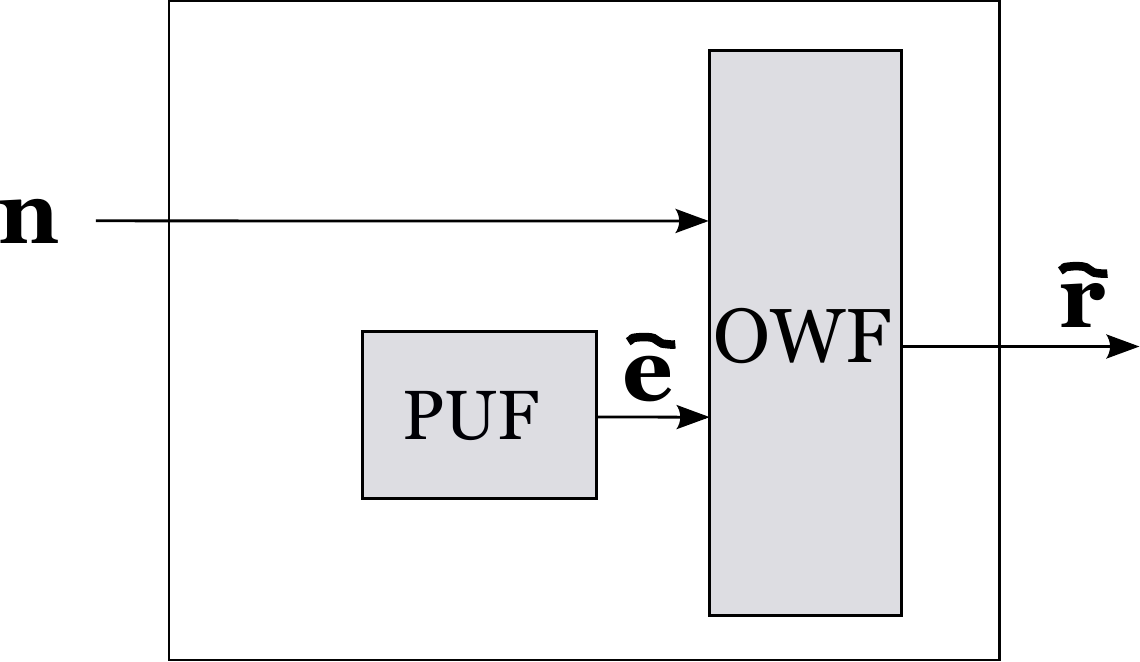}
	\caption{TREVERSE unilateral instantiation, TREVERSE-A, on the prover. The nonce $\mathbf{n}$ is sent by the server. PUF response is required to be iid, e.g., for the SRAM PUF and ROPUF with each RO used only once.}
	\label{fig:TREVERSE-I}
\end{figure} 

\subsubsection{TREVERSE-A}
We propose TREVERSE-A illustrated in Figure~\ref{fig:TREVERSE-I} as a prover architecture, which is comparable with a prover implementation in the classical authentication using a digital key stored in the secure NVM---see Figure~\ref{fig:serverNVM}. 
We make two assumptions for PUFs that utilize this architecture. 
\begin{itemize}
\item We assume the simulatable PUF on the prover generates responses that are information theoretically independent. In other words, each CRP is generated from a spatially separate physical structure.
\item We assume that prover generated PUF response is from a hardwired and, therefore, fixed challenge.
\end{itemize}
Examples of Simulatable PUFs that are appropriate include the SRAM PUF and the ROPUF\footnote{Recent work~\cite{delvaux2017security} takes initial steps into taking bias and spatial correlation into account e.g., for SRAM PUF~\cite{delvaux2017security} (Chapter 4.3.5) when evaluating the entropy loss bound of a PUF key generator. However, due to the complexity to formally develop an easy-to-use tight bound, the iid property of responses is commonly assumed for PUFs such as SRAM PUF and ROPUF.}. 
For instance, in ROPUF, we can generate $\floor{\frac{k}{2}}$ independent response bits out of $k$ ROs. These $\floor{\frac{k}{2}}$ bits can be readout entirely as key material. For an SRAM PUF, a start address can be hardwired and the subsequent SRAM cells readout as the PUF response $\tilde{\bf e}$. 

\vspace{0.3cm}
\noindent{\bf Protocol:} The unilateral authentication protocol with TREVERSE-A as follows:
\begin{enumerate}
\item The server issues a nonce $\mathbf{n}$ and transmits it to the prover.
\item The prover reads out the fixed PUF response $\tilde{\bf e}$ and transmits the output $\tilde{\bf r}$=\textsf{OWF}($\tilde{\bf e}$, $\bf n$) to the server.
\item The server performs TREVERSE using Algorithm~\eqref{Algorithm:TAE}. If the server sees that the computed response $\bf r\leftarrow$\textsf{OWF}(${\bf e}^t$, $\bf n$) is identical to $\tilde{\bf r}$, the prover is considered authenticated by the server.
\end{enumerate}

TREVERSE-A is, to the best of our knowledge, {\it the first PUF authentication instantiation that is same to the classical digital key based unilateral authentication illustrated in Fig.~\ref{fig:serverNVM}}. 
In the next section, we describe a prover PUF architecture that eschews the constraint on the PUF property necessary for TREVERSE-A.

\begin{figure}
	\centering
	\includegraphics[trim=0 0 0 0,clip,width=0.25\textwidth]{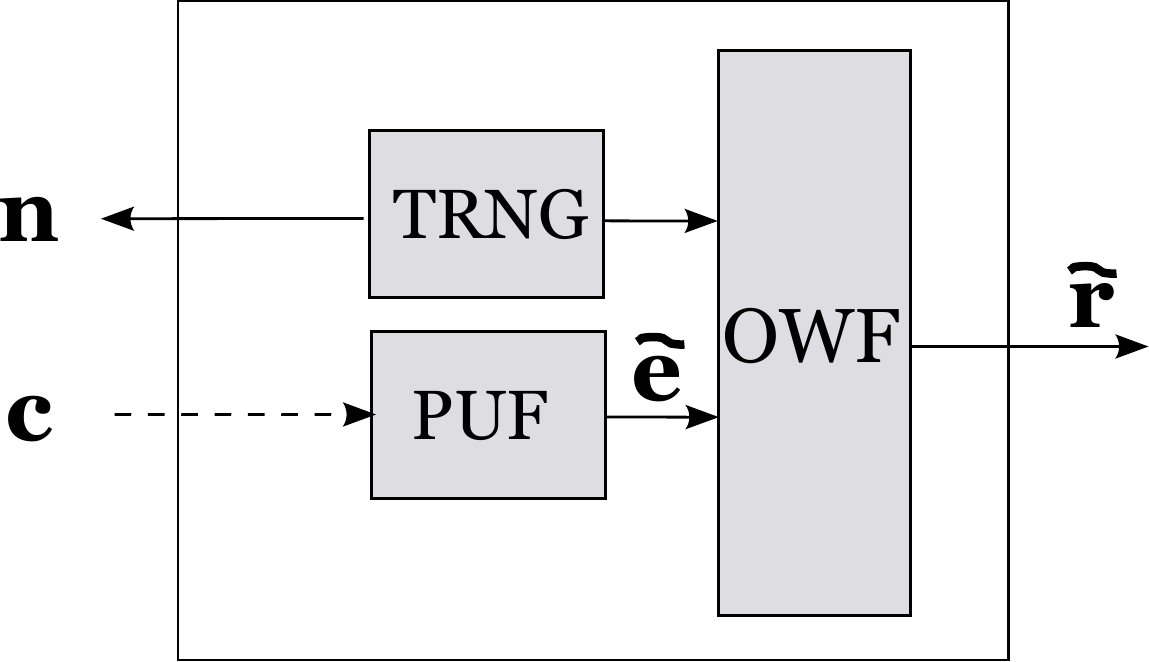}
	\caption{TREVERSE unilateral instantiation, TREVERSE-B, on the prover. The nonce $\mathbf{n}$ is generated by the prover and sent back to the server. There is no constraint on the PUF property. The dotted arrow means that the $\tilde{\bf e}$ can be refreshed by the server issued $\bf c$ on demand, or the $\tilde{\bf e}$ can also be fixed given a fixed $\bf c$.}
	\label{fig:InstantiationII}
\end{figure} 
\subsubsection{TREVERSE-B}\label{sec:TREVERSE-B}

We propose the token architecture we refer to as TREVERSE-B to provide a generic prover implementation for any given simulatable PUF. The prover instantiation is depicted in  Fig.~\ref{fig:InstantiationII}. This architecture provides following advantages:

\begin{itemize}
\item It can be employed with various PUF types: LAPUFs, ROPUFs and SRAM PUFs. Although the later two are applicable to TREVERSE-A.
\item It allows the server to refresh the prover response or secret  $\title{\bf e}$ according to the server issued challenge $\bf c$ on demand between sessions to benefit from the large CRP space of PUFs such as LAPUFs. 
\item It does not strictly require that PUF responses are information theoretically independent. 
\end{itemize}

\vspace{0.3cm}
\noindent{\bf Protocol:} The unilateral authentication protocol with TREVERSE-B follows:
\begin{enumerate}
\item The server issues a challenge $\mathbf{c}$ and transmits it to the prover.
\item The prover applies $\mathbf{c}$ to gain the PUF response $\tilde{\bf e}$ where $\tilde{\bf e}\leftarrow$\textsf{PUF}($\tilde{\bf c})$. The prover \textsf{TRNG} generates a nonce $\mathbf{n}$. Then the prover transmits output $\tilde{\bf r}\leftarrow$\textsf{OWF}($\tilde{\bf e}$,$\bf n$) and the nonce $\bf n$ to the server.

\item The server performs TREVERSE using Algorithm~\ref{Algorithm:TAE}. If the server computed response $\bf r\leftarrow$\textsf{OWF}(${\bf e}^t$, $\bf n$) is identical to $\tilde{\bf r}$, the prover is considered authenticated by the server.
\end{enumerate}

\vspace{1.2mm}
Considering that a PUF generally produces a 1-bit response to a given challenge, a linear feedback shift register (LFSR) or a monotonic counter can be utilized to expand a seed challenge $\bf c$ into required number of challenges to obtain a $k$-bit response $ \bf \tilde{e}$. The nonce generator on a token can be a true random number generator (TRNG) or a well designed pseudo random generator (PRNG) such as the one employed by Yu {\it et al.}~\cite{yulockdown}. The TRNG can be derived from available PUF resources via unreliable PUF responses~\cite{ranasinghe2005random,holcomb2009power,van2012efficient,wang2012flash,zheng2017true}, for example, from APUFs~\cite{ranasinghe2005random}, ROPUFs~\cite{zheng2017true} and SRAM PUFs~\cite{holcomb2009power,van2012efficient}. Notably, as we will discuss in Section~\ref{Sec:Comparison}, most of secure state-of-the-art PUF based authentication mechanisms rely on a TRNG.

\subsection{Mutual Authentication}\label{Sec:mutualAuth}
The TREVERSE-B prover architecture also enables mutual authentication by adding several steps, as detailed below. 

\vspace{0.3cm}
\noindent{\bf Protocol:} The mutual authentication protocol with TREVERSE-B:
\begin{enumerate}

\item The server issues a challenge $\mathbf{c}$ and transmits it to the prover.
\item The prover applies challenge $\mathbf{c}$ to readout PUF response $\tilde{\bf e}$ where $\tilde{\bf e}\leftarrow$\textsf{PUF}($\tilde{\bf c})$. The prover \textsf{TRNG} generates a nonce $\mathbf{n_1}$. Then the prover transmits output $\tilde{\bf r}\leftarrow$\textsf{OWF}($\tilde{\bf e}$,${\bf n}_1$) and the nonce ${\bf n}_1$ to the server.

\item The server performs TREVERSE using Algorithm~\ref{Algorithm:TAE}. If the server computed response $\bf r\leftarrow$\textsf{OWF}(${\bf e}^t$,$\bf n_1$) is identical to $\tilde{\bf r}$, the prover is considered authenticated by the server otherwise the session is aborted.

\item The server acknowledges the prover if accepted. The prover issues nonce ${\bf n}_2$ and transmits it to the server and computes a $\tilde{\bf r}_2\leftarrow$\textsf{OWF}($\tilde{\bf e}$,${\bf n}_2$).

\item Upon receipt of nonce $\bf n_2$, the server computes ${\bf r}_2\leftarrow$\textsf{OWF}(${\bf e}^t$, ${\bf n}_2$), and transmits ${\bf r}_2$ to the prover. Here, ${\bf e}^t$=$\tilde{\bf e}$ by way of step~2 in the protocol. 

\item The prover accepts the authenticity of the server if and only if $\tilde{\bf r}_2$=${\bf r}_2$, otherwise the server is rejected and the mutual authentication is aborted.
\end{enumerate}

\subsection{Security Analysis}\label{Sec:SecurityAnalysis}
We have looked at the problem achieving classical dynamic authentication and mutual authentication with noisy PUFs. We analyze the security of TREVERSE protocols in the archetypal setting of  two parties communicating over an insecure channel attempting to authenticate each party. The two parties are attempting to achieve the security task of authentication or mutual authentication using the insecure communication medium via a TREVERSE protocol.

\subsubsection{Adversary Model}
We adopt an adversary model commonly used with analyzing PUF based security mechanisms~\cite{ruhrmair2013puf,ruhrmair2010modeling,rostami2014robust,yulockdown}. In summary: i) an adversary is allowed to eavesdrop on the communication channel; and ii) arbitrarily apply challenges via the publicly accessible interface to observe the prover response $\tilde{\bf r}$. We assume that \textsf{SimPUF} enrollment is performed by the server in a secure environment using one-time privileged access and such access is prohibited afterwards. As in previous work~\cite{ruhrmair2013puf,ruhrmair2010modeling,rostami2014robust,yulockdown}, we focus on brute-force attacks, replay attacks, and modeling attacks. We also discuss physical attacks.

\subsubsection{Brute-force Attacks}

The probability of correctly guessing a $k$-bit response $ \bf \tilde{e} $ is expressed as:
\begin{equation}\label{Eq:BruteForce}
{\mathbb{P}}=(\max\{\tau, 1-\tau\})^k
\end{equation}
where the $\tau$ is the response bias---to be `1'/`0'. Notably, modern PUFs usually have low bias where $\tau$ is normally close to 0.5~\cite{maes2013physically}. The LAPUF evaluated in this work has a $\tau$ of 50.05\%. We can see that when $k$ is larger than, e.g., 80, the brute-force attack success probability becomes extremely small. Therefore, brute-force attacks are extremely unlikely to succeed.

\subsubsection{Replay Attacks}
In a replay attack scenario, an adversary attempts to authenticate a fraudulent prover to the server by exploiting previously recorded information from protocol sessions. First, consider TREVERSE-A architecture based protocols. The nonce sent by the server is refreshed each session to prevent replay attacks. Second, in TREVERSE-B architecture based protocols, the nonce generated by the token is refreshed for each session; Therefore, replay attacks are prevented.

\subsubsection{Modeling Attacks}\label{Sec:modelingAttack}
Essentially, modeling attacks are dependent on the type of PUF
Weak PUFs such as ROPUFs and SRAM PUFs with information theoretically independent responses are inherently immune from modeling attacks~\cite{ruhrmair2013puf}. 
As a consequence, both TREVERSE-A and TREVERSE-B when employing these PUFs are immune to modeling attacks.
However, PUFs with responses that are correlated with each other, are potentially vulnerable to modeling attacks. A typical characteristic of this PUF type is a very large CRP space provided with limited implementation area. Therefore, such PUF is related to the TREVERSE-B prover architecture. Specifically, we consider most studied LAPUFs.

In the TREVERSE-B instantiation, the response $\tilde{\bf e}$ is hidden behind the \textsf{OWF} trapdoor function. Therefore, conventional modeling attacks exploiting, for example, support vector machine (SVM) and logistic regression (LR) machine learning algorithms~\cite{ruhrmair2013puf,ruhrmair2010modeling} needing access to both challenge $\bf c$ and response $\tilde{\bf e}$ are prevented. Reliability based modeling attacks that exploit reliability information of CRPs, where a direct relationship between a challenge and a response is no longer required~\cite{becker2015pitfalls} are also prevented. This is because an adversary is unable to discover bit-specific reliability information. Attempts to ascertain bit specific reliability information of a challenge-response pair will not succeed since the on-chip nonce prevents inferring the reliability of a PUF response $\tilde{\bf e}$ by observing the output  $\tilde{\bf r}$. Therefore, all known modeling attacks are inapplicable to the TREVERSE-B instantiation. Interestingly, based on the above analysis, we can also conclude that the bit-specific reliability information as well as the PUF response in a TREVERSE-B instantiation is hidden from an adversary in a computationally intractable problem---a trapdoor function.

Notably, composite PUFs, especially those built upon the APUF~\cite{sahoo2017multiplexer,nguyen2018interpose}, can increase modeling attack resilience without using an \textsf{OWF} construct as in TREVERSE. Although these composite PUFs can be adopted as a device PUF because TREVERSE is independent of the underlying device PUF type as long as a simulatable PUF can be obtained, composite PUFs usually result in more noisy response bits. Therefore using a composite PUF will increase the number of unreliable responses TREVERSE is required to trial. As TREVERSE by design relying on an \textsf{OWF} to prevent modeling attacks, a common LAPUF instead of a composite PUF built upon LAPUFs is more preferable to avoid the need to deal with an unnecessarily high number of unreliable response bits.

\subsubsection{Invasive attacks}
Although we do not specifically consider a single PUF construction but describe TREVERSE as a new lightweight method for PUF based authentication, we report on potential invasive attacks and highlight countermeasures as well as the difficulty of mounting such attacks on PUFs. 
In comparison with digitally stored keys, a PUF provides inherent tamper resistance to invasive attacks in comparison with keys stored in NVM~\cite{herder2014physical,suh2007physical,delvaux2015survey}. The key material hidden in the physical structure of PUF circuitry is harder to readout compared to keys stored in NVM. In addition, the process of tampering can destroy a PUF. 

An attacker delayering the IC (integrated circuit) and probing PUF circuits to extract information to construct derived PUF secrets is harder without affecting the PUF CRP behavior or even destroying it once the PUF layout is constructed carefully. One example of such construction proposal is the controlled PUF ~\cite{gassend2002controlled,gassend2008controlled} with a control logic and an APUF where the APUF prevents invasive attacks on the control logic and the control logic protects the APUF from protocol level attacks, and another experimentally validated construction is the capacitive PUF~\cite{tuyls2006read,immler2018b}. For hybrid attacks combining timing and power side channel information with machine learning  usually requires on-chip peripheral circuits~\cite{ruhrmair2014efficient} to measure side channel information of the response. However, those circuits appear to be unavailable in resource-constraint devices. In addition, these attacks can be eliminated through careful circuit design techniques, e.g., dynamic and differential CMOS logic~\cite{ruhrmair2014efficient}.
More importantly, the required power and timing measurement on a directly exposed LAPUF response $\bf e$ in~\cite{ruhrmair2014efficient} is  prevented in a TREVERSE-A or TREVERSE-B prover architecture, because the response $\bf e$ is now hidden from an adversary through the transformation through the one way function \textsf{OWF}. Attacks using photonic side channel information~\cite{tajik2014physical} requires specialized laboratory equipment and professional skills; such photonic emission based attacks can be eliminated by circuit design techniques such as the use of interconnect meshes~\cite{boit2016ic}.

\section{Formalizing Authentication Capability}\label{Sec:AuthenCap}

The TREVERSE authentication inevitably leads to the question of investigating the two important capabilities: i) the false acceptance rate (FAR)---the probability of the server accepting an illegitimate prover; and ii) false rejection rate (FRR)---probability of falsely rejecting a legitimate prover. In the following, we formally derive these metrics by considering a bit specific reliability model. 

\subsection{False Acceptance Rate}\label{Sec:FAR} 
The TREVERSE authentication has zero tolerance for errors outside the $k - m$ reliable response bits bound. 
Recall, that after each trial response ${\bf e}^t$ construction, the server employs an exact match criterion; that is $\tilde{\mathbf{e}}^t=\mathbf{e}$.
Thus, for a given $ m $ highly unreliable bit selections and $ k $ PUF response bits, at most $m$ bits in fixed positions can be different from the server enrolled response. Therefore, FAR can be expressed as the probability of any other token generating $k-m$ bits of a given token to a chosen challenge $\bf c$. Given response bias $\tau$, FAR can be evaluated as:
\begin{equation}\label{Eq:FAR}
{\rm FAR}=(\max\{\tau,1-\tau\})^{k-m}
\end{equation}
Supposing that the $k-m$ is large---it is indeed the case, e.g., up to 80 according to our evaluation in Section~\ref{Sec:AugAuthen}, this success is extremely small in practice.

\subsection{False Rejection Rate}
A legitimate prover response is falsely rejected by a server if at least one of $k-m$ bits of a PUF response $\tilde{\bf e}$ from a legitimate token mismatch with the selected $k-m$ bits of the enrolled response $\bf e$ at the server---see Algorithm~\eqref{Algorithm:TAE}. Then, we can see that formally deriving an accurate false rejection rate requires a model of the response error distribution capable of capturing the bit specific nature of response reliability. The error distribution will allow us to estimate the probability of at least one of $k-m$ bits of a legitimate PUF response $\tilde{\bf e}$ mismatching with the $k-m$ bits of the enrolled response $\bf e$ at the server. 

Inferencing error probabilities based on a bit specific reliability model can be found in prior studies~\cite{maes2009soft,maes2013accurate,herder2017trapdoor}. We will use the formulation in \cite{herder2017trapdoor} to build a response model as in~\cite{maes2009soft,maes2013accurate} because we want a method of relating the bit response model to physical variables that can be directly observable and measurable: in particular, the confidence information of a bit.

\begin{figure} [t]
\centering
\includegraphics[trim=0 0 0 0,clip,width=0.35\textwidth]{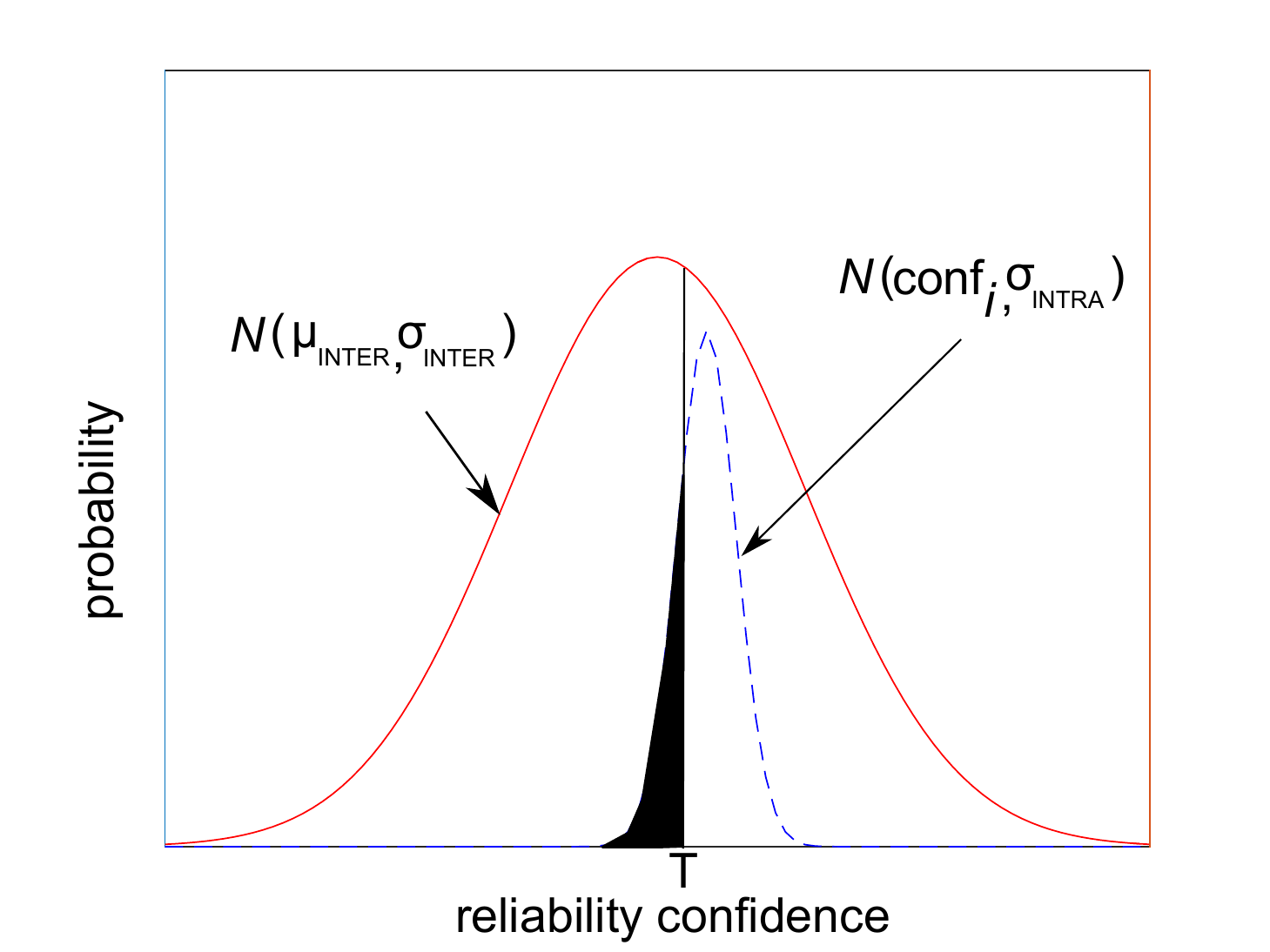}
\caption{Probability density of reliability confidence information. The solid red curve is the reliability confidence distribution of a large population of response bits during enrollment. Ideally, this follows the normal distribution $ N\sim ({\mu}_{\rm INTER},{\sigma}_{\rm INTER}) $. The dashed blue curve is the probability density distribution of reliability confidence of a given bit under different environmental conditions and over multiple evaluations; this distribution is modeled as a normal distribution $ N\sim ({\rm conf},{\sigma}_{\rm INTRA}) $.}
\label{fig:InterIntraDist}
\end{figure}

\vspace{2mm}

\noindent\textbf{Response Model}: We describe a PUF response using the response confidence information. We let the random variable $ \mathcal{C}_{\rm enroll} $ represent the response reliability confidence from which a value ${\rm conf}_i$ is obtained for a given PUF response during enrollment. For a population of response bits $\bf e$, we assume the random variable $ \mathcal{C}_{\rm enroll} $ is described by $ N\sim ({\mu}_{\rm INTER},{\sigma}_{\rm INTER})$.  We illustrate this distribution in Fig.~\ref{fig:InterIntraDist}. Our assumption of normality is validated in the experimental results presented in~\cite{maes2009soft,maes2013accurate,herder2017trapdoor}. When a PUF response is reevaluated, operating conditions such as electrical noise influence the measurement. We model this noise using the random variable $\mathcal{N}$. When a response $e_i$ is reevaluated at the $j$-th instance, a value $n_i^{(j)}$ sampled from $\mathcal{N}$ influences the confidence information of a bit. We define a reevaluated bit response $e_i$ as follows~\footnote{Using $ \tilde{e}_i=1 $ if $ {\rm conf}_i>T $ and $ \tilde{e}_i=0 $ if $ {\rm conf}_i \le T$ in~\eqref{Eq:PeCDF}  will yield the same results}:
\begin{equation}\label{Eq:reponse-model}
    \tilde{e}_i = 
    \begin{cases}
     1 & \text{if } {{\rm conf}_i + n_i^{(j)}}<T \\
     0 & \text{else}
    \end{cases}
\end{equation}
with $T$ a threshold parameter. In practice, this $T$ is usually set to be 0. For example, in ROPUF, if subtraction of two ROs frequencies is lower than 0, then producing response `1', otherwise, `0'. In APUF, if time difference between top and bottom is lower than 0, then producing response `1', otherwise, `0'. Therefore, following this common empirical setting, we set the threshold $T$ to be 0 henceforth.

Here we assume $\mathcal{N}$ is a normal distribution described by $ N\sim ({\mu_{\rm INTRA}},{\sigma}_{\rm INTRA})$. 
The assumption of normality we employ is experimentally validated in~\cite{maes2009soft,maes2013accurate}. Notably, the response model we follow is described in~\cite{maes2009soft,maes2013accurate} with the exception that we define a response model based on using bit confidence information.
\vspace{3mm}




\noindent\textbf{Response Distribution}: Then for a response $e_i$ we define the one-probability---Eq. (2) in~~\cite{maes2009soft}, the probability of response $ \tilde{e}_i $ being one under the $j$th reevaluation---is defined as:

\begin{equation}\label{Eq:FRR1}
p_{\tilde{e}_i}={\rm Pr}(\tilde{e}_i^{(j)}=1)
\end{equation}

Considering our definition of response bit in (\refeq{Eq:reponse-model}) under the assumption of a normal distribution for $\mathcal{N}$, we can write the one-probability as: 

\begin{equation}\label{Eq:FRR2}
p_{\tilde{e}_i}=\Phi\big( \frac{T-{\rm conf_i}}{{\sigma}_{\rm INTRA}}\big),
\end{equation}

where $ \Phi $ is the cumulative distribution function (CDF) of the standard normal distribution. The probability $p_{\tilde{e}_i} $ can be illustrated by the shaded region in Fig.~\ref{fig:InterIntraDist}. Here, $ p_{\tilde{e}_i} $ of a given response is a sample from the random variable $ P_{\tilde{e}} $ denoting the possible one-probability of all response bits. As in~\cite{maes2009soft,maes2013accurate}, we define the CDF of $ P_{\tilde{e}} $ as: 

\begin{equation}\label{Eq:PeCDF}
\begin{split}
{\rm CDF}_{P_{\tilde{e}}}(x)=&{\rm Pr}(p_{\tilde{e}}\le x) 
={\rm Pr}(P_{\tilde{e}}\le x)
\\
=&{\rm Pr}\bigg({\Phi}\bigg(\frac{T-{\mathcal{C}_{\rm enroll}}}{{\sigma}_{\rm INTRA}}\bigg)\le x\bigg)\\
=&{\Phi}(\lambda_1 {\Phi}^{-1}(x)+\lambda_2).
\end{split}
\end{equation}


In~\eqref{Eq:PeCDF}, $ \lambda_1=\frac{{\sigma}_{\rm INTRA}}{{\sigma}_{\rm INTER}} $ and $ \lambda_2=\frac{\mu_{\rm INTER} }{\sigma_{\rm INTER} }$. The $T$ is set to be 0 in~\eqref{Eq:PeCDF} for simplification, this setting actually always gives conservative estimation of the FRR. We are now able to express the one-probability of a population of response bits using four parameters that can be obtained from direct measurements. Recall in Section~\ref{Sec:ResponseReliability} we discussed the extraction of bit specific confidence information for different PUF types during enrollment. Consequently, we can obtain $\sigma_{\rm INTER}$ and $\mu_{\rm INTER}$ from direct measurements. In~\cite{herder2017trapdoor}, it was recognized that $\sigma_{\rm INTRA}$ and $\mu_{\rm INTRA}$ can be measured from the distribution of change in confidence information from enrollment with respect to a change in operating conditions. Consequently, ${\rm Pr}(\mathcal{C}_{\rm ref}-\mathcal{C}_{\rm enroll})$, where {\rm ref} is the new operating condition, can be used to estimate the distribution parameters of $\mathcal{N}$.

{\bf Remark:} To simplify the formalization of the Eq.~\eqref{Eq:PeCDF}, by assuming $T$ to be 0, we have assumed the response is uniformly and randomly distributed---probability of being `1'/`0' is 50\%. It's worth to mention that such an assumption provides a conservative assessment~\cite{herder2017trapdoor}, which is indeed experimentally validated in Section~\ref{Sec:valROPUF}.
\vspace{3mm}

\noindent\textbf{Response Error Distribution}: Now, we are able to find response error distribution based on the one-probability as described in~\cite{maes2009soft,maes2013accurate}. Recall that the enrolled response generated by a \textsf{SimPUF} is $e$ and we assume this to be the correct response. Now, the error probability given a bit $ \tilde{e}_i $ reevaluated on the $j$th occasion can be defined as:

\begin{equation}\label{Eq:FRR4}
p_{{\rm err}_i}=\underset{j}{\rm Pr}(\tilde{e}_i^j\ne e_i).
\end{equation}

The error probability $ p_{{\rm err}_i} $ given response $ \tilde{e}_i $ is a sample of a random variable $ P_{{\rm err}_i} $ of a population of response bits. Then, the CDF of $ P_{{\rm err}_i} $ can be defined and expressed as in~\cite{maes2009soft,maes2013accurate}:

\begin{equation}\label{Eq:FRR5}
\begin{split}
&{\rm CDF}{P_{{\rm err}}}(x) \\
=&{\rm Pr}(p_{{\rm err}_i}\le x)\\
=&{\rm Pr}(P_{\rm err}\le x)\\
=&{\rm CDF}P_{\tilde{e}}(x)+1-{\rm CDF}P_{\tilde{e}}(1-x)\\
=&{\Phi}(\lambda_1 {\Phi}^{-1}(x)+\lambda_2)+1-{\Phi}(\lambda_1 {\Phi}^{-1}(1-x)+\lambda_2).
\end{split}
\end{equation}

\begin{figure} [t]
\centering
\includegraphics[trim=0 0 0 0,clip,width=0.40\textwidth]{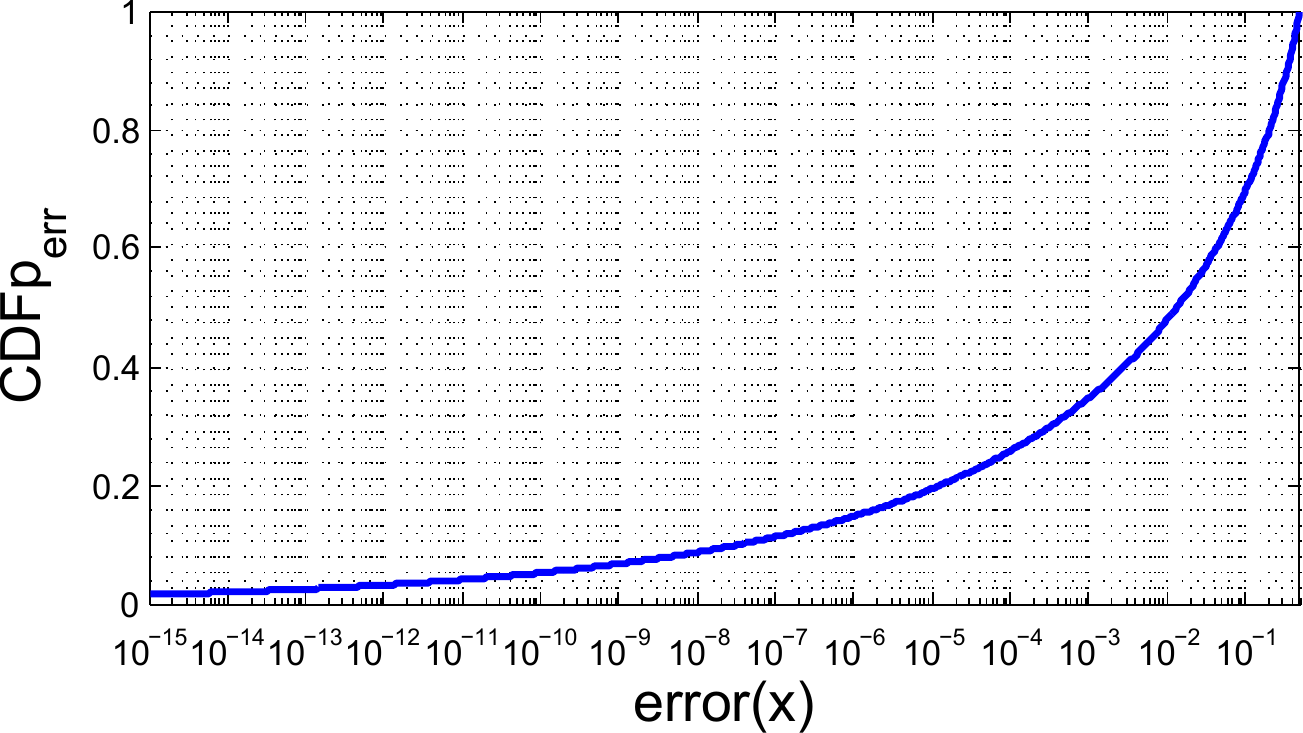}
\caption{Plot of $ {\rm CDF}{p_{{\rm err}}} $(x) in Eq(\ref{Eq:FRR5}) as a function of error probability $ x $.}
\label{fig:CDFPerr}
\end{figure}

We illustrate response error distribution $ {\rm CDF}{P_{{\rm err}}}(x) $ using an example from ROPUF data. In Fig.~\ref{fig:CDFPerr}\footnote{Here, $ \lambda_1=0.3231 $ and $ \lambda_2=-0.3477$ are drawn from the measurements given in Table.~\ref{tab:ROIntraInter}.} in Appendix shows the $ {\rm CDF}{P_{{\rm err}}}(x) $ as a function of error probability $x$. One important observation we can obtain from the cumulative distribution $ {\rm CDF}{P_{{\rm err}}}(x) $ is the determination of the percentage of response bits satisfying a error probability of less than $ x $. For example, just over 10\% of responses have an error probability is less than $10^{-7}$. 
\vspace{3mm}

\noindent\textbf{False Rejection Rate}: 
The number of errors in a $ k $-bit response no longer follows a binomial distribution but a Poisson-binomial distribution~\cite{maes2013accurate}. This is because the one-probability of a response bit is no longer constant and therefore cannot be modeled using a binomial distribution. Given $ k $ response bits, their error probabilities can be represented as $ {\bf p}_{\rm err} $ = $ (p_{{\rm err}_1},...,p_{{\rm err}_k}) $. As highlighted in~\cite{maes2013accurate}, generating the random distribution of bit failures using the bit specific response error model is difficult to achieve analytically since it involves the k-dimensional
distribution of ${\bf p}_{\rm err}$. However, as demonstrated in~\cite{maes2013accurate}, we are able to efficiently simulate $k$ bit responses by randomly sampling $k$ probabilities from the cumulative distribution $ {\rm CDF}{P_{{\rm err}}}(x) $ by using inverse transform sampling.  
Given TREVERSE authentication tolerates $m$ most unreliable bits: i) we sort randomly derived $ {\bf p}_{\rm err} $ in descending order and obtain $ {\bf p}_{\rm err}^{S} $; and ii)  exclude the first $ m $ elements in $ {\bf p}_{\rm err}^{S} $ with lowest reliability confidence to obtain $ {\bf p}_{\rm err}^{S(k-m)} $ with length of $ k-m $. 
We can now express FRR as:
\begin{equation}\label{Eq:FRR}
{\rm FRR}=1-F_{\rm PB}(t=0;{\bf p}_{\rm err}^{S(k-m)}),
\end{equation}
with $ F_{\rm PB} $(t;$ {\bf p}_{\rm err}^{S(k-m)} $) the Poisson-binomial cumulative distribution function~\cite{maes2013accurate}. We use the setting $t=0$ since $ k-m $ bits must exhibit no errors. In practice, randomly sampling a large number (we use 1,000) of $ k $-bit responses and repeatedly evaluating the corresponding sampled $ {\bf p}_{\rm err} $ using (\refeq{Eq:FRR}) yield a large random sample of FRR. The mean of FRR of those evaluations is adopted. 
		

\subsection{Validating False Rejection Model}\label{Sec:FRRModelValidation}
This section validates the bit-specific reliability model derived FRR by showing that the statistical FRR in~(\ref{Eq:FRR}) fits empirical results. The validations are from both ROPUF and LAPUF.
\subsubsection{PUF dataset}
There are three public ROPUF datasets: Virginia Tech~\cite{maiti2010large}, FPL2017~\cite{wild2017fair} and HOST2018~\cite{hesselbarth2018large}. Both FPL2017 and HOST2018 are only evaluated under a limited range of operating conditions---only varying temperature; therefore, ROPUFs show a much lower worst-case unreliability $\epsilon$: 3.57\% for FPL2017 and 3.06\% for HOST2018. In contrast, Virginia Tech ROPUFs exhibit a 9.66\% worst-case $\epsilon$. As we are interested in evaluating TREVERSE under a worst-case setting, we chose the older Virginia dataset for comprehensive validations and select the latest HOST2018 dataset for a complementary evaluation detailed in Appendix.~\ref{app:HOST2018}.

As for the Virginia ROPUF dataset, five ROPUFs are implemented across five Spartan3E S500 FPGA boards. Each FPGA implements one ROPUF that consists of 512 ROs. Details of the constructions are given in~\cite{maiti2010large}. The dataset contains each RO's frequency measurements. Each RO frequency measurement is repeated 100 times under the operating voltages of 0.96~V, 1.08~V, 1.20~V, 1.32~V, and 1.44~V at a fixed temperature of $25\celsius $ to capture supply voltage influences. Similarly, each RO frequency is also evaluated 100 times under $35\celsius $, $45\celsius $, $55\celsius $, and $65\celsius $, with a fixed supply voltage of 1.20~V, to reflect influence from temperature changes.  

\subsubsection{Validation with an ROPUF}\label{Sec:valROPUF}


There are $ {512\choose 2}= 130816$ possible combinations to select a pair of ROs out of 512 ROs in one ROPUF~\footnote{Note that we can use each RO only once to extract a 256-bit response to make the response generation independent. The reason of not doing so here is to facilitate the experimental demonstration with a large CRP sample.}. Therefore, one ROPUF yields 130816 CRPs. The reliability of all five ROPUFs are evaluated. Worst unreliability will result in worst FRR, thus, we are interested in the ROPUF instance that exhibits the {\it worst} BER or $ \epsilon $, which is summarized in Table.~\ref{tab:ROIntraInter} (second column) under varying operating conditions. The BER is dominantly influenced by the supply voltage, which is in well agreement with other reports~\cite{maiti2010large}. 
		
\begin{figure}
	\centering
	\includegraphics[trim=0 0 0 0,clip,width=0.40\textwidth]{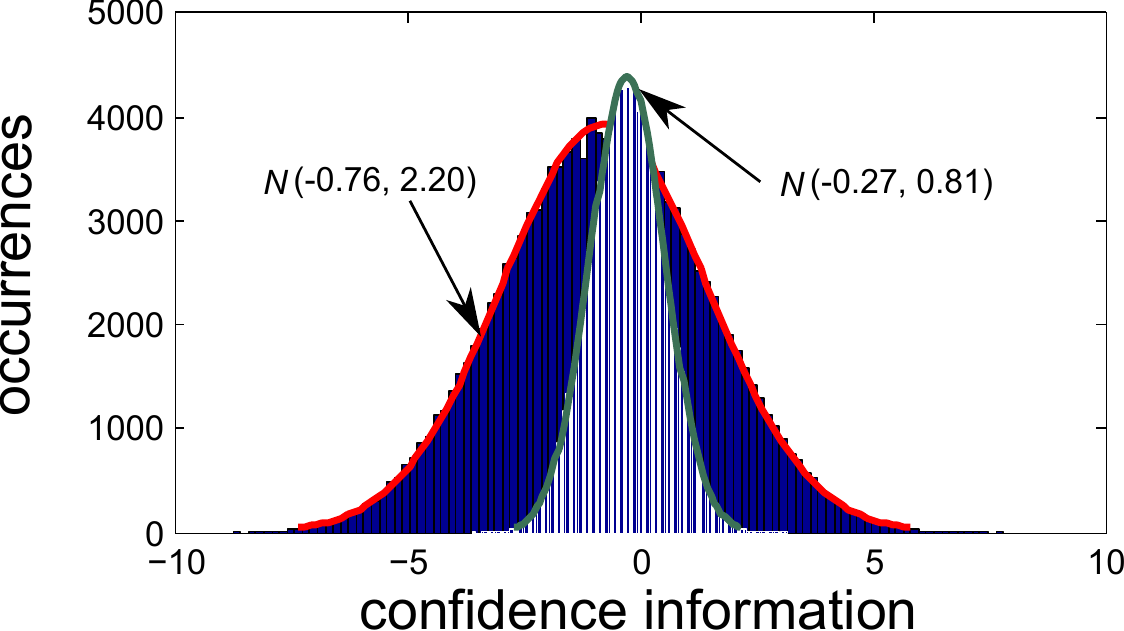}
	\caption{Measured ${\sigma}_{\rm INTRA} $ and ${\sigma}_{\rm INTER} $. The shown ${\sigma}_{\rm INTRA}=0.81 $ is evaluated under the worst-corner (0.96~V, $ 25\celsius $). While the ${\sigma}_{\rm INTER}=2.2 $ is evaluated under nominal operating condition of (1.20~V, $ 25\celsius $).}
	\label{fig:ROIntraInter}
\end{figure}

The reliability confidence of the ROPUF is the frequency subtraction of the pairwise ROs. For all 130816 response bits, distribution of frequency subtraction of these response bits are evaluated under different operating conditions. Same to the observation in~\cite{herder2017trapdoor}, the mean and variance of the distribution changes only slightly under differing operating conditions. Thus, the $ {\mu}_{\rm INTER} $ and $ {\sigma}_{\rm INTER} $ measured under the nominal operating condition (1.20~V, $25\celsius$) is used, as shown in Fig.~\ref{fig:ROIntraInter}. Notably, the $ {\mu}_{\rm INTER} $ is not ideally equal to 0 that thereby induces a severe bias, in particular, response `1' to be 36.65\%. To measure $ {\sigma}_{\rm INTRA} $ that is the variance introduced by including noise and voltage, temperature deviation from that of nominal operating condition, the frequency subtraction given all response bits are measured under the nominal operating condition (1.20~V, $25\celsius$) treated as a {\it reference}, then frequency subtraction given all response bits are measured {\it again} under a deviating operating condition, e.g., (0.96~V, $25\celsius$). The change between these two measurements is assessed. 
The standard deviation is recognized as $ {\sigma}_{\rm INTRA} $. In Fig.~\ref{fig:ROIntraInter}， the $ {\sigma}_{\rm INTRA} $ evaluated under the worst-corner of (0.96~V, $ 25\celsius $) is shown. Table.~\ref{tab:ROIntraInter} lists $ \lambda_1=\frac{{{\sigma}_{\rm INTRA}}}{{\sigma}_{\rm INTER}} $ evaluated under different operating conditions and $ \lambda_2=\frac{{{\mu}_{\rm INTER}}}{{\sigma}_{\rm INTER}} $. 
		
\begin{table}
	\centering 
	\caption{$ \lambda_1 $ and $ \epsilon $ measured under different operating conditions and the $ \lambda_2 $. Referenced operating condition is (1.20~V, $25\celsius$).}
			\resizebox{0.30\textwidth}{!}{
	\begin{tabular}{c|| c || c || c || c} %
		\toprule 
		\toprule 
				
		Operating condition & $ \epsilon $ & $\sigma_{\rm INTRA}$ & $ \lambda_1 $ & $ \lambda_2 $ \\ 
		\midrule
		($ 25\celsius $, 0.96~V) & 9.66\% & 0.8006 & 0.3672 & -0.3477\\
		($ 25\celsius $, 1.08~V) & 4.68\% & 0.4248 & 0.1933 & -0.3477\\
		($ 25\celsius $, 1.20~V) & 0.98\% & 0.0523 & 0.0239 & -0.3477\\
		($ 35\celsius $, 1.20~V) & 1.82\% & 0.1627 & 0.0728 & -0.3477\\
		($ 45\celsius $, 1.20~V) & 1.88\% & 0.1569 & 0.0700 & -0.3477\\
		($ 55\celsius $, 1.20~V) & 2.08\% & 0.1741 & 0.0795 & -0.3477\\
		($ 65\celsius $, 1.20~V) & 2.17\% & 0.1933 & 0.0881 & -0.3477\\
		($ 25\celsius $, 1.32~V) & 4.70\% & 0.4729 & 0.2151 & -0.3477\\
		($ 25\celsius $, 1.44~V) & 6.42\% & 0.7182 & 0.3231 & -0.3477\\
		\bottomrule
	\end{tabular}
			}
	\label{tab:ROIntraInter} 
\end{table}

\vspace{0.3cm}
{\bf Results:} Empirical and statistical results of FRR are shown in Table~\ref{tab:AuthenTabDefaultROPUF}. They agree well. 
 The FRR decreases as the $ m $ increases. In addition, the FRR is minimizing when the $ \bf \tilde{e} $ is regenerated under an operating condition that is close to the nominal operating condition. In other words, {\it the FRR gets worse as $\sigma_{\rm INTRA}$ goes up}.
		
One may note that the statistical results show a conservative assessment of the FRR in comparison with the empirical results. The reason is that the response has a bias, response `1' to be 36.65\%, in this ROPUF case. In other words, Confidence in Fig.~\ref{fig:confidence} does not follow a normal distribution with a {\it mean value of zero}, instead $\mu_{\rm INTER}$ deviates from zero, which explains the conservative assessments~\cite{herder2017trapdoor}. 

\begin{table}[t]
	\centering 
	\caption{FRR of the ROPUF under different operating conditions and $ m $ settings, where $ k=64 $. Referenced operating condition is (1.2 V, $25\celsius$).}
	\resizebox{0.50\textwidth}{!}{
	\begin{tabular}{c|| c || c || c || c || c}
		\toprule 
		\toprule 
		&\multicolumn{1}{c}{$ 25\celsius $, 0.96~V} & \multicolumn{1}{c}{$ 25\celsius $, 1.08~V}& \multicolumn{1}{c}{$ 65\celsius $, 1.20~V}& \multicolumn{1}{c}{$ 25\celsius $, 1.32~V}& \multicolumn{1}{c}{$ 25\celsius $, 1.44~V} \\
		\cmidrule(l){2-2} \cmidrule(l){3-3} \cmidrule(l){4-4}\cmidrule(l){5-5}\cmidrule(l){6-6}
		$ m $ & FRR & FRR & FRR & FRR & FRR \\ 
		\midrule
		$ 12 $ & 85.12\%; 90.11\% & 21.23\%; 37.21\% & 1.04\%; 4.95\% & 20.57\%; 46.16\% & 48.95\%; 83.48\% \\ 
		$ 14 $ & 77.72\%; 84.23\% & 13.69\%; 25.98\% & 0.37\%; 4.08\% & 12.78\%; 34.63\% & 37.30\%; 75.10\% \\ 
		$ 16 $ & 69.35\%; 77.89\% & 8.12\%; 17.09\% & 0.14\%; 3.52\% & 7.41\%; 25.71\% & 26.97\%; 66.32\% \\ 
		$ 18 $ & 60.11\%; 69.80\% & 4.68\%; 10.78\% & 0.03\%; 3.19\% & 4.12\%; 17.08\% & 18.08\%; 56.87\% \\ 
		$ 20 $ & 50.64\%; 61.17\% & 2.43\%; 6.36\% & 0.01\%; 2.89\% & 2.23\%; 11.04\% & 11.53\%; 46.36\% \\ 
		$ 22 $ & 41.02\%; 51.01\% & 1.29\%; 3.93\% & 0.01\%; 2.59\% & 1.13\%; 7.06\% & 7.26\%; 37.05\% \\ 
		$ 24 $ & 32.08\%; 43.10\% & 0.63\%; 2.05\% & 0\%; 2.33\% & 0.54\%; 4.07\% & 4.21\%; 29.38\% \\ 
		$ 26 $ & 24.75\%; 33.93\% & 0.23\%; 1.27\% & 0\%; 2.10\% & 0.28\%; 2.58\% & 2.20\%; 20.93\% \\ 
		\midrule
		\bottomrule 
	\end{tabular}}
	\label{tab:AuthenTabDefaultROPUF} 
	\begin{tablenotes}
	\item{The FRR from empirical evaluations and FRR statistical analyses based on Eq(\ref{Eq:FRR}) are listed for comparison, where the format is (empirical; statistical).}
	\end{tablenotes}
\end{table}
			
\subsubsection{Validation with a LAPUF}\label{Sec:ExpTREVERSE-B}
We use the $k$-sum ROPUF as a representative of the LAPUF to validate the TREVERSE-B. 

\vspace{0.30cm}
{\bf LAPUF dataset description:}
Frequency measurements of all ROs of Virginia Tech's ROPUF are leveraged to form $k$-sum ROPUF. We evaluated five $ k $-sum ROPUFs; each of them is constructed in one FPGA board by using 128 ROs---$ k=64 $. The frequency summation and consequent comparison are post-processed using MATLAB. Among five evaluated $ k $-sum ROPUFs, the most noisy $ k $-sum ROPUFs with worst-case BER of 14.53\% occurred at (0.96~V, $ 25\celsius $), 
while the (1.20~V, $ 25\celsius $) acts as the nominal operating corner. For convenience, we will henceforth refer to the $ k $-sum ROPUF as LAPUF.

\vspace{0.30cm}
{\bf Extraction of $\lambda_1$ and $\lambda_2$ from LAPUF model:}
The reliability confidence information of the LAPUF is {\it predicted} by the LAPUF model that serves as \textsf{SimPUF}. We use 10,000 challenges to extract $\lambda_1$ and $\lambda_2$.
			
The reliability confidence of the response bit of the LAPUF is the frequency subtraction (difference) between the top and bottom RO rows. By predicting frequency differences given all response bits through the LAPUF model under the nominal operating condition and plotting all frequency differences, the standard variance is recognized as $ {\sigma}_{\rm INTER} $ and the mean is the $ {\mu}_{\rm INTER} $.
			
To extract $ {\sigma}_{\rm INTRA} $, the frequency differences given all response bits are {\it predicted} by the LAPUF model trained by CRPs evaluated under the nominal condition (1.20~V, $25\celsius$) as a {\it reference}, then frequency differences for all the same response bits are predicted again by the LAPUF model trained with CRPs evaluated under a {\it differing} operating condition, e.g., (0.96~V, $25\celsius$). The change between these two evaluations is calculated. By plotting 10,000 frequency changes, a distribution is obtained. Then its standard deviation is recognized as the $ {\sigma}_{\rm INTRA} $. Once the ${\sigma}_{\rm INTER} $, ${\mu}_{\rm INTER} $ and ${\sigma}_{\rm INTRA} $ are acquired, the $\lambda_1$ and $\lambda_2$ can be directly determined. 

\vspace{0.3cm}
{\bf Results:}
Empirically and statistically evaluated FRR of the LAPUF are shown in Table~\ref{tab:AuthenTabDefaultLAPUF}. 
Instead of the empirical FRR  always being smaller than the statistical FRR as is the case for the ROPUF in Table~\ref{tab:AuthenTabDefaultROPUF}, they almost perfectly agree with each other for the LAPUF.
Recall that the conservative statistical FRR for ROPUF is induced by the response bias 36.65\% of the ROPUF, while the response bias of the investigated LAPUF is equal to 50.05\%. Therefore, the statistical FRR accurately reflects the empirical FRR {\it even though the statistical FRR evaluation of the LAPUF is built upon learned LAPUF models}. 
\begin{table} 
	\centering 
	\caption{FRR of the LAPUF under different operating conditions and $ m $ settings, where $ k=64 $. Referenced operating condition (1.20~V, $25\celsius$).}
	\resizebox{0.50\textwidth}{!}{
		\begin{tabular}{c|| c || c || c || c || c} %
			\toprule 
			\toprule 
			&\multicolumn{1}{c}{$ 25\celsius $, 0.96~V} & \multicolumn{1}{c}{$ 25\celsius $, 1.08~V}& \multicolumn{1}{c}{$ 65\celsius $, 1.20~V}& \multicolumn{1}{c}{$ 25\celsius $, 1.32~V}& \multicolumn{1}{c}{$ 25\celsius $, 1.44~V} \\%
			\cmidrule(l){2-2} \cmidrule(l){3-3} \cmidrule(l){4-4}\cmidrule(l){5-5}\cmidrule(l){6-6}
			$ m $ & FRR & FRR & FRR & FRR & FRR \\ 
			\midrule
			$ 12 $ & 98.73\%; 98.00\% & 59.59\%; 56.87\% & 6.02\%; 6.21\% & 44.31\%; 39.21\% & 75.20\%; 73.37\% \\ 
			$ 14 $ & 97.55\%; 96.37\% & 48.38\%; 45.14\% & 2.60\%; 3.72\% & 32.42\%; 30.10\% & 65.82\%; 62.74\% \\ 
			$ 16 $ & 95.57\%; 93.83\% & 36.98\%; 34.97\% & 1.03\%; 2.63\% & 22.73\%; 20.50\% & 55.72\%; 52.83\% \\ 
			$ 18 $ & 93.16\%; 90.37\% & 27.01\%; 26.02\% & 0.51\%; 2.08\% & 15.10\%; 14.32\% & 45.56\%; 42.71\% \\ 
			$ 20 $ & 89.56\%; 85.35\% & 19.06\%; 17.37\% & 0.19\%; 1.81\% & 9.33\%; 8.42\% & 34.95\%; 33.56\% \\ 
			$ 22 $ & 84.79\%; 80.29\% & 12.64\%; 12.52\% & 0.05\%; 1.62\% & 5.49\%; 5.09\% & 26.21\%; 24.52\% \\ 
			$ 24 $ & 79.25\%; 72.50\% & 8.17\%; 7.57\% & 0.05\%; 1.44\% & 3.07\%; 2.89\% & 18.80\%; 17.74\% \\ 
			$ 26 $ & 72.50\%; 66.00\% & 4.85\%; 4.82\% & 0.02\%; 1.29\% & 1.64\%; 1.69\% & 13.14\%; 12.28\% \\ 
						
			\midrule
			\bottomrule 
		\end{tabular}}
		\label{tab:AuthenTabDefaultLAPUF} 
		\begin{tablenotes}
			\item{The FRR format is same with that of Table~\ref{tab:AuthenTabDefaultROPUF}.}
		\end{tablenotes}
\end{table}

\subsubsection{Summarize}
Our extensive 
analysis of the ROPUF and LAPUF validate the derived FRR, which reflects the empirical FRR well. It is worth to note that the statistical FRR is a bit conservative evaluation of the empirical FRR when the PUF is with a moderate bias. 
\subsubsection{Remark}
To accurately assess the FRR, two crucial parameters $ \lambda_1 $ and $ \lambda_2 $ (as function of $ {\sigma}_{\rm INTER} $, $ {\sigma}_{\rm INTRA} $, $ {\mu}_{\rm INTER} $, with $ \lambda_1=\frac{{\sigma}_{\rm INTRA}}{{\sigma}_{\rm INTER}} $ and $ \lambda_2=\frac{\mu_{\rm INTER} }{\sigma_{\rm INTER} }$) are required. As we have experimentally showcased, determination of $ {\sigma}_{\rm INTER} $, $ {\sigma}_{\rm INTRA} $, $ {\mu}_{\rm INTER} $ is easy and only a one-time task to the server at the PUF provisioning phase. To be precise, it is just enrolling two \textsf{SimPUF}s given the same PUF but under two operating conditions---one under nominal operating condition, the other under (expected) worse-case operating corner.
\section{Authentication Capability Evaluations}\label{Sec:AugAuthen}
The previous section formalizes the FAR and FRR, and validates the accuracy of the FRR that is derived from an accurate PUF reliability model. In this section we 
analyze the authentication capability 
for both ROPUF and LAPUF.

From a practical perspective it is very important to have a small $ m $ because $ m $ stands for the computation overhead of the server. The smaller $m$, the less computation for the server to authenticate a single token. In this section, we present two simple, efficient and {\it compatible} methods to implement TREVERSE authentication in practice to achieve industry accepted false rejection rates. 

\subsection{$ d $-Authentication}
The first method is $ d $-authentication. In one {\it authentication session}, the server issues $ d $ challenges $ \{{\bf c}_{1},...,{\bf c}_{d} \}$. The prover sequentially returns $ d $ outputs $ \{ {\bf \tilde{r}}_1,...,{\bf \tilde{r}}_d \} $. 
The server verifies the authenticity of each received output in sequential order. Authentication succeeds once a received output is accepted after which the server stops checking the rest. 
If {\it none} of the $ d $ received outputs can pass the authentication, then this authentication session is rejected. By assuming each received output is independent, the FRR of $ d $-authentication, $\rm FRR_{d}$, is given as:
\begin{equation}\label{Eq:FRRd}
{\rm FRR}_{d}={\rm FRR}^d
\end{equation}
Detailed results of ${\rm FRR}_{d}$ of ROPUF and LAPUF are in the Appendix.

${\rm FAR}_d$ is computed as:
\begin{equation}
{\rm FAR}_d=d\times{\rm FAR}.
\end{equation}
We can see that $ d $-authentication {\it linearly} increases FAR while {\it exponentially} minimizes FRR. 
				
\subsection{Multiple-Reference}\label{Sec:MultiRef}
The second method explores multiple reference responses to enhance the authentication capability. Overall, at the provisioning phase, multiple responses $ {\bf e}_{\rm ref} $ and their corresponding $ {\bf conf}_{\rm ref} $ under discrete operating corners subject to the same challenges applied to the same PUF are enrolled. In other words, instead of enrolling one \textsf{SimPUF} under only one operating condition---nominal condition, multiple \textsf{SimPUF}s are enrolled under discrete operating corners. 
				
Taking ROPUF as an example to ease understanding, during the enrollment phase, the frequencies of all ROs are measured under two operating corners: ($ 25\celsius $, 1.08~V) and ($ 25\celsius $, 1.32~V). As a consequence, the $ {\bf e}_{\rm ref_1},~ {\bf e}_{\rm ref_2}$ and their corresponding $ {\bf conf}_{\rm ref_1},~{\bf conf}_{\rm ref_2}$ are obtained, where ref$_1$ and ref$_2$ are operating corners of (25\celsius,1.08~V) and (25\celsius,1.32~V), respectively. During the authentication, for each received $ \bf \tilde{r} $, TREVERSE authentication is performed on $ \bf \tilde{r} $ based on both $ {\bf e}_{\rm ref_1},~ {\bf e}_{\rm ref_2}$ at the same time. If any one of two TREVERSE authentications passes---$ {\bf \tilde{r}}={\bf r}_{\rm ref_1} $ or $ {\bf \tilde{r}}={\bf r}_{\rm ref_2} $, the authentication succeeds. This helps decreasing the FRR significantly. 

By assuming each referenced response is independent\footnote{Empirical results of ${\rm FRR}_{mr}$ of ROPUF and LAPUF in the Appendix show that this independence assumption is appropriate.}
the FRR when multiple-reference is utilized, denoted as FRR$_{\rm mr}$, is formalized as:
\begin{equation}\label{Eq:FRRM}
{\rm FRR}_{\rm mr}=\prod_{i=1}^{M} {\rm FRR}_{\rm ref_i}
\end{equation}
with ${\rm FRR}_{\rm ref_i}$ is the FRR when a single referenced operating corner is used for TREVERSE authentication. $M$ is the number of referenced operating corners enrolled. 

As for the FAR$_{mr}$, it only {\it linearly} increases in $M$:
\begin{equation}\label{Eq:FARM}
{\rm FAR}_{\rm mr}={\rm FAR}\times M
\end{equation}

\subsection{Merging $d$-Authentication and Multiple-Reference}\label{sec:mergingDM}

\begin{algorithm}[t]
	\small
	\caption{$d$-authentication and multiple-reference}
	\label{Algorithm:Augment}
	\begin{algorithmic}[1]
    \State \Comment Server has enrolled multiple SimPUFs, with number of $M$, under $M$ different operating conditions for the same PUF.
	\Procedure{$\mathbf{Augment Authenticate}$~} {$M$ SimPUFs}
    \State $\mathit{authState}\leftarrow$ Fail
	\For{$ i=1:d $}
	\State Send challenge ${\bf c}_i$ to prover and receive $\tilde{\bf r}_i=$\textsf{OWF}($\tilde{\bf e}_i,{\bf n}_i$) from the prover
		\For{$ j=1:M $}
		\State using SimPUF$_j$ to perform TREVERSE according to Algorithm~\ref{Algorithm:TAE}
		\If {Success}
		\State \Return $authState\leftarrow$Success \Comment{Stop executing---escape from both the inner and out for loop}
		\EndIf
		\EndFor
	\EndFor 
	\EndProcedure		
	\Statex
\end{algorithmic}
\vspace{-0.4cm}%
\end{algorithm}
The above two augmented authentication methods are compatible with each other. They can be implemented together and the steps are detailed in Algorithm~\ref{Algorithm:Augment}. Generally, if the TREVERSE fails by using a single output $\tilde{\bf r}$ from the prover, the server can ask the prover to refresh a new $\tilde{\bf r}$ according to the server issued new challenge $\bf c$. The number of refreshment is preset to be no more than $d$. For each $\tilde{\bf r}_i, i\in {1,...,d}$, the server sequentially uses one of $M$ enrolled SimPUFs---each SimPUF is enrolled under a different operating condition given the same PUF---to perform TREVERSE according to the algorithm~1. If it successes, then the server stops to i) use the rest SimPUFs to perform TREVERSE and ii) ask further $\tilde{\bf r}$ refreshments from the prover. Otherwise, the server continues to use the SimPUFs to perform TREVERSE and ask for further $\tilde{\bf r}$ refreshments. If after i) $d$ times of $\tilde{\bf r}$ refreshments and ii) performing TREVERSE through all $M$ SimPUFs for each $\tilde{\bf r}$ refreshment, the authentication is still not successful, then this authentication session fails.

When both methods are implemented together, the FRR and FAR are termed as FRR$_{\rm M,d}$ and FAR$_{\rm M,d}$ and expressed as:
\begin{equation}\label{Eq:FRRMd}
{\rm FRR}_{\rm M,d}=(\prod_{i=1}^{M} {\rm FRR}_{\rm ref_i})^d. 
\end{equation}
				
\begin{equation}
{\rm FAR}_{\rm M,d}={\rm FAR}\times M\times d. 
\end{equation}
				
$M$ stands for the number of referenced responses and $d$ is the number of authentication rounds used during one authentication session. ${\rm FRR}_{\rm ref_i}$ is given by Eq(\ref{Eq:FRR}) and ${\rm FAR}$ is given by Eq(\ref{Eq:FAR}).

When an exhaustive trial and error approach is adopted, the maximum number of trials is equal to
\begin{equation}
N_{\rm worst}=2^m\times M \times d.
\end{equation} 

\begin{figure}
	\centering
	\includegraphics[trim=0 0 0 0,clip,width=0.3\textwidth]{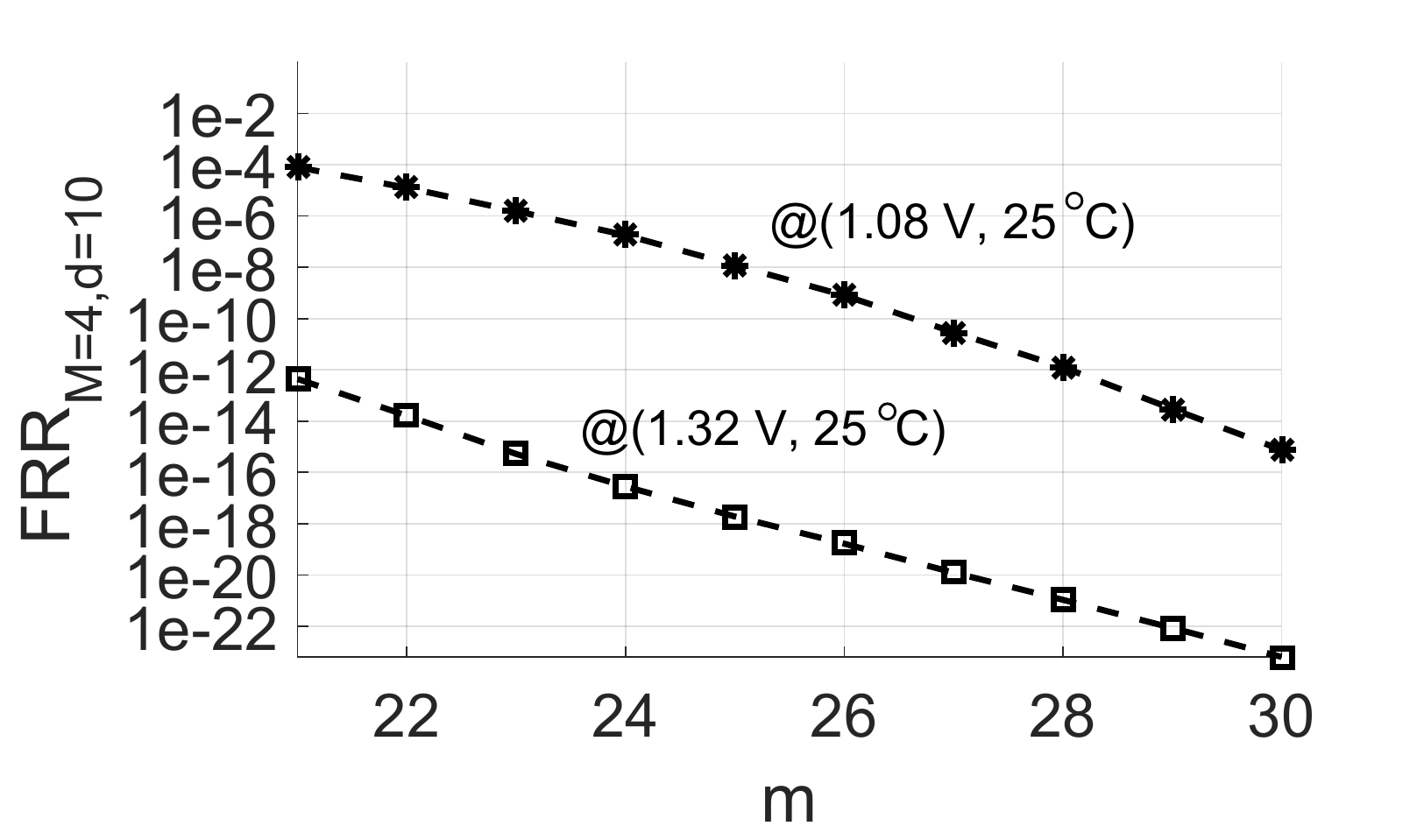}
	\caption{The ${\rm FRR}_{\rm M,d}$ of ROPUF with ($M=4$, $d=10$ and $k=110$). Four reference responses are from (0.96 V, 25$\celsius$), (1.20 V, 25$\celsius$), (1.20 V, 65$\celsius$) and (1.44 V, 25$\celsius$).}
	\label{fig:FRR_ROPUF_4_10}
\end{figure}

\begin{figure}
	\centering
	\includegraphics[trim=0 0 0 0,clip,width=0.3\textwidth]{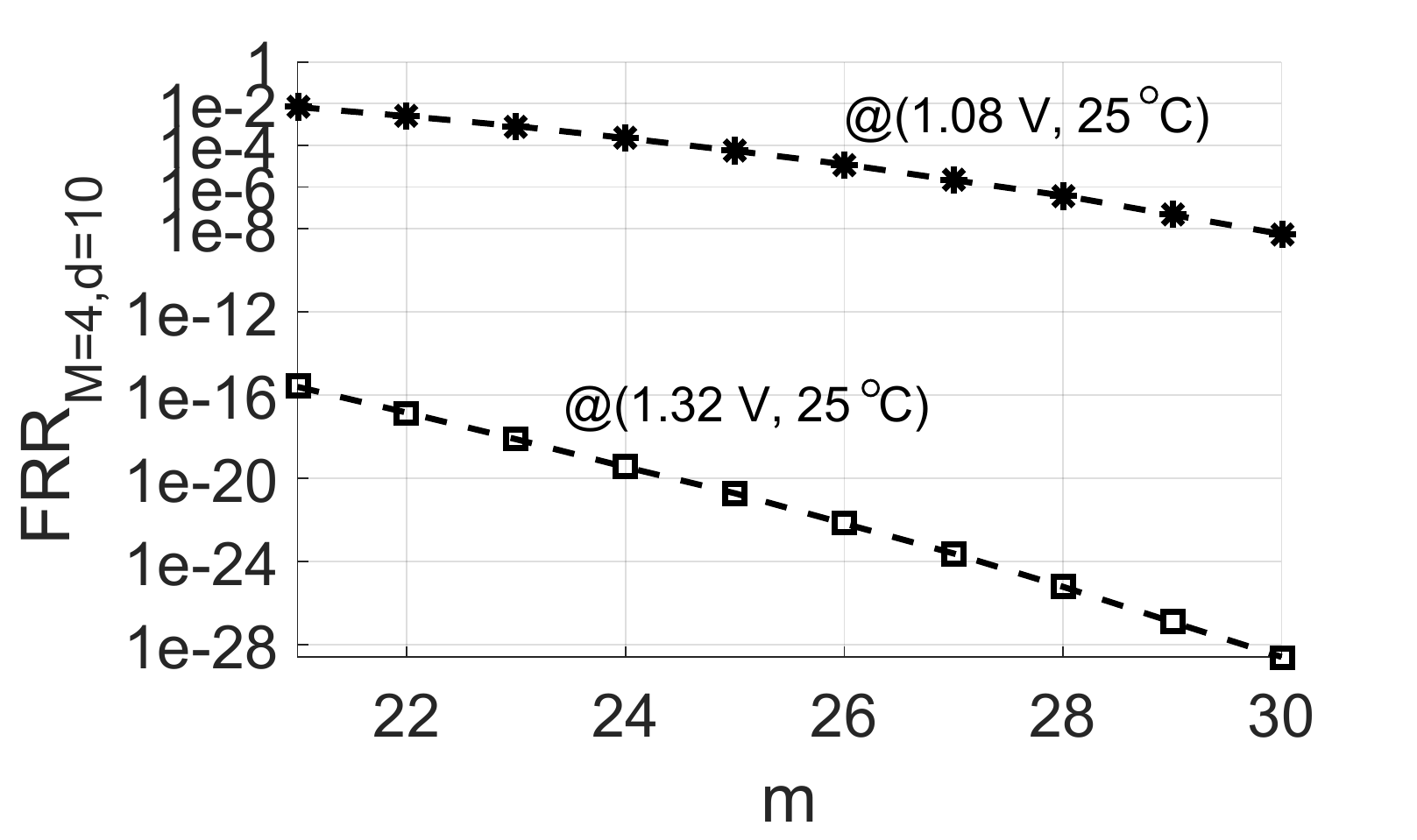}
	\caption{The ${\rm FRR}_{\rm M,d}$ of LAPUF with ($M=4$, $d=10$ and $k=110$). Four reference responses are from (0.96 V, 25$\celsius$),  (1.20 V, 25$\celsius$), (1.20 V, 65$\celsius$) and (1.44 V, 25$\celsius$).}
	\label{fig:FRR_LAPUF_4_10}
\end{figure}
Both ROPUF and LAPUF are extensively tested when $d$-authentication and multiple-reference are employed together. The ${\rm FRR}_{\rm M,d}$ of ROPUF and LAPUF are detailed in Fig~\ref{fig:FRR_ROPUF_4_10} and Table~\ref{fig:FRR_LAPUF_4_10}, respectively, where ($M=4$, $d=10$ and $k=110$). The four reference responses are from (0.96 V, $25\celsius$),  (1.20 V, $25\celsius$), (1.20 V, $65\celsius$) and (1.44 V, $25\celsius$)\footnote{We are limited by fine-grained reference in the tested public dataset. Once more fine-grained reference are employed in practice, the $m$ will be greatly decreased given the fixed FRR. Herein, we constrain our tests on the public dataset.}. Under such setting, the worst case $N_{\rm worst}$ is around $2^{27}$ and $2^{29}$ for ROPUF to achieve ${\rm FRR}_{\rm M,d} < 10^{-3}$ and ${\rm FRR}_{\rm M,d} < 10^{-6}$, respectively. The $N_{\rm worst}$ is around $2^{29}$ and $2^{31}$ for LAPUF to achieve ${\rm FRR}_{\rm M,d} < 10^{-3}$ and ${\rm FRR}_{\rm M,d} < 10^{-6}$ because the LAPUF is more noisy than the ROPUF---14.53\% worst $\epsilon$ of LAPUF versus 9.66\% worst $\epsilon$ of ROPUF.

\section{Comparison and Discussion}\label{Sec:Comparison}
\subsection{Comparison}
Table~\ref{tab:AuthenComp} includes six finalists out of 21 PUF-based authentication protocols examined by Delvaux (Chapter 5.4)~\cite{delvaux2017security}. Whereas protocols that i) do not offer {\it any} resistance to both
noise and machine learning attacks, ii) are vulnerable to conventional protocol attacks and iii) are not suitable for implementation purposes, are excluded. We notice that Slender PUF~\cite{rostami2014robust} and Lockdown~\cite{yulockdown} can resist conventional modeling attacks e.g., SVM, logistic regression, but are vulnerable to other forms of modeling attacks~\cite{tobisch2015scaling,becker2015pitfalls}. Delvaux~\cite{delvaux2017machine} recently reviewed the other 5 up-to-date PUF-based authentication protocols, which we also include. We compare TREVERSE with the PUF based authentications listed in Table~\ref{tab:AuthenComp}.

From Table~\ref{tab:AuthenComp}, we can see without \textsf{OWF}, it is extremely hard, if not impossible, to resist modeling attacks (e.g., the entries from Slender to LHS-PUF) unless the lockdown technique (detailed in Section~\ref{sec:related-smiliarity-measure}) is deployed to explicitly limit the CRPs available by the adversary. Without adequate CRP for training, an accurate model is infeasible to be learned, thus, preventing modeling attacks. However, lockdown inevitably sacrifices practicality of the PUF authentication because only a limited number of authentications, e.g., 1000 can be securely issued. Afterwards, the PUF integrated token has to be disposed to maintain security. 

When \textsf{OWF} is used (e.g., the entries from Sadeghi to Controlled), ECC logic and associated helper data are always required. Unfortunately, the ECC logic is very expensive, see experimental validation in Section~\ref{sec:lightweight}. Moreover, the helper data has to be carefully handled to avoid information leakage and helper data manipulation attacks~\cite{delvaux2015helper,robust2017becker}. Regarding to the PUF-FSM that does remove the ECC logic, it is, however, still vulnerable to modeling attack~\cite{delvaux2017machine} because of the response reliability information (highly reliable responses) is leaked.

Overall, we can see that the on-chip TRNG is a common building block in the PUF based authentication. Indeed, without it, mutual authentication realization is impossible.

According to Delvuax's aftermath in~\cite{delvaux2017machine}: "A fairly conservative approach to craft a PUF-based authentication protocol is to convert a noisy response into a stable secret key and then use a keyed cryptographic algorithm to perform the authentication". Our TREVERSE, to a large extent, falls into the keyed cryptographic approach but we counter-intuitively remove the requirement of expensive ECC logic and the worries on securely handling of the helper data. In comparison with the lockdown, we eschew the fetter of limited authentications by exploiting a \textsf{OWF} that is still lightweight and further significantly enhances the security.

Overall, TREVERSE, for the first time, ultimately enables a generic PUF based authentication mechanism to be comparable with the classical digital key based dynamic authentication as shown in Fig.~\ref{fig:serverNVM} since the noisy PUF response is oblivious to the prover and is treated as a digital key. We would like to remind that whenever mutual authentication is built upon a digital key, the on-chip TRNG is also in need in order to provide token side freshness.

\begin{table}
	\centering 
	\caption{PUF based Authentication Comparison}
	\resizebox{0.40\textwidth}{!}{
	\begin{tabular}{c| c | c | c | c | c | c | c | c | c} %
					
		& \rotatebox{90}{token authenticity} & \rotatebox{90}{server authenticity} & \rotatebox{90}{TRNG} & \rotatebox{90}{ECC logic} & \rotatebox{90}{\textsf{OWF}} & \rotatebox{90}{Authentication rounds} &  \rotatebox{90}{PUF independent} & \rotatebox{90}{Modeling resistance} & \rotatebox{90}{Comparable to Class. Authen.$^a$}  \\ 
		\midrule
		Slender~\cite{rostami2014robust} & \checkmark & $\times$ & \checkmark & $\times$ & $\times$ & $\infty$ & $\times$ & $\times$ & N/A \\
        Noise bifur.~\cite{yu2014noise} & \checkmark & $\times$ & \checkmark & $\times$ & $\times$ & $\infty$ & $\times$ & $\times$ & N/A \\
        PolyPUF~\cite{konigsmark2016polypuf} & \checkmark & $\times$ & \checkmark & $\times$ & $\times$ & $\infty$ & $\times$ & $\times$ & N/A \\
        OB-PUF~\cite{gao2016obfuscated} & \checkmark & $\times$ & \checkmark & $\times$ & $\times$ & $\infty$ & $\times$ & $\times$ & N/A \\
        RPUF~\cite{ye2016rpuf} & \checkmark & $\times$ & \checkmark & $\times$ & $\times$ & $\infty$ & $\times$ & $\times$ & N/A \\
        LHS-PUF~\cite{idriss2017lightweight} & \checkmark & \checkmark & \checkmark & $\times$ & $\times$ & $\infty$ & $\times$ & $\times$ & N/A \\
         Lockdown I~\cite{yulockdown} & \checkmark & $\times$ & $\times$ & $\times$ & $\times$ & $<d$ & \checkmark & \checkmark & N/A \\
        Lockdown II~\cite{yulockdown} & \checkmark & \checkmark & \checkmark & $\times$ & $\times$ & $d$ & \checkmark & \checkmark & N/A \\
        \midrule
        Sadeghi~\cite{sadeghi2010enhancing} & \checkmark & $\times$ & \checkmark & \checkmark & \checkmark & $\infty$ & \checkmark & \checkmark & $\times$ \\
        Rever. FE II~\cite{maes2013physically} & \checkmark & \checkmark & \checkmark & \checkmark & \checkmark & $\infty$ & \checkmark & \checkmark & $\times$ \\
         Controlled~\cite{gassend2002controlled} & \checkmark & $\times$ & $\times$ & \checkmark & \checkmark & $\infty$ & $\times$ & \checkmark & $\times$ \\
         PUF-FSM~\cite{gao2018puf} & \checkmark & \checkmark & \checkmark & $\times$ & \checkmark & $\infty$ & $\times$ & $\times$ & $\times$ \\
       TREVERSE-A & \checkmark & $\times$ & $\times$ & $\times$ & \checkmark & $\infty$ & $\times$ & \checkmark & \checkmark\\
        TREVERSE-B & \checkmark & \checkmark & \checkmark & $\times$ & \checkmark & $\infty$ & \checkmark & \checkmark & \checkmark \\
		\bottomrule
	\end{tabular}}
      \begin{tablenotes}
      \small
      \item $^a$ Is the PUF authentication instantiation on the token comparable to the classical key based entity/client dynamic authentication as depicted in Fig.~\ref{fig:serverNVM}?. In this context, to be fair, we only compare with those PUF key based authentication as all of them require \textsf{OWF}.
    \end{tablenotes}
	\label{tab:AuthenComp} 
\end{table}
\subsection{Discussion}
\subsubsection{Server and Prover Overhead}\label{sec:lightweight}
We evaluate the overhead of the TREVERSE from the server and prover side, respectively.

\begin{figure}[h]
	\centering
	\includegraphics[trim=0 0 0 0,clip,width=0.40\textwidth]{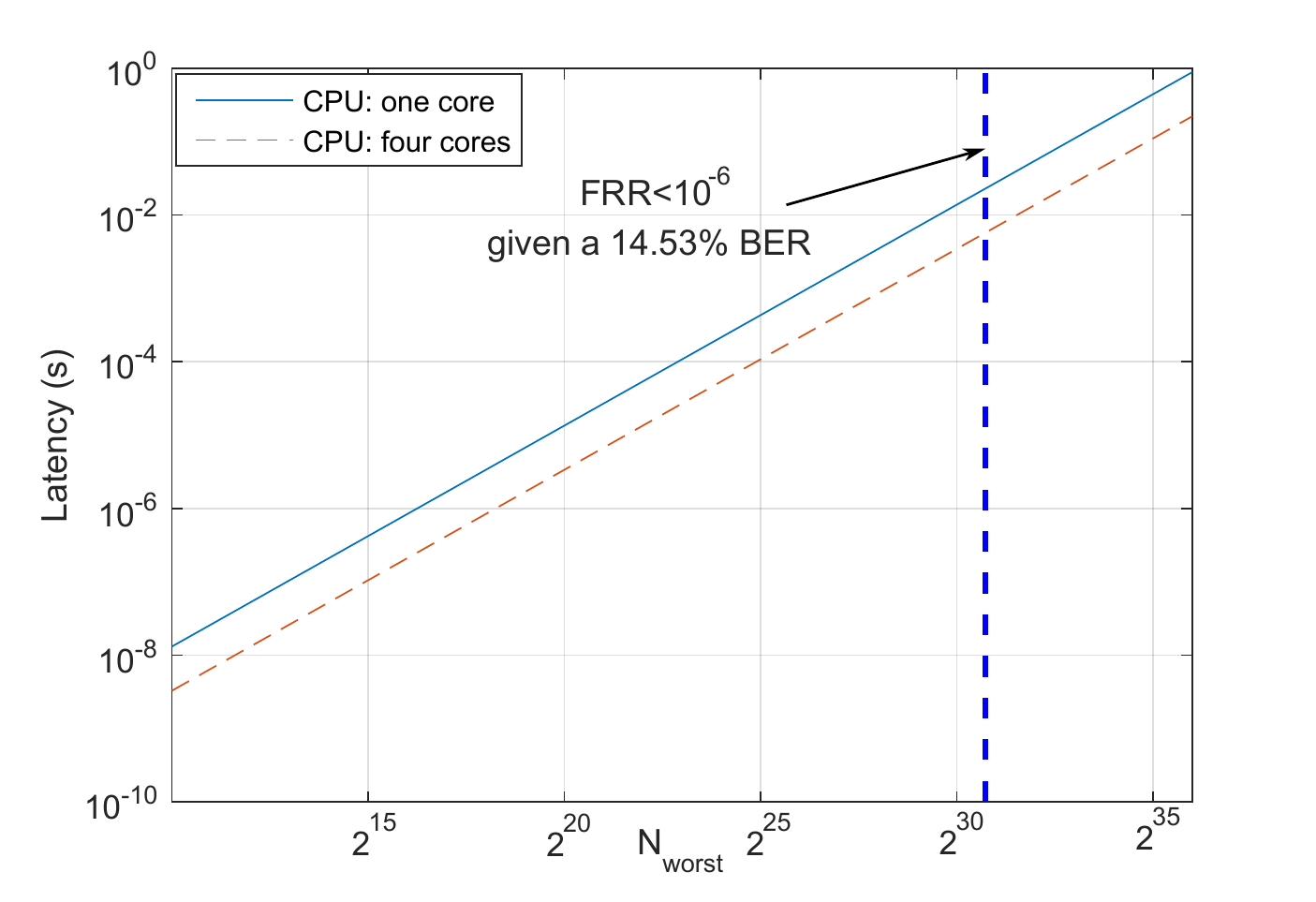}
	\caption{Estimated server time latency $T_s$ as a function of the worst-case trial numbers $N_{\rm worst}$ under a common personal computer computational power settings, detailed in Appendix~\ref{App:serverSet}. }
	\label{fig:LatencyServer}
\end{figure}

\noindent{\bf Server Overhead:} The server setting is detailed in Appendix~\ref{App:serverSet}. Generally, the time $T_s$ required for the server to complete one TREVERSE authentication session can be estimated by:
\begin{equation}
    T_s = N_{\rm worst} \times k \times \frac{1}{8} \times \frac{1}{{\rm Hash}_{\rm speed}} \times  \frac{1}{N_{\rm GPUcore}} \times  \frac{1}{ N_{\rm CPUcore}} 
\end{equation}
where $N_{\rm worst}$ the number of trials that the server needs to perform in the worst case, $k$ the length of response $\bf e$, and $N_{\rm GPUcore}$ and $N_{\rm CPUcore}$ the number of GPU cores and the CPU cores, respectively, equipped by the server. The Fig.~\ref{fig:LatencyServer} illustrates the $T_s$ as a function of $N_{\rm worst}$ given the settings of $k=128$, ${\rm Hash}_{\rm speed} = 648$ MiBps, number of GPU cores of 1920, and number of CPU core of one and four, respectively. As an example, to tolerate a 14.53\% BER of the LAPUF while maintaining the FRR $< 10^{-6}$, $2^{31}$ trials in worst case is required by the server, which only needs 28 ms and 7 ms when employing a CPU with a single core and four cores, respectively.

\noindent{\bf Prover ECC vs TREVERSE Overhead:} We choose a MSP430FR5969 microcontroller and BCH code as error correction code (ECC) to evaluate the prover overhead in terms of clock cycles, detailed prover settings and descriptions are in Appendix~.\ref{App:proverSetting}. In general, the ECC enabled authentication builds upon either through reverse fuzzy extractor (RFE) that employs the ECC encoding on-chip or fuzzy extractor (FE) that employs the ECC decoding on-chip. Based on the settings detalied in Appendix~.\ref{App:proverSetting}.
the time latency for the RFE in the prover side (MSP430FR5969 microcontroller) costs 2.55 s according to the Table~\ref{tab:FEandRFE} if it is required to achieve a $10^{-6}$ FRR given a BER of 14.53\%. Please note that for such resource-constraint provers, the time latency of implementing the FE that employs ECC decoding tends to be unacceptable, e.g., could take more than 100 s.

According to the evaluation results in Section~\ref{sec:mergingDM}, the prover may run $d=10$ times successively in the worst case under the $d$-authentication to lower the FRR $<10^{-6}$ tolerating the worst-case BER of 14.53\% of the LAPUF. Therefore, the clock overhead for the prover in the worst case is performing $d=10$ times hash operations, which cost 1,047,230 clock cycles, taking about 1 s. We can see that the TREVERSE greatly outperforms the ECC based authentication---reducing by 60\%---even when the lightweight RFE~\cite{van2012reverse} rather the common FE is employed.

Besides the overhead, the TREVERSE inherently removes potential security vulnerabilities induced from ECC and associated helper data as detailed in the next Section~\ref{Sec:security}.

In fact, the server can choose to increase $m$, thus, number of trials, to further decrease the $d$ needed in the prover side, which could reduce the worst-case overhead (clock cycles) of the prover. In addition, we can see that in practice, the latency bottleneck seems usually located in the prover side rather the server side. Therefore, it is desirable to reduce the prover side overhead as much as possible considering the fact that the server side can be powerful and flexibly configured, which is what the TREVERSE is motivated and designed for.




\subsubsection{Security}\label{Sec:security}
In comparison with PUF authentication without \textsf{OWF}, our TREVERSE is straight forwardly secure against modeling attacks because of the hardness of inverting the \textsf{OWF}. In comparison with PUF authentication with \textsf{OWF} which requires ECC logic and thus helper data, our TREVERSE is also more secure because there is no ECC and associated helper data involved.

Helper data manipulation (HDM) attacks~\cite{delvaux2015helper,robust2017becker} have demonstrated various error correction code and the decoding strategy of the code's implementation are vulnerable. A generic countermeasure against recent Becker's HDM attacks does not yet exist, appears to be an open challenge, especially to the robust fuzzy extractor~\cite{robust2017becker}. In this context, we are only aware, to date, that the linear BCH code based syndrome decoding is proven to be immune to the HDM attack~\footnote{{\it A detailed discussion and a proof that syndrome based decoding is immune from the HDM attacks presented by Becker can be found in Section 6.1 of the article in~\cite{robust2017becker}} }. Therefore, this is the reason why we opt for evaluating the ECC overhead based on the BCH code based syndrome decoding in Section~\ref{sec:lightweight}.

Since TREVERSE requires no ECC, consequently, no helper data, helper data manipulation attacks~\cite{delvaux2015helper,robust2017becker} are inherently eschewed.

\subsubsection{Generic}
Firstly, the TREVERSE authentication is generic to all PUF types as long as the PUF has its corresponding \textsf{SimPUF}. Secondly, the validated two simple yet efficient and complementary augmentation methods---$ d $-authentication and multiple-reference---to enhance TREVERSE authentication capability are also independent on PUF types.
\subsubsection{Server-Aided}
The TREVERSE fully takes advantage of a resource-rich server. It is the server enrolling and storing the \textsf{SimPUF} during the enrollment, then grading the PUF response reliability confidence, and the server carries out the trials and checks during the authentication phase. In addition, when multiple referenced response technique is used to significantly augment the authentication capability, the server takes all the overhead without bringing any overhead to the token even if more referenced responses are used.

\subsubsection{Countering Aging or Fault Induced Errors}
TREVERSE inherently eschews security concerns associated with ECC associated helper data. However, if there are drastic variations in the PUF response specific reliability model due to ageing, it is possible for TREVERSE to result in increased key failure rates.

Notably, ageing is PUF design specific, there exists methods of countering aging related behaviour changes of a given PUF design---see~\cite{maes2014countering} for SRAM PUFs and see~\cite{liu2017acro} for ROPUFs. Therefore, we believe these existing counter measures can be adopted to combat ageing related degradation on TREVERSE performance.

\begin{figure}[h]
	\centering
	\includegraphics[trim=0 0 0 0,clip,width=0.35\textwidth]{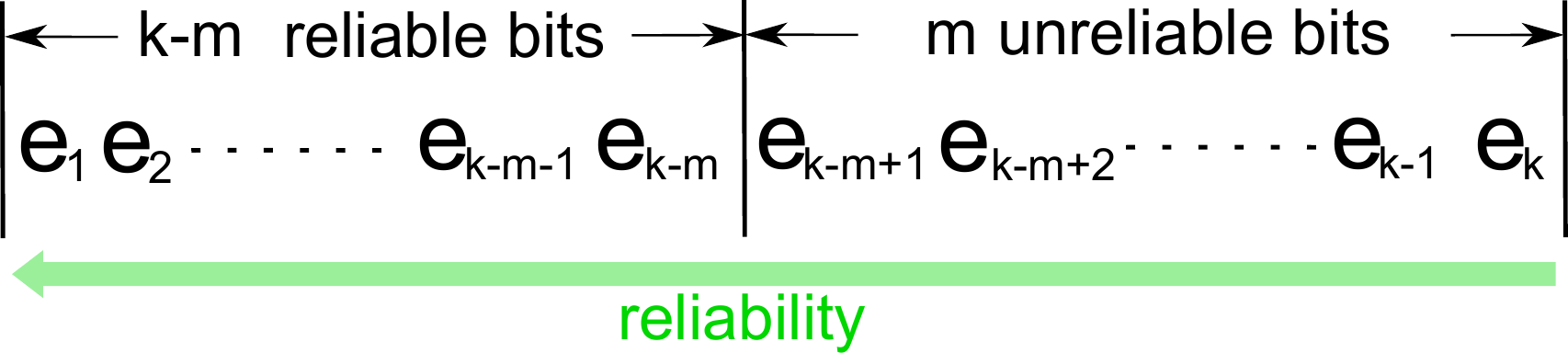}
	\caption{Detection-update trial-and-error performed periodically or when abnormally increased rejection rate occurs as a general approach to counter aging or fault induced flipped bits in reliable bits.}
	\label{fig:agingAuthen}
\end{figure}

Nonetheless, by slightly modifying the trial-and-error method used in TREVERSE, we can handle the rare errors that occur in the most reliable bits. We refer to this mechanism as the detection-update trial-and-error, where the detection-update concept is similar to~\cite{kirkpatrick2010software}. As illustrated in Fig.~\ref{fig:agingAuthen}, we suppose that the server has already sorted the $k$ response bits according to reliability---the $e_1$ with the highest reliability and $e_k$ with the lowest reliability. Besides trying all error patterns for all the $m$ unreliable bits, the server can periodically, considering the rate of aging of the hardware, or when abnormally high rejection rates are observed, iteratively flip one or two bits within $k-m$ reliable bits. If any bits within this $k-m$ reliable bits are found to be flipped on the PUF integrated device, the server can detect it through iterative trial and error checks. Consequently, the server can update the bit value on the server side within this $k-m$ reliable bits to compensate for aging induced errors on the PUF device.

In this context, the number of trials performed by the server becomes ${k-m \choose n_{\rm ag}}\times 2^m $ with $n_{\rm ag}$ the number of errors induced by {\it aging or unexpected faults}. Notably, the value $n_{\rm ag}$ will be very small, since the server can pick a short period, e.g. once within a  one month, to conduct the update to ensure changes due to any ageing affects are minimal. This periodical detection-update trial-and-error will bring an extra ${k-m \choose n_{\rm ag}}$ multiple to the trials in comparison with the original exhaustive search over the $m$ unreliable bits. However, it is worth noting that such a detection-update trial-and-error is not needed for normal authentication sessions but, rather, performed periodically. Moreover, when performing detection-update trial-and-error, the server can first reduce $m$ to decrease the trials because we know that $2^m$ trials are only needed in the worst case. For example, given $k=110$, the server can set $m=27$ resulting in $2^{27}$ trials in the worst case for trial-and-error, for the detection-update trial-and-error the server can reduce $m$ to be 22 and set $n_{\rm ag}=1$, giving ${88 \choose 1}\times 2^{22} < 2^{29}$ trials in the worst case.

This detection-update and trial-and-error mechanism brings extra management overhead to the server. However, the errors resulting from aging or faults are expected to be infrequent. For example, Maes {\it et al.}~\cite{maes2012experimental} reported a maximum unreliability increase of 3.9\% from 4.5 years ageing for ROPUFs, here, approximately 1 bit is flipped out of 128 bits per year due to aging. Thus, the extra management overhead is slight. In addition, TREVERSE is designed to utilize a resource-rich server that can be flexibly configured and is computationally powerful; therefore, such a light overhead increase is acceptable in order to achieve a lightweight and secure implementation on the resource-constraint device.

Overall, we have provided methods for addressing ageing at the protocol level and the device level, accounting for ageing by considering an adaptation of the SimPUF is both a challenging and interesting direction for our future work. We believe such a method can allow TREVERSE to function without the slight increase in sever overhead to deal with ageing or the use of device level methods to combat ageing related influences on PUF response.
\section{Related Work}\label{Sec:Related}

PUF based authentication mechanisms must directly deal with noise inherent to the source of entropy used for deriving keys. The solutions to address a noisy PUF response broadly fall into three approaches: i) PUF re-engineering to enhancing response reliability (see Section~\ref{sec:related-enhancing-reliability}); ii) Reconciling response errors (See Section ~\ref{sec:related-reconcile-errors}); iii) Response similarity measures (see Section~\ref{sec:related-smiliarity-measure}). The method selected for dealing with noisy responses dictates the authentication protocol, cost of implementation on the token, the range of PUF primitives that can be used with the authentication protocol as well the security and practicality of the mechanism. We discuss related work based on the above methods.

TREVERSE explicitly employs the bit specific nature of PUF responses as a trapdoor. A recent and closely related study in~\cite{herder2017trapdoor} employed a response reconciliation method requiring on-chip based measurements of bit specific reliability, which is discussed in Section~\ref{sec:realted-fuzzy-trapdoor}.  

\subsection{PUF re-engineering: enhancing response reliability} \label{sec:related-enhancing-reliability} 
The aim of PUF re-engineering is i) to design intrinsic reliable PUF or ii) special PUFs that facilitate winnowing high reliable responses.

A digital PUF~\cite{bhargava2013high,xu2015digital,miao2016lrr} is typically an intrinsic reliable PUF to achieve a noisy free PUF. In general, a digital PUF response is not susceptible to environmental parameter fluctuations. As an example, intentional design defects can be utilized to induce faults in a digital circuit, hence dramatically impacting its logic functionality. Instead of process variations, these faults can be used as a `fingerprint' for a digital PUF~\cite{xu2015digital,miao2016lrr}. However, a digital PUF always requires dedicated design steps such as layout consideration~\cite{miao2016lrr} and even special fabrication steps such as hot carrier injection~\cite{bhargava2013high}, time-dependent dielectric breakdown~\cite{chuang2017physically}. Recent works also utilize properties of emerging nano elements such as memristor~\cite{che2014non} and phase change memory~\cite{khan2017phase} to realize intrinsic reliable PUFs. However, such emerging PUF constructions are not likely to be deployed in practice in the near future.

In~\cite{bhargava2014efficient}, Bhargava {\it et al.} construct a so called sense amplifier PUF (SAPUF), which uses each sense amplifier (SA) cell to generate one response---architecture is alike SRAM PUF but replaces each SRAM cell with SA cell. In this SAPUF, using built-in self-test, the reliability of the PUF response can be determined by the magnitude of the unavoidable offset voltage of a SA introduced from the manufacturing randomness. The larger the offset voltage, the more reliable the SAPUF response. In~\cite{bhargava2014efficient}, a bitmap (location mask) of reliable responses are employed as helper data only readable but not writable by an attacker; consequently, the helper data needs to be stored on-chip.

Overall, the PUF re-engineering is constructing a special PUF rather than a general approach that tackes  reliability issues of existentially easy to fabricate popular silicon PUFs. While TREVERSE is a general framework for any PUF type with a simulatable PUF.

\subsection{Reconciling Response Errors}\label{sec:related-reconcile-errors}
In literature, stabilizing response errors usually targets the PUF based key generation application---works targeting authentication may choose~\cite{sutar2018d} simple error correction code, mainly for weak PUFs with limited CRP space. The PUF key generator to derive a reliable key from noisy PUF responses usually comprises two components: secure sketch and entropy accumulator. Both together are usually referred to as a fuzzy extractor~\cite{maes2012pufky,delvaux2015helper}. The secure sketch is responsible for reconciling noisy responses, which has two prevalent constructions: code-offset and syndrome based~\cite{delvaux2015helper}. Regardless of the specific construction, the secure sketch is a pair of randomized procedures: the sketching procedure takes a response $\bf r$ as an input, then computes the enrolled key and the helper data; the recovery procedure takes both the helper data and a reevaluated response ${\bf r}'$ to reconstruct the enrolled key. The set of responses might not be ideally uniformly distributed, thereby, the entropy accumulator compresses the input response into a cryptographic key with uniform distribution relying on a random oracle, OWF. 

As experimentally validated in Section~\ref{sec:lightweight}, although the \textsf{OWF} implementation (entropy accumulator in fuzzy extractor) in practice can be lightweight, the ECC logic (secure sketch in the fuzzy extractor) is usually very expensive. Therefore, the fuzzy extractor is hard to be directly mounted to secure resource tight IoT devices. In addition, the helper data that is needed when using a fuzzy extractor has to be carefully handled to eliminate attacks introduced by itself.

Our TREVERSE resolves the above deficiencies as the ECC logic and helper data are no longer needed.

{\bf Reverse Fuzzy Extractor:} In general, sketching and recovery procedures are realized via ECC encoding and decoding, respectively. As the ECC decoding is computationally more intensive than encoding, one can embed the ECC encoding logic (ie. a computationally lighter sketching procedure) in the resource-restricted PUF. The computationally harder decoding logic is performed by a resource-rich server. This arrangement is called a reverse fuzzy extractor~\cite{van2012reverse}. The reverse fuzzy extractor exploits the server to do intensive computation.

We use the `reverse' term to {\it only} refer to the exploitation of a resource-rich server to perform authentications. Neither ECC decoder nor encoder exist in TREVERSE.

Our TREVERSE removes the ECC logic, therefore, it is lightweight. The TREVERSE also removes helper data, therefore, it has better security because potential attacks such as helper data manipulation attack~\cite{delvaux2015helper,delvaux2015survey,robust2017becker} and reliability-based modeling attacks~\cite{becker2014active,becker2015pitfalls} are inherently avoided.

{\bf Fuzzy Extractor with trapdoor:}
\label{sec:realted-fuzzy-trapdoor}
Herder {\it et al.}~\cite{herder2017trapdoor} propose the computationally-secure fuzzy extractor to extract cryptographic keys from the PUF. In this context, the response reliability confidence information is, for the first time, treated as a trapdoor to build up a stateless key generator, while previous works do take bit-specific reliability into consideration to improve the efficiency of PUF key derivation but public such information. In other words, the response reliability information now is never exposed but {measured internally} within the PUF key generator on demand---discarded after internal usage. In this context, the ROPUF is chosen for experimentally validations. Because the ROPUF's response reliability confidence that can be directly acquired via subtracting frequencies of two ROs meets with the trapdoor requirement. However, for most PUF constructions the response reliability confidence is hard to be directly and easily measured on-chip. For instance, the APUF's response reliability is not measurable on-chip unless using expensive peripheral circuits. Therefore, the stateless key generator exploiting the trapdoor reliability information is not, or at least difficult, applicable to other PUF structures.
		
Firstly, the TREVERSE does not require the token to measure the response reliability on-chip, which is suitable for a wide range of PUF types including LAPUFs, ROPUFs, and SRAM PUFs---as long as the PUF has its corresponding \textsf{SimPUF}. Secondly, in~\cite{herder2017trapdoor}, the token has to take all the computation burden to reconstruct a stateless key that is targeted for key generation. Lightweight appears not the concern. Conversely, TREVERSE is the first to utilize the bit-specific reliability as trapdoor to realize lightweight and secure authentication by moving as much as computation from the token to the server.
		
\begin{figure}
\centering
\includegraphics[trim=0 0 0 0,clip,width=0.35\textwidth]{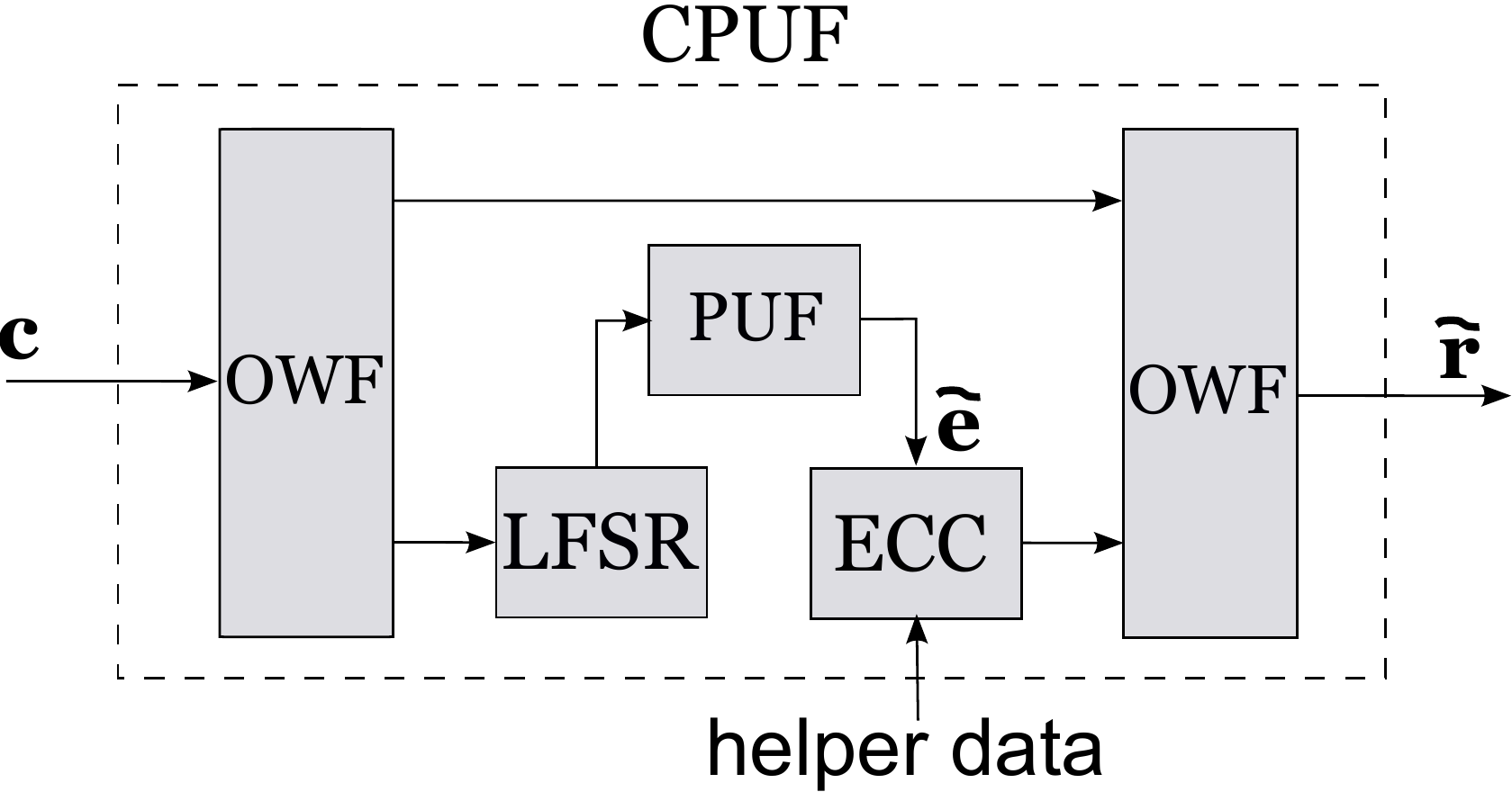}
\caption{Generalized controlled PUF (CPUF) construction. Noting in practice, only one \textsf{OWF} logic implementation is required to save area overhead, while this \textsf{OWF} is sequentially used twice.}
\label{fig:CPUF}
\end{figure}

\subsection{Similarity Measures} 
\label{sec:related-smiliarity-measure}
For strong PUF authentication, the server compares the received response vector with its enrolled response vector given the same challenge. Based on the similarity, specifically, Hamming distance (HD) that majority work relies on due to its simplicity---there are other silimarity metrics such as using fault tolerant
ring weight scheme~\cite{yan2016puf}, between the received and enrolled response, the server accepts the authenticity of the prover on the condition that the HD is lower than a threshold, otherwise, rejects. To be specific, such HD based authentication is only suitable for the strong PUF with a very large CRP (challenge) space to prevent fully CRP characterization within e.g., years. Notably, the practical HD-based lightweight authentication is always built upon the APUF or its variants such as XOR-APUF~\cite{ruhrmair2013puf}, Slender PUF~\cite{rostami2014robust}, obfuscated APUF~\cite{yu2014noise,gao2016obfuscated}. Unfortunately, though might be lightweight, it has been shown that the security of the HD based authentication is very difficult to be guaranteed in presence of modeling attacks. According to recent survey on HD-based authentication~\cite{ruhrmair2013puf,becker2015gap,becker2015pitfalls} in Delvuax thesis in 2017~\cite{delvaux2017security}, all these HD-based authentication are broken under modeling attacks except the lockdown PUF.

The lockdown PUF restricts the maximum number of LAPUF CRPs that can be acquired by an adversary~\cite{yulockdown}. This is achieved by an explicit CRP release. 
Generally, the token and the server together determine challenges presented to the LAPUF. Responses are divided into two subparts, the first subpart needs to be firstly provided by the server. The later subpart is visible only when the first subpart response sent from the server is close enough to that generated by the token. 
Thus, an adversary is unable to obtain new CRP materials without authorization from the server. Once a number of CRPs have been issued, the token has to be disposed and never used again, which results into a usage trade-off of limited authentication rounds, e.g., typically 1000 times~\cite{yulockdown}. Unlike other HD-based authentications, the lockdown, in essence, {\it not intends to invent an architecture to increase the complexity of modeling attacks by the adversary}. In the other way round, it limits the ability of obtaining an adequate number of CRPs for training by the adversary to prevent modeling attacks.

We notice one PUF authentication based on HD comparison occasionally uses the `trial and error' inexplicitly~\cite{ozturk2008towards}. Therefore, we briefly introduce it to show their trial and error is used in a totally different manner~\cite{ozturk2008towards}. This design has been shown impractical and insecure, we refer in-depth analysis to~\cite{delvaux2017security} (Chapter 5.3.4). Generally, the token implementation has two strong PUFs, specifically, APUFs: an inner APUF and an outer APUF. The server learns models of both APUFs during enrollment phase in order to emulate response given a random challenge. The inner PUF becomes not accessible once it is deployed in field. There is an initial state $\bf s$ with length $n$ of inner APUF that acts as a challenge seed. Noting $\bf s$ in~\cite{ozturk2008towards} requires a secure NVM, which undermines the PUF value~\cite{delvaux2017security}.  When authentication starts, this $\bf s$ is expanded into $n$ subchallenges to generate a concatenated response ${\bf e}_1$. This ${\bf e}_1$ is then XORed with the challenge $\bf c$ issued by the server and then is applied to the outer APUF to produce response ${\bf e}_2$ sent to server for authentication. Meanwhile, the $\bf s$ is updated by ${\bf e}_1$. In~\cite{ozturk2008towards}, it is claimed that the ${\bf e}_1$ can be always stable, e.g., by using major voting technique, and is therefore the same as the $\tilde{{\bf e}}_1$ emulated by the server. In the worst case, it is assumed that there is 1 out of $n$ errors. This error will desynchronize the authentication. In this context, {\it the server iteratively flips each bit of $\tilde{{\bf e}}_1$ to try  authentication for synchronization}. To this end, we can see this occasional trial and error usage is totally different from ours.

		

\subsection{Summary}
TREVERSE distinguishes from the above approaches by constructively exploiting bit-specific reliability to tackle noisy PUF responses through a trial-and-error approach at the server.

\section{Conclusion and Future Work}\label{Sec:Conclusion}
						
The developed TREVERSE fully leverages a computational resource-rich server to ensure a (mutual) lightweight token realization. The TREVERSE allows the prover to directly processes noisy PUF responses via \textsf{OWF}, it ultimately discards expensive ECC logic, thus, removing the necessity of the helper data that is exploitable by an adversary. Through implementing  $ d $-authentication and multiple-reference that complement each other to exponentially enhance the authentication capability, we are able to reduce the FRR to be less than $10^{-6}$ by tolerating BER up to 9.66\% and 14.53\% for ROPUF and LAPUF while maintaining  trial computations being practically acceptable to the server, even a personal computer. 
						
i) The current TREVERSE utilizes an exhaustive search over least reliable bits, which is not the best optimization. Future work can investigate other efficient search approach to further improve the authentication efficiency. ii) One limitation of the current experimental evaluation is that the aging effect is not included. We leave this as a future work. We believe that the aging effect can be easily addressed, especially by utilizing more references via the multiple-reference technique. iii) Another future work can investigate the TREVERSE instantiation through the AISC APUF instance that is also a typical representative of the LAPUF. 
iv) It is also valuable to test a TREVERSE instantiation using intrinsic SRAM PUFs. In this context, the key is to expedite the bit reliability information enrollment, ideally by evading the exhaustively repeated physical measurements. 

\section{Acknowledgment}
We thank Yang Su for evaluating part of results in Table~\ref{tab:BCH}.

\clearpage

\appendices

\section{Results of d-Authentication.}
In Tables~\ref{tab:AuthenTabdAuthROPUF} and~\ref{tab:AuthenTabdAuthLAPUF}, we give empirical and statistical evaluations on the ${\rm FRR}_d$ of ROPUF and LAPUF respectively. When the $d=10$, the ${\rm FRR}_d$ is significantly reduced. In practice, considering that the PUF operating condition vary not too much, e.g., no more than 10\% voltage deviation (1.08~V-1.32~V), the $ d $-authentication can already minimize the ${\rm FRR}_d$ to be acceptable. For example, when the $m=16$, the ${\rm FRR}_d$ is always less than $10^{-3}$.
\begin{table} 
	\centering 
		\caption{${\rm FRR}_d$ of the ROPUF under different operating conditions and $ m=12,14,16 $ settings, where $ k=64 $ and $ d=10 $.}
		\resizebox{0.50\textwidth}{!}{
			\begin{tabular}{c|| c || c || c || c || c} %
				\toprule 
				\toprule 
				&\multicolumn{1}{c}{$ 25\celsius $, 0.96~V} & \multicolumn{1}{c}{$ 25\celsius $, 1.08~V}& \multicolumn{1}{c}{$ 65\celsius $, 1.20~V}& \multicolumn{1}{c}{$ 25\celsius $, 1.32~V}& \multicolumn{1}{c}{$ 25\celsius $, 1.44~V} \\%
				\cmidrule(l){2-2} \cmidrule(l){3-3} \cmidrule(l){4-4}\cmidrule(l){5-5}\cmidrule(l){6-6}
				$ m $ & ${\rm FRR}_d$ & ${\rm FRR}_d$ & ${\rm FRR}_d$ & ${\rm FRR}_d$ & ${\rm FRR}_d$ \\ 
				\midrule
				$ 12 $ & 20.60\%; 35.30\% & 0\%;  5.09$\times 10^{-5}$ & 0\%; $<10^{-9}$ & 0\%; 4.39$\times 10^{-4}$ & 0.10\%; 16.44\% \\ 
				$ 14 $ & 8.00\%; 17.97\% & 0\%; 1.4$\times 10^{-6}$ & 0\%; $<10^{-9}$ & 0\%; 2.48$\times 10^{-5}$ & 0\%; 5.71\% \\ 
				$ 16 $ & 2.50\%; 8.22\% & 0\%; 2.13$\times 10^{-8}$ & 0\%; $<10^{-9}$ & 0\%; 1.27$\times 10^{-6}$ & 0\%; 1.65\% \\ 
			\midrule
			\bottomrule 
		\end{tabular}}
								\label{tab:AuthenTabdAuthROPUF} 
		\begin{tablenotes}
			\item{The ${\rm FRR}_d$ based on empirical evaluations and statistical analyses based on Eq(\ref{Eq:FRRd}) are listed for comparison. The format is (empirical;statistical).}
		\end{tablenotes}
\end{table}
							\begin{table} 
								\centering 
								\caption{${\rm FRR}_d$ of the LAPUF under different operating conditions and $ m=12,14,16 $ settings, where $ k=64 $ and $ d=10 $.}
								\resizebox{0.50\textwidth}{!}{
									\begin{tabular}{c|| c || c || c || c || c} %
										\toprule 
										\toprule 
										&\multicolumn{1}{c}{$ 25\celsius $, 0.96~V} & \multicolumn{1}{c}{$ 25\celsius $, 1.08~V}& \multicolumn{1}{c}{$ 65\celsius $, 1.20~V}& \multicolumn{1}{c}{$ 25\celsius $, 1.32~V}& \multicolumn{1}{c}{$ 25\celsius $, 1.44~V} \\%
										\cmidrule(l){2-2} \cmidrule(l){3-3} \cmidrule(l){4-4}\cmidrule(l){5-5}\cmidrule(l){6-6}
										$ m $ & ${\rm FRR}_d$ & ${\rm FRR}_d$ & ${\rm FRR}_d$ & ${\rm FRR}_d$ & ${\rm FRR}_d$ \\ 
										\midrule
										$ 12 $ & 88.20\%; 88.09\%  & 0.10\%; 1.09\%   & 0\%; $<10^{-9}$  & 0.10\%; 0.10\%  & 7.00\%; 6.74\%  \\ 
										$ 14 $ & 78.00\%; 77.47\%  & 0\%; 0.16\% & 0\%; $<10^{-9}$  & 0.10\%; 6.76$\times10^{-5}$  & 2.20\%; 1.82\%  \\ 
										$ 16 $ & 64.20\%; 63.30\%  & 0\%; 1.44$\times10^{-4}$ & 0\%; $<10^{-9}$  & 0\%; 2.44$\times10^{-6}$  & 0.40\%; 0.34\% \\ 
										\midrule
										\bottomrule 
									\end{tabular}}
									\label{tab:AuthenTabdAuthLAPUF} 
									\begin{tablenotes}
										\item{The ${\rm FRR}_d$ based on empirical evaluations and statistical analyses based on Eq(\ref{Eq:FRRd}) are listed for comparison. The format is (empirical;statistical).}
									\end{tablenotes}
								\end{table}

								\begin{table}
									\centering 
									\caption{$ \lambda_{\rm 1; ref} $ and $ \epsilon_{\rm ref} $ of ROPUF under different operating conditions and the $ \lambda_{\rm 2;ref} $.}
									\resizebox{0.50\textwidth}{!}{
										\begin{tabular}{c|| c || c || c || c || c || c} %
											\toprule 
											\toprule 
											
											Operating condition & $ \epsilon_{\rm ref_1} $ & $ \lambda_{\rm 1; ref_1} $ & $ \lambda_{\rm 2;ref_1} $ & $ \epsilon_{\rm ref_2} $ & $ \lambda_{\rm 1; ref_2} $ & $ \lambda_{\rm 2;ref_2} $\\ 
											\midrule
											($ 25\celsius $, 0.96~V) & 5.20\% & 0.4477 & -0.3276 & 14.20\% & 0.5029 & -0.3613\\
											($ 25\celsius $, 1.08~V) & 0.96\% & 0.0324 & -0.3276 & 9.30\% & 0.3533 & -0.3613\\
											($ 65\celsius $, 1.20~V) & 6.39\% & 0.2389 & -0.3276 & 3.92\% & 0.1961 & -0.3613\\
											($ 25\celsius $, 1.32~V) & 9.29\% & 0.4579 & -0.3276 & 0.76\% & 0.0266 & -0.3613\\
											($ 25\celsius $, 1.44~V) & 11.09\% & 0.5808 & -0.3276 & 2.24\% & 0.1082 & -0.3613\\
											\bottomrule
										\end{tabular}
									}
									\label{tab:ROPUFIntraInterTwoRef} 
								\end{table}

								\begin{table*} 
									\centering 
									\caption{${\rm FRR}_M$ of the ROPUF under different operating conditions and $ m=8,10,12,14,16 $ settings, where $ k=64 $ with two references ($ 25\celsius $, 1.08~V) and $ 25\celsius $, 1.32~V.}
									\resizebox{0.95\textwidth}{!}{
										\begin{tabular}{c|| c || c || c} %
											\toprule 
											\toprule 
											&\multicolumn{1}{c}{$ 25\celsius $, 0.96~V} & \multicolumn{1}{c}{$ 65\celsius $, 1.20~V}& \multicolumn{1}{c}{$ 25\celsius $, 1.44~V} \\%
											\cmidrule(l){2-2} \cmidrule(l){3-3} \cmidrule(l){4-4}
											$ m $ & (${\rm FRR}_{\rm ref_1}$;${\rm FRR}_{\rm ref_2}$;${\rm FRR}_M$) & (${\rm FRR}_{\rm ref_1}$;${\rm FRR}_{\rm ref_2}$;${\rm FRR}_M$) & (${\rm FRR}_{\rm ref_1}$;${\rm FRR}_{\rm ref_2}$;${\rm FRR}_M$)\\ 
											\midrule
											$ 8 $ & (53.32\%; 99.66\%; 52.45\%), (69.33\%; 99.60\%; 69.05\%)  & (71.83\%; 30.62\%; 23.06\%), (79.52\%; 63.31\%; 50.34\%)  & (97.86\%; 4.85\%; 4.93\%), (99.24\%; 19.72\%; 19.57\%) \\ 
											$ 10 $ & (39.11\%; 99.23\%; 39.43\%), (54.13\%; 99.15\%; 53.67\%)  & (59.57\%; 18.42\%; 12.27\%), (68.53\%; 49.87\%; 34.18\%) & (95.76\%; 1.78\%; 1.68\%), (98.47\%; 10.38\%; 10.22) \\ 
											$ 12 $ & (26.90\%; 98.43\%; 26.64\%), (42.20\%; 98.35\%; 41.50\%)  & (47.52\%; 10.14\%; 5.81\%), (56.92\%; 37.47\%; 21.33\%)  & (92.50\%; 0.69\%; 0.35\%), (97.31\%; 6.22\%; 6.05\%) \\ 
											$ 14 $ & (17.18\%; 97.29\%; 17.25\%), (31.15\%; 97.22\%; 30.28\%)  & (36.16\%; 4.90\%; 2.64\%), (45.98\%; 26.50\%; 12.18\%) & (87.85\%; 0.17\%; 0.06\%), (95.14\%; 3.75\%; 3.57\%) \\ 
											$ 16 $ & (10.26\%; 95.47\%; 10.63\%), (21.99\%; 95.03\%; 20.90\%)  & (26.06\%; 2.41\%; 1.00\%), (36.52\%; 17.40\%; 6.35\%) & (82.02\%; 0.07\%; 0.02\%), (92.36\%; 2.64\%; 2.44\%) \\ 
											\midrule
											\bottomrule 
										\end{tabular}}
										\label{tab:AuthenTabMultAuthROPUF} 
										\begin{tablenotes}
											\item{Before `,' shows empirical results, after `,' shows statistical results.}
										\end{tablenotes}
									\end{table*}					
									\begin{table*}
										\centering 
										\caption{${\rm FRR}_M$ of the LAPUF under different operating conditions and $ m=8,10,12,14,16 $ settings, where $ k=64 $ with two references ($ 25\celsius $, 1.08~V) and ($ 25\celsius $, 1.32~V).}
										\resizebox{0.95\textwidth}{!}{
											\begin{tabular}{c|| c || c || c} %
												\toprule 
												\toprule 
												&\multicolumn{1}{c}{$ 25\celsius $, 0.96~V} & \multicolumn{1}{c}{$ 65\celsius $, 1.20~V}& \multicolumn{1}{c}{$ 25\celsius $, 1.44~V} \\%
												\cmidrule(l){2-2} \cmidrule(l){3-3} \cmidrule(l){4-4}
												$ m $ & (${\rm FRR}_{\rm ref_1}$;${\rm FRR}_{\rm ref_2}$;${\rm FRR}_M$) & (${\rm FRR}_{\rm ref_1}$;${\rm FRR}_{\rm ref_2}$;${\rm FRR}_M$) & (${\rm FRR}_{\rm ref_1}$;${\rm FRR}_{\rm ref_2}$;${\rm FRR}_M$)\\ 
												\midrule
												$ 8 $ & (82.32\%; 99.99\%; 83.77\%), (82.21\%; 99.95\%; 82.16\%)  & (95.25\%; 53.22\%; 49.29\%), (94.18\%; 45.47\%; 42.84\%)  & (99.85\%; 11.61\%; 10.39\%), (99.75\%; 13.73\%; 13.70\%) \\ 
												$ 10 $ & (72.70\%; 99.98\%; 74.53\%), (72.82\%; 99.90\%; 72.75\%)  & (91.63\%; 38.99\%; 34.18\%), (90.29\%; 31.60\%; 28.53\%)  & (99.67\%; 5.22\%; 4.63\%), (99.40\%; 7.91\%; 7.86\%) \\ 
												$ 12 $ & (61.63\%; 99.98\%; 63.92\%), (61.24\%; 99.90\%; 61.19\%)  & (86.16\%; 26.89\%; 22.72\%), (83.50\%; 20.45\%; 17.08\%)  & (99.37\%; 2.02\%; 1.80\%), (98.81\%; 5.00\%; 4.94\%) \\ 
												$ 14 $ & (50.23\%; 99.97\%; 51.84\%), (49.36\%; 99.89\%; 49.31\%)  & (79.36\%; 17.31\%; 13.86\%), (76.41\%; 13.05\%; 9.97\%)  & (98.85\%; 0.72\%; 0.66\%), (97.70\%; 3.70\%; 3.61\%) \\ 
												$ 16 $ & (39.58\%; 99.94\%; 41.15\%), (38.19\%; 99.87\%; 38.14\%)  & (71.68\%; 10.27\%; 7.56\%), (67.77\%; 8.01\%; 5.43\%)  & (97.87\%; 0.22\%; 0.21\%), (96.27\%; 3.06\%; 2.95\%) \\ 
												\midrule
												\bottomrule 
											\end{tabular}}
											\label{tab:AuthenTabMultAuthLAPUF} 
											\begin{tablenotes}
												\item{Before `,' shows empirical results, after `,' shows statistical results.}
											\end{tablenotes}
										\end{table*}
                                        
\section{Results of Multiple-Reference.}
\begin{figure*}
	\centering
	\includegraphics[trim=0 0 0 0,clip,width=0.90\textwidth]{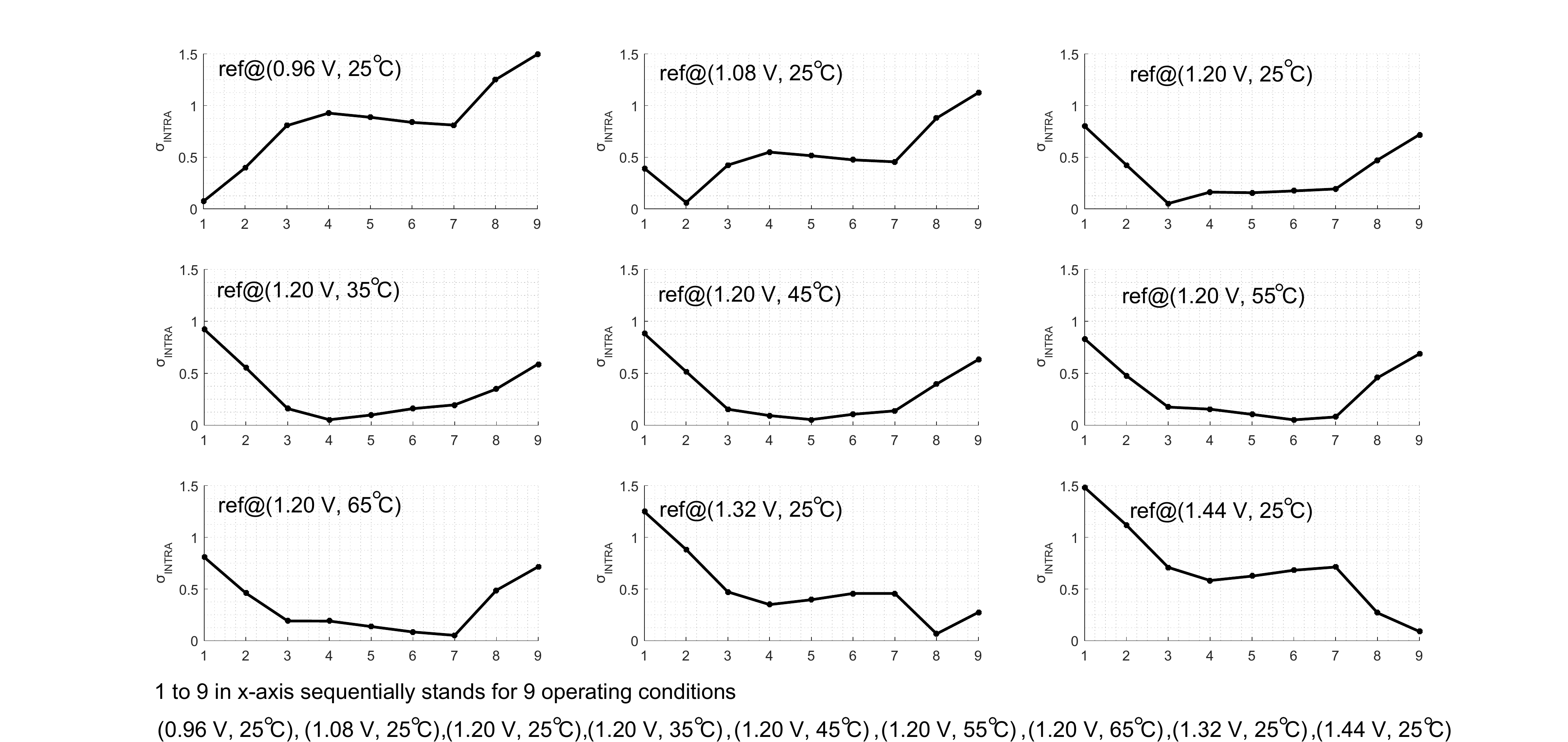}
	\caption{ROPUF's $\sigma_{\rm INTRA}$ under nine varying operating conditions given a specific referenced operating condition.}
	\label{fig:sigma_INTRA9OCplot}
\end{figure*}

We plot the ROPUF's $\sigma_{\rm INTRA}$ under nine varying operating conditions given a specific referenced operating condition in Fig.~\ref{fig:sigma_INTRA9OCplot}. Reminding that given a referenced response, higher $\sigma_{\rm INTRA}$ for the regenerated response, higher the FRR. 

In Table~\ref{tab:ROPUFIntraInterTwoRef}, the $ \lambda_1 $, $ \lambda_2 $ and $ \epsilon $ of the ROPUF based on two referenced operating corners, ref$_1$ of ($ 25\celsius $, 1.08~V) and ref$_2$ of ($ 25\celsius $, 1.32~V), are experimentally evaluated. 
										
In Tables~\ref{tab:AuthenTabMultAuthROPUF} and~\ref{tab:AuthenTabMultAuthLAPUF} , by using two-reference, we experimentally and statistically evaluate the ${\rm FRR}_{\rm ref_1}$, ${\rm FRR}_{\rm ref_2}$ and consequently ${\rm FRR}_{\rm mr}$ for ROPUF and LAPUF, respectively. It is obvious that the ${\rm FRR}_{\rm mr}$ is significantly reduced. For example, in Table~\ref{tab:AuthenTabDefaultROPUF}, the FRR of the ROPUF is up to 90\% when a single referenced operating corner is used under the settings of $m=12, k=64$. In Table~\ref{tab:AuthenTabMultAuthROPUF}, the FRR$_{\rm mr}$ is substantially reduced to 27\% when two-reference is exploited under the same $m$ and $n$ settings.		

\section{Validation with HOST2018 Dataset}\label{app:HOST2018}
\subsection{HOST2018 Dataset}
In HOST 2018, Robert {\it et al.}~\cite{hesselbarth2018large} released the latest ROPUF dataset collected from 217 Xilinx Artix-7 XC7A35T FPGAs, each containing a total of 6,592 ROs, comprised of six different routing paths with 550 to 1,696 instances per type. It is suggested by this work~\cite{hesselbarth2018large} that the ROPUF response generation should be from comparison of ROs under the same routing path in order to avoid bias. Therefore, we use 1,600 ROs mapped to the upper left of the FPGA (see Table~1 in~\cite{hesselbarth2018large}) under the same routing path. To be aligned with our test for Virginia dataset in Section~\ref{Sec:FRRModelValidation}, we only use the first 512 ROs for our TREVERSE evaluations. Out of 217 FPGAs, 50 of them are tested under six varying temperature corners: $5\celsius$, $15\celsius$, $25\celsius$, $35\celsius$, $45\celsius$, $55\celsius$~\cite{hesselbarth2018large}. Under each temperature, the measurement is repeated 101 times.


\subsection{ROPUF Validation}
Using the setting described in~Section~\ref{Sec:valROPUF}, 130,816 responses are generated from one ROPUF consisting of 512 ROs implemented in one FPGA. By using $25\celsius$ as the reference, the $55\celsius$ setting provides the worst operating corner exhibiting the worst unreliability; $\epsilon$. Herein, we have sorted the worst $\epsilon$ of all 50 ROPUFs, and used the worst five ROPUFs exhibiting the highest $\epsilon$ to represent the worst-case for evaluations. To be precise, FPGA IDs of those ROPUFs are 39, 31, 47, 10, 50, and these demonstrate a worst case $\epsilon$ of 3.06\%, 2.71\%, 2.59\%, 2.51\% and 2.48\%, respectively, under $55\celsius$. Bias for these five ROPUFs are 53.25\%, 51.71\%, 55,76\%, 55.75\%, 56.09\%.


We have evaluated FRR from both empirical and statistical---formal derivation expressed in equation~\ref{Eq:FRR})---tests. It is clear from Table~\ref{tab:AuthenROPUFHost2018} (ROPUF39 under operating temperature of $5\celsius$ and $55\celsius$) and \ref{tab:AuthenROPUFHost2018Five} (five ROPUFs under worst operating temperature of $55\celsius$), that the statistical results from our formalized model support the empirical results obtained from the dataset. As expected, the FRR from the statistical analysis is always higher than the empirically obtained FRR since the FRR determined from the statistical model is a conservative estimate. This result agrees with our conclusion in Section~\ref{Sec:FRRModelValidation}.

We have evaluated FRR from both empirical and statistical (formal equation  Eq(\ref{Eq:FRR})) tests. It is clear that, from Table~\ref{tab:AuthenROPUFHost2018} (ROPUF39 under operating temperature of $5\celsius$ and $55\celsius$) and \ref{tab:AuthenROPUFHost2018Five} (five ROPUFs under worst operating temperature of $55\celsius$), statistical results from our formalized equation match empirical results. Under expectation, the statistical FRR is a conservative estimation, since it is always higher than the empirical FRR, which again agrees with our conclusion in Section~\ref{Sec:FRRModelValidation}.

\subsection{LAPUF Validation}
We have also evaluated the FRR of LAPUF adopting the settings described in Section~\ref{Sec:ExpTREVERSE-B}. Here we use the same five FPGA boards selected previously to obtain five LAPUFs. Table~\ref{tab:AuthenLAPUFHost2018Five} summarizes the FRR of both empirical and statistical analysis---the worst $\epsilon$ is 3.10\% for FPGA ID = 39. We can see that the statistical model is able to provide a conservative estimate of the FRR.


\subsection{FRR with d-Authentication}
\begin{figure}
	\centering
	\includegraphics[trim=0 0 0 0,clip,width=0.40\textwidth]{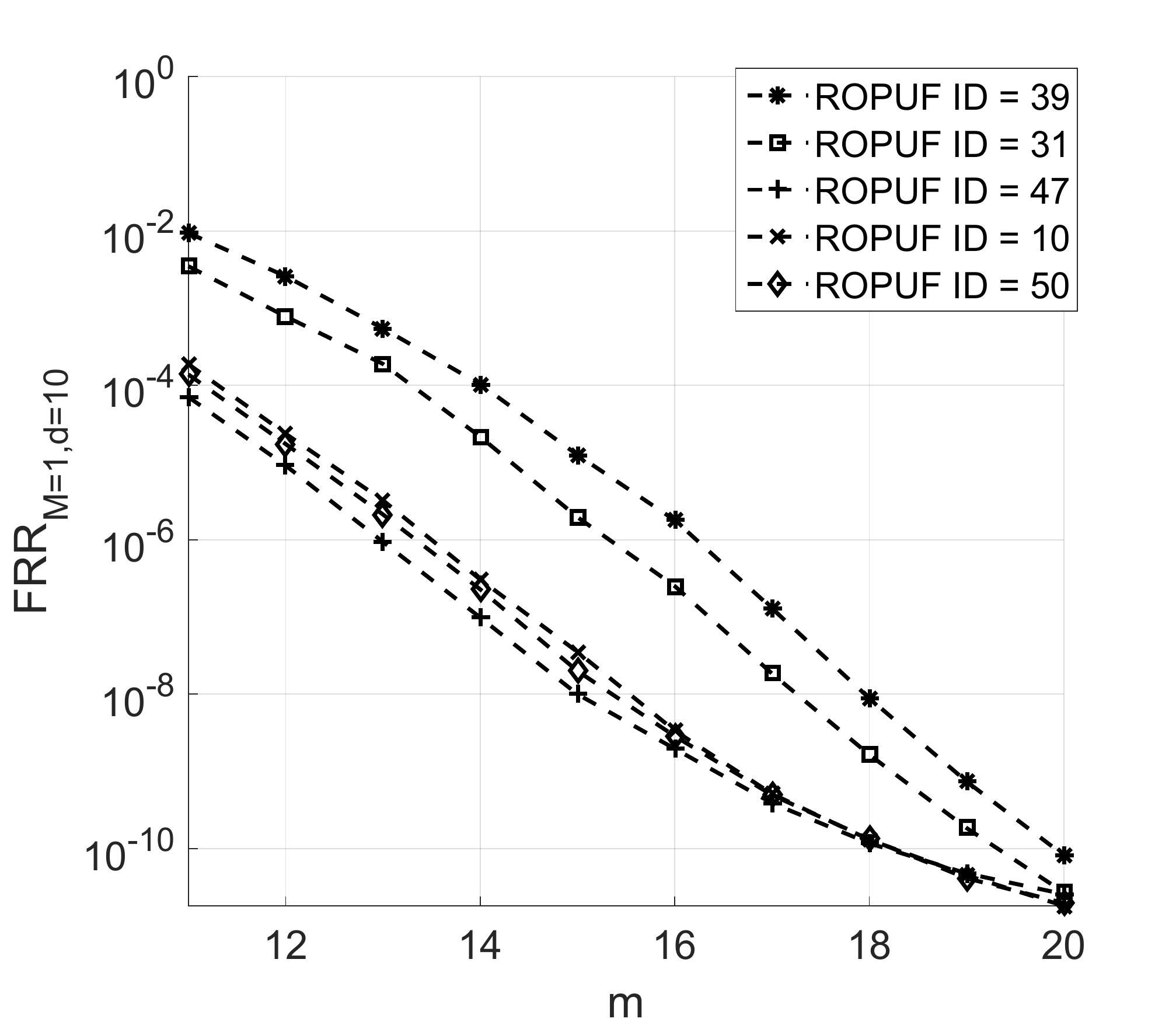}
	\caption{The ${\rm FRR}_{\rm M,d}$ of the five nosiest ROPUF (HOST2018 dataset). Given the lower worst-case $\epsilon$ observed in this dataset compared to that from Virginia Tech~\cite{maiti2010large}, only a single reference response (M=1) from 25$\celsius$ is needed. Here, $M=1$, $d=10$ and $k=110$.}
	\label{fig:FRR5ROPUF}
\end{figure}

\begin{figure}
	\centering
	\includegraphics[trim=0 0 0 0,clip,width=0.40\textwidth]{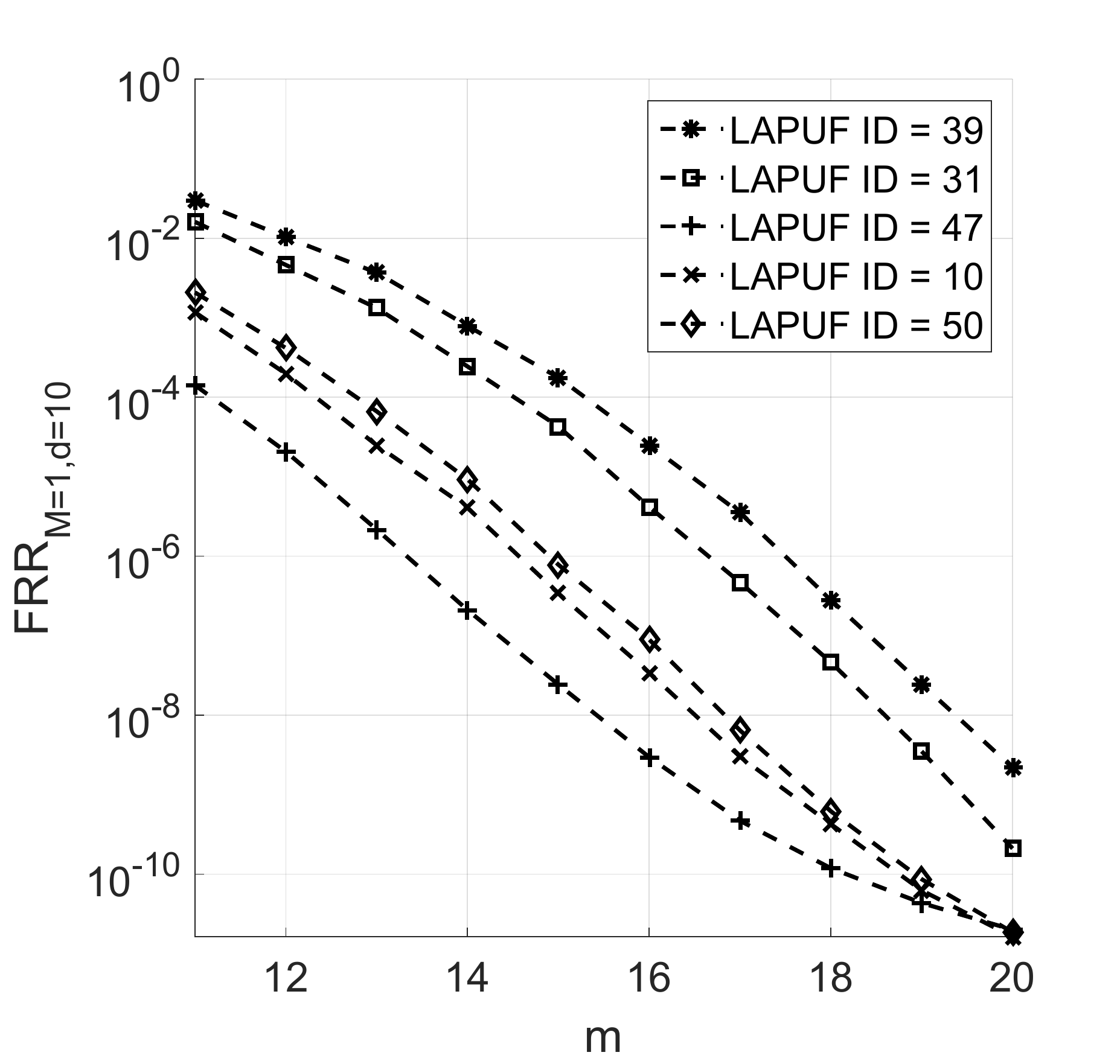}
	\caption{The ${\rm FRR}_{\rm M,d}$ of five LAPUFs (HOST2018 dataset) with ($M=1$, $d=10$ and $k=110$). Given the lower worst case $\epsilon$ observed in this dataset compared to that from Virginia Tech~\cite{maiti2010large}, only a single reference response (M=1) from 25$\celsius$ is needed.}
	\label{fig:FinalFRR5LAPUF}
\end{figure}

Fig.~\ref{fig:FRR5ROPUF} and~\ref{fig:FinalFRR5LAPUF} detail the ${\rm FRR}_{\rm M,d}$ of ROPUFs and LAPUFs when only a single SimPUF is enrolled---$M = 1$---and $d=10$. For all five ROPUFs and LAPUFs, a small number of trials with $m = 17$ and $m = 18$ is able to ensure ${\rm FRR}_{\rm M,d} < 10^{-6}$. This is because the $\epsilon$ of both ROPUF and LAPUF are small. Therefore, the server only needs to perform at most $2^{18} \times 10$ trials. This number of trial is significantly less than the results---$2^{29}$ for ROPUF and $2^{31}$ for LAPUF in see Section~6.3---obtaiend from the validation using the Virginia Tech dataset exhibiting a worst case $\epsilon$ of 9.66\% for ROPUF and 14.53\% for LAPUF.

\begin{table}[t]
	\centering 
	\caption{FRR of the ROPUF (HOST2018 dataset) from FPGA ID 39 under $5\celsius$ and $55\celsius$ operating conditions and $ m $ settings, where $ k=110 $. The referenced operating condition is $25\celsius$. This is the nosiest ROPUF out of the 50 tested ROPUFs.  }
	\resizebox{0.35\textwidth}{!}{
	\begin{tabular}{c|| c|| c}
		\toprule 
		\toprule 
		&\multicolumn{1}{c}{$ 5\celsius $} ($\epsilon$ = 2.15\%) & \multicolumn{1}{c}{$ 55\celsius $} ($\epsilon$ = 3.06\%) \\
		\cmidrule(l){2-2} \cmidrule(l){3-3}
		$ m $ & FRR & FRR\\ 
		\midrule
		$ 11 $ & 13.82\%; 23.99\% & 35.99\%; 62.71\% \\ 
		$ 12 $ & 10.28\%; 19.75\% & 30.73\%; 55.11\% \\ 
		$ 13 $ & 7.65\%; 17.33\% & 26.13\%; 47.18\% \\ 
		$ 14 $ & 5.58\%; 15.40\% & 21.51\%; 39.94\% \\ 
		$ 15 $ & 4.05\%; 13.98\% & 17.60\%; 32.35\% \\ 
		$ 16 $ & 3.05\%; 13.07\% & 13.99\%; 26.62\% \\ 
		$ 17 $ & 2.20\%; 12.42\% & 11.01\%; 20.46\% \\ 
		$ 18 $ & 1.50\%; 11.89\% & 8.72\%; 15.65\% \\ 
		$ 19 $ & 1.03\%; 11.49\% & 6.91\%; 12.21\% \\ 
		$ 20 $ & 0.75\%; 11.11\% & 5.53\%; 9.80\% \\ 
		\midrule
		\bottomrule 
	\end{tabular}}
	\label{tab:AuthenROPUFHost2018} 
	\begin{tablenotes}
	\item{The FRR from empirical evaluations and FRR statistical analyses based on Eq(\ref{Eq:FRR}) are listed for comparison, where the format is (empirical; statistical).}
	\end{tablenotes}
\end{table}

\begin{table}[t]
	\centering 
	\caption{FRR of the five ROPUFs (HOST2018 dataset) exhibiting the worst unreliability $\epsilon$  under the operating conditions of $55\celsius$ and $ m $ settings, where $ k=110 $. Referenced operating condition is at $25\celsius$. }
	\resizebox{0.50\textwidth}{!}{
	\begin{tabular}{c|| c|| c || c || c || c}
		\toprule 
		\toprule 
		&\multicolumn{1}{c} {FPGA ID = 39} & \multicolumn{1}{c} {FPGA ID = 31} & \multicolumn{1}{c} {FPGA ID = 47}  & \multicolumn{1}{c} {FPGA ID = 10} & \multicolumn{1}{c} {FPGA ID = 50} \\
		\cmidrule(l){2-2} \cmidrule(l){3-3} \cmidrule(l){4-5} \cmidrule(l){5-5} \cmidrule(l){6-6}
		$ m $ & FRR ($\epsilon = 3.06\%$) & FRR ($\epsilon = 2.71\%$) & FRR ($\epsilon = 2.59\%$) & FRR ($\epsilon = 2.51\%$) & FRR ($\epsilon = 2.48\%$)\\ 
		\midrule
		$ 11 $ & 35.99\%; 62.71\% & 26.06\%; 56.79\% & 21.10\%; 38.44\% & 20.06\%; 42.40\%  & 17.51\%; 41.11\%\\ 
		$ 12 $ & 30.73\%; 55.11\% & 21.26\%; 48.89\% & 16.32\%; 31.33\% & 16.14\%; 34.45\%  & 13.31\%; 33.41\%\\ 
		$ 13 $ & 26.13\%; 47.18\% & 16.78\%; 42.43\% & 12.24\%; 24.95\% & 12.48\%; 28.25\%  & 10.25\%; 27.08\%\\ 
		$ 14 $ & 21.51\%; 39.94\% & 13.20\%; 34.14\% & 9.36\%; 19.91\% & 9.66\%; 22.31\%  & 7.47\%; 21.66\%\\ 
		$ 15 $ & 17.60\%; 32.35\% & 10.40\%; 26.86\% & 6.90\%; 15.84\% & 7.35\%; 17.97\%  & 5.49\%; 16.97\%\\ 
		$ 16 $ & 13.99\%; 26.62\% & 8.16\%; 21.86\% & 5.13\%; 13.46\% & 5.44\%; 14.21\%  & 3.94\%; 13.98\%\\ 
		$ 17 $ & 11.01\%; 20.46\% & 6.01\%; 16.85\% & 3.75\%; 11.44\% & 4.17\%; 11.77\%  & 2.70\%; 11.72\%\\ 
		$ 18 $ & 8.72\%; 15.65\% & 4.41\%; 13.23\% & 2.65\%; 10.16\% & 3.04\%; 10.26\%  & 1.95\%; 10.29\%\\ 
		$ 19 $ & 6.91\%; 12.21\% & 3.35\%; 10.63\% & 1.67\%; 9.28\% & 2.26\%; 9.23\%  & 1.34\%; 9.15\%\\ 
		$ 20 $ & 5.53\%; 9.80\% & 2.60\%; 8.77\% & 1.16\%; 8.73\% & 1.47\%; 8.42\%  & 0.90\%; 8.50\%\\ 
		\midrule
		\bottomrule 
	\end{tabular}}
	\label{tab:AuthenROPUFHost2018Five} 
	\begin{tablenotes}
	\item{The FRR from empirical evaluations and FRR statistical analyses based on Eq(\ref{Eq:FRR}) are listed for comparison, where the format is (empirical; statistical).}
	\end{tablenotes}
\end{table}

\begin{table}[t]
	\centering 
	\caption{FRR of the five LAPUF (HOST2018 dataset) under worst operating conditions of $55\celsius$ and $ m $ settings, where $ k=110 $. Referenced operating condition is $25\celsius$.}
	\resizebox{0.50\textwidth}{!}{
	\begin{tabular}{c|| c|| c || c || c || c}
		\toprule 
		\toprule 
		&\multicolumn{1}{c} {FPGA ID = 39} & \multicolumn{1}{c} {FPGA ID = 31} & \multicolumn{1}{c} {FPGA ID = 47}  & \multicolumn{1}{c} {FPGA ID = 10} & \multicolumn{1}{c} {FPGA ID = 50} \\
		\cmidrule(l){2-2} \cmidrule(l){3-3} \cmidrule(l){4-5} \cmidrule(l){5-5} \cmidrule(l){6-6}
		$ m $ & FRR ($\epsilon = 3.10\%$) & FRR ($\epsilon = 2.93\%$) & FRR ($\epsilon = 2.52\%$) & FRR ($\epsilon = 2.66\%$) & FRR ($\epsilon = 2.71\%$)\\ 
		\midrule
		$ 11 $ & 44.53\%; 70.40\% & 39.21\%; 66.16\% & 20.40\%; 41.22\% & 27.52\%; 50.91\%  & 27.25\%; 53.92\%\\ 
		$ 12 $ & 38.77\%; 63.28\% & 33.44\%; 58.39\% & 15.45\%; 34.01\% & 22.07\%; 42.62\%  & 21.94\%; 45.91\%\\ 
		$ 13 $ & 32.78\%; 57.10\% & 28.53\%; 51.55\% & 11.64\%; 27.14\% & 17.86\%; 34.64\%  & 17.43\%; 38.21\%\\ 
		$ 14 $ & 27.69\%; 48.93\% & 23.79\%; 43.52\% & 8.69\%; 21.49\% & 14.27\%; 28.89\%  & 13.76\%; 31.38\%\\ 
		$ 15 $ & 22.75\%; 42.15\% & 19.68\%; 36.57\% & 6.57\%; 17.32\% & 11.11\%; 22.55\%  & 10.61\%; 24.51\%\\ 
		$ 16 $ & 18.84\%; 34.65\% & 15.74\%; 28.97\% & 4.57\%; 14.03\% & 8.67\%; 17.92\%  & 7.94\%; 19.73\%\\ 
		$ 17 $ & 15.64\%; 28.48\% & 12.38\%; 23.28\% & 3.29\%; 11.67\% & 6.48\%; 14.05\%  & 5.91\%; 15.16\%\\ 
		$ 18 $ & 12.84\%; 22.13\% & 9.96\%; 18.47\% & 2.38\%; 10.17\% & 4.79\%; 11.56\%  & 4.37\%; 11.99\%\\ 
		$ 19 $ & 10.32\%; 17.34\% & 7.72\%; 14.27\% & 1.69\%; 9.19\% & 3.59\%; 9.54\%  & 3.18\%; 9.85\%\\ 
		$ 20 $ & 7.96\%; 13.61\% & 6.06\%; 10.77\% & 1.11\%; 8.51\% & 2.59\%; 8.32\%  & 2.20\%; 8.45\%\\ 
		\midrule
		\bottomrule 
	\end{tabular}}
	\label{tab:AuthenLAPUFHost2018Five} 
	\begin{tablenotes}
	\item{The FRR from empirical evaluations and FRR statistical analyses based on Eq(\ref{Eq:FRR}) are listed for comparison, where the format is (empirical; statistical).}
	\end{tablenotes}
\end{table}

\section{Server Setting}\label{App:serverSet}
Table~\ref{tab:hash} summarizes the hash software implementation overhead in a resource-constraint PUF device. From Table~\ref{tab:hash}, we can see that the BLAKE2s consuming least clock cycle, exhibiting best performance. Therefore, we choose BLAKE2s to be implemented in the prover side for overhead estimation. Table~\ref{tab:GPU} summarizes three common GPUs in the market. Table~\ref{tab:hashIntel} summarizes the hash function throughput/speed in the Skylake Intel CPU with a single core (Core i5-6600, 3310MHz)---latest CPU shall have a much better performance, which stands for the hash throughout of the server side. Specifically, for the chosen Blake2s, its speed of Blake2s is 648 MiBps in the server side according to Table~\ref{tab:hashIntel}.

\section{Prover Setting:}\label{App:proverSetting} 
We assume for most low-end devices where the PUF is preferable, the error correction and hash operation are performed through software implementation using microcontroller due to unavailability of dedicated hardware implementation, e.g., on the FPGA platform.

Herein, we choose a MSP430FR5969 microcontroller that is usually embedded within the low-end Internet of Thing (IoT) device to evaluate the hash overhead and ECC overhead through software implementation. Thus, the overhead is measured by clock cycles. The hash software implementation overhead has been summarized in Table~\ref{tab:hash}. As for the error correction code, we choose the BCH code, the reason is detailed in Section~\ref{Sec:security}. In addition, we choose to implement the error correction encoding rather than the decoding in the device side, since the encoding is more lightweight than the decoding---ECC encoding and the hash form the so-called reverse fuzzy extractor (RFE)~\cite{van2012reverse}. This can be clearly observed from Table~\ref{tab:BCH} that summarizes the software implementation overhead of BCH encoding and  decoding, regarding to the same BCH code size ($n_1, k_1, t_1$) or same error correction capability. Here, $n_1$ is the codeword length, $k_1$ is the code size, $t_1$ is the errors that can be corrected within this $n_1$-bit block.

It is not common to use a single large BCH block to perform error correction, which is typically split into small processing blocks to reduce implementation
complexity/overhead~\cite{delvaux2015helper,gao2018lightweight}. Herein, for the given example of choosing a small BCH($n_1$,$k_1$,$t_1$) block, to gain a security level around 110 (this number is to align with our final evaluation employing a 110-bit response, in Section 6.3), around $L= \frac{110}{k_1}$ BCH($n_1$,$k_1$,$t_1$) blocks are required. The failure rate of recovering a $n_1$-bit response using a BCH($n_1$,$k_1$,$t_1$) code given a BER is expressed:
\begin{equation}
\mathbb{P}_{1}= 1 - \textsf{F}_B(t_1;n_1,{\rm BER}),
\end{equation}

Given $L$ BCH($n_1$,$k_1$,$t_1$) blocks are employed, the failure rate of all those BCH blocks is expressed: 
\begin{equation}
\mathbb{P}_{\rm fail}= 1 - (1-\mathbb{P}_{1})^L.
\end{equation}

Table~\ref{tab:FEandRFE} summarizes the key failure rate $\mathbb{P}_{\rm fail}$ given different smaller BCH($n_1$,$k_1$,$t_1$) blocks are selected when the BER is 14.93\%. This table also correspondingly evaluates the overhead of the RFE that employs the BCH encoding and the FE that employs the BCH decoding. The overhead is measured by number of clock cycles.

The clock frequency of the low-end MCU is usually several decades of MHz. For the  WISP4.1DL CRFID device we evaluated, it's maximum frequency is 16MHz. However, this CRIFD is batter-less and needs to save power, the clock frequency is configured to be 1MHz in practice. Therefore, one clock cycle takes 1 us to execute.

\begin{table}[h]
	\centering 
	\caption{ Hash Overhead, evaluated using a MSP430FR5969 microcontroller embedded within the CRFID transponder, which is an intermittently powered batteryless device resembling a typical resource-constraint IoT device setting.}
	\resizebox{0.4\textwidth}{!}{
	\begin{tabular}{c|| c || c } %
		\toprule 
		\toprule 
		
		name & digest size & clock cycles\\ 
		\midrule
	DM-SPECK64 & 64 bits & 178,448  \\
	BMW-256 & 256 bits & 150,046   \\
	SHA1 & 160 bits & 159,969  \\
	BLAKE2s-256 & 256 bits & 106,482\\
	BLAKE2s-128 & 128 bits & 104,723\\
	SHA3-256 & 256 bits & 584,126 \\				
		\bottomrule
	\end{tabular}
    }
	\label{tab:hash} 
		\begin{tablenotes}
		\small
		\item{Results are from~\cite{su2019secucode,su2019hash}. }
	\end{tablenotes}
\end{table}


\begin{table}[h]
	\centering 
	\caption{Hash Function Speed on Skylake Intel CPU using a single core (Core i5-6600, 3310MHz). }
	\resizebox{0.5\textwidth}{!}{
	\begin{tabular}{c|| c} %
		\toprule 
		\toprule 
		
		name & speed in mebibyte per second (MiBps)  \\ 
		\midrule
	Blake2b & 947 \\
	SHA-1 & 909 \\
	Blake2s & 648 \\
	MD5 & 632 \\
	SHA-512 & 623 \\
	SHAKE-128 & 445 \\	
	SHA-256 & 413 \\
	SHA3-256 & 367 \\
	SHA3-512 & 198 \\
		\bottomrule
	\end{tabular}
    }
	\label{tab:hashIntel} 
		\begin{tablenotes}
		\small
		\item{Reported in https://blake2.net/.}
	    \end{tablenotes}
\end{table}

\begin{table}[h]
	\centering 
	\caption{ Three Commonly Used GPU Specifications.}
	\resizebox{0.5\textwidth}{!}{
	\begin{tabular}{c|| c || c || c || c} %
		\toprule 
		\toprule 
		
		name & cores & clock speed & memory & price \\ 
		\midrule
	NVIDIA Titan Xp & 3840 & 1.6GHz  & 12GB GDDR5X & \$1,200  \\
	NVIDIA GTX1070 & 1920 & 1.68GHz  & 8GB GDDR5 & \$380 \\
	AMD Radeon RX 580 & 2304 & 1.26GHz  & 8GB GDDR5 & \$300 \\
		\bottomrule
	\end{tabular}
    }
	\label{tab:GPU} 
        \begin{tablenotes}
        \small
		\item{To perform acceleration using GPU, CUDA can be adopted.}
	    \end{tablenotes}	
\end{table}

\begin{table}[h]
	\centering 
	\caption{BCH code encoding and decoding overhead. }
		\begin{tabular}{c|| c || c } %
			\toprule 
			\toprule 
			
			($ n_1 $,$ k_1 $,$ t_1 $) &  \begin{tabular}{@{}c@{}} encoding  \\ clock cycles \end{tabular} & \begin{tabular}{@{}c@{}} decoding  \\ clock cycles \end{tabular} \\ \midrule 
			(255,123,19) & 930,093  & 2,515,163   \\
			(255,63,30) & 680,087  & 4,116,796   \\
			(255,47,42) & 583,024  & 6,102,010   \\
			(255,37,45) & 476,744  & 6,582,507   \\
			(255,29,47) & 377,220  & 6,976,341 \\
			(255,21,55) & 294,783  & 8,345,992  \\
			(255,9,63) & 115,709  & 8,380,790  \\	\midrule
			(511,28,111) & 559,150  & fail  \\			(511,19,119) & 408,127  & fail  \\	
			\bottomrule
		\end{tabular}
		\label{tab:BCH} 	
		\begin{tablenotes}
        \item For (511,28,111) and (511,19,119) decoding, we failed to implement it in the CRFID as the computation is too heavy for the resource-constraint CRFID to handle. Part of these evaluation results are from~\cite{gao2018lightweight} while the rest are newly evaluated. 
		\end{tablenotes}
\end{table}	

\begin{table}[h]
	\centering 
	\caption{ Clock cycles required of fuzzy extractor (FE) that employs BCH decoding and reverse fuzzy extractor (RFE)~\cite{van2012reverse} that employs BCH encoding to tolerate a BER of 14.53\%. }
	\resizebox{0.50\textwidth}{!}{
		\begin{tabular}{c|| || c || c || c || c} %
			\toprule 
			\toprule 
			
			($ n_1 $,$ k_1 $,$ t_1 $) &  block num. & Key Failure Rate  & RFE & FE \\ 
			\midrule
			(255,21,55) & 5  & $4.6\times 10^{-3}$ & 1,578,638 clocks & 41,834,683 clocks   \\
			(255,9,63) & 12  & $7.73\times 10^{-5}$ & 1,493,231 clocks & 100,674,203 clocks \\
			\midrule
			(511,28,111) & 4  & $1.95\times 10^{-5}$ & 2,341,323 clocks & fail \\
			(511,19,119) & 6 &  $3.22\times 10^{-7}$ & 2,553,485 clocks & fail  \\			
			\bottomrule
		\end{tabular}
        }
		\label{tab:FEandRFE} 
		\begin{tablenotes}
        \item The key failure rate can be viewed as the false rejection rate.
		\end{tablenotes}
\end{table}


\begin{thebibliography}{10}
	\providecommand{\url}[1]{#1}
	\csname url@samestyle\endcsname
	\providecommand{\newblock}{\relax}
	\providecommand{\bibinfo}[2]{#2}
	\providecommand{\BIBentrySTDinterwordspacing}{\spaceskip=0pt\relax}
	\providecommand{\BIBentryALTinterwordstretchfactor}{4}
	\providecommand{\BIBentryALTinterwordspacing}{\spaceskip=\fontdimen2\font plus
		\BIBentryALTinterwordstretchfactor\fontdimen3\font minus
		\fontdimen4\font\relax}
	\providecommand{\BIBforeignlanguage}[2]{{%
			\expandafter\ifx\csname l@#1\endcsname\relax
			\typeout{** WARNING: IEEEtran.bst: No hyphenation pattern has been}%
			\typeout{** loaded for the language `#1'. Using the pattern for}%
			\typeout{** the default language instead.}%
			\else
			\language=\csname l@#1\endcsname
			\fi
			#2}}
	\providecommand{\BIBdecl}{\relax}
	\BIBdecl
	
	\bibitem{gassend2002silicon}
	B.~Gassend, D.~Clarke, M.~Van~Dijk, and S.~Devadas, ``Silicon physical random
	functions,'' in \emph{CCS}, 2002, pp. 148--160.
	
	\bibitem{suh2007physical}
	G.~E. Suh and S.~Devadas, ``Physical unclonable functions for device
	authentication and secret key generation,'' in \emph{DAC}, 2007, pp. 9--14.
	
	\bibitem{zhang2017xor}
	L.~Zhang, C.~Wang, W.~Liu, M.~O'Neill, and F.~Lombardi, ``{XOR} gate based
	low-cost configurable {RO} {PUF},'' in \emph{ISCAS}.\hskip 1em plus 0.5em
	minus 0.4em\relax IEEE, 2017, pp. 1--4.
	
	\bibitem{rahman2016aging}
	M.~T. Rahman, F.~Rahman, D.~Forte, and M.~Tehranipoor, ``An aging-resistant
	{RO-PUF} for reliable key generation,'' \emph{IEEE Trans. Emerg. Topics
		Comput.}, vol.~4, no.~3, pp. 335--348, 2016.
	
	\bibitem{holcomb2007initial}
	D.~E. Holcomb, W.~P. Burleson, and K.~Fu, ``{Initial {SRAM} state as a
		fingerprint and source of true random numbers for {RFID} tags},'' in
	\emph{Proceedings of the Conference on RFID Security}, 2007.
	
	\bibitem{cao2015low}
	Y.~Cao, L.~Zhang, C.-H. Chang, and S.~Chen, ``A low-power hybrid {RO PUF} with
	improved thermal stability for lightweight applications,'' \emph{IEEE Trans.
		Comput.-Aided Design Integr. Circuits Syst.}, vol.~34, no.~7, pp. 1143--1147,
	2015.
	
	\bibitem{kim2018dram}
	J.~S. Kim, M.~Patel, H.~Hassan, and O.~Mutlu, ``The {DRAM} latency {PUF}:
	Quickly evaluating physical unclonable functions by exploiting the
	latency-reliability tradeoff in modern commodity dram devices,'' in
	\emph{HPCA}.\hskip 1em plus 0.5em minus 0.4em\relax IEEE, 2018, pp. 194--207.
	
	\bibitem{sutar2018d}
	S.~Sutar, A.~Raha, D.~Kulkarni, R.~Shorey, J.~Tew, and V.~Raghunathan,
	``D-{PUF}: An intrinsically reconfigurable {DRAM} {PUF} for device
	authentication and random number generation,'' \emph{ACM Transactions on
		Embedded Computing Systems (TECS)}, vol.~17, no.~1, p.~17, 2018.
	
	\bibitem{sutar2018memory}
	S.~Sutar, A.~Raha, and V.~Raghunathan, ``Memory-based combination pufs for
	device authentication in embedded systems,'' \emph{IEEE Trans. Multi-Scale
		Comput. Syst.}, vol.~4, no.~4, pp. 793--810, 2018.
	
	\bibitem{gao2015emerging}
	Y.~Gao, D.~C. Ranasinghe, S.~F. Al-Sarawi, O.~Kavehei, and D.~Abbott,
	``Emerging physical unclonable functions with nanotechnology,'' \emph{\rm
		IEEE Access}, vol.~4, pp. 61--80, 2016.
	
	\bibitem{herder2014physical}
	C.~Herder, M.-D. Yu, F.~Koushanfar, and S.~Devadas, ``{Physical unclonable
		functions and applications: A tutorial},'' \emph{Proc. IEEE}, vol. 102, pp.
	1126--1141, 2014.
	
	\bibitem{delvaux2017security}
	J.~Delvaux, ``Security analysis of {PUF}-based key generation and entity
	authentication,'' Ph.D. dissertation, Shanghai Jiao Tong University, China,
	2017.
	
	\bibitem{gunlu2018privacy}
	O.~G{\"u}nl{\"u} and G.~Kramer, ``Privacy, secrecy, and storage with multiple
	noisy measurements of identifiers,'' \emph{IEEE Trans. Inf. Forensics
		Security}, vol.~13, no.~11, pp. 2872--2883, 2018.
	
	\bibitem{yu2017pervasive}
	M.-D.~M. Yu and S.~Devadas, ``Pervasive, dynamic authentication of physical
	items,'' \emph{Communications of the ACM}, vol.~60, no.~4, pp. 32--39, 2017.
	
	\bibitem{delvaux2015survey}
	J.~Delvaux, R.~Peeters, D.~Gu, and I.~Verbauwhede, ``A survey on lightweight
	entity authentication with strong pufs,'' \emph{ACM Computing Surveys
		(CSUR)}, vol.~48, no.~2, p.~26, 2015.
	
	\bibitem{delvaux2017machine}
	J.~Delvaux, ``Machine-learning attacks on {PolyPUFs}, {OB-PUFs}, {RPUFs},
	{LHS-PUFs}, and {PUF鈥揊SMs},'' \emph{IEEE Trans. Inf. Forensics Security},
	vol.~14, no.~8, pp. 2043--2058, 2019.
	
	\bibitem{bhargava2013high}
	M.~Bhargava and K.~Mai, ``A high reliability {PUF} using hot carrier injection
	based response reinforcement,'' in \emph{CHES}.\hskip 1em plus 0.5em minus
	0.4em\relax Springer, 2013, pp. 90--106.
	
	\bibitem{xu2015digital}
	T.~Xu and M.~Potkonjak, ``Digital {PUF} using intentional faults,'' in
	\emph{IEEE Int. Symp. Quality Electronic Design}, 2015, pp. 448--451.
	
	\bibitem{miao2016lrr}
	J.~Miao, M.~Li, S.~Roy, and B.~Yu, ``{LRR-DPUF}: Learning resilient and
	reliable digital physical unclonable function,'' in \emph{ICCAD}, 2016, pp.
	1--8.
	
	\bibitem{bhargava2014efficient}
	M.~{B}hargava and K.~Mai, ``An efficient reliable {PUF}-based cryptographic key
	generator in 65nm {CMOS},'' in \emph{DATE}, 2014, p.~70.
	
	\bibitem{maes2016secure}
	R.~Maes, V.~van~der Leest, E.~van~der Sluis, and F.~Willems, ``Secure key
	generation from biased {PUFs}: extended version,'' \emph{Journal of
		Cryptographic Engineering}, vol.~6, no.~2, pp. 121--137, 2016.
	
	\bibitem{yu2014noise}
	M.-D. Yu, D.~M'Raihi, I.~Verbauwhede, and S.~Devadas, ``A noise bifurcation
	architecture for linear additive physical functions,'' in \emph{HOST}, 2014,
	pp. 124--129.
	
	\bibitem{rostami2014robust}
	M.~Rostami, M.~Majzoobi, F.~Koushanfar, D.~S. Wallach, and S.~Devadas, ``Robust
	and reverse-engineering resilient {PUF} authentication and key-exchange by
	substring matching,'' \emph{IEEE Trans. Emerg. Topics Comput.}, vol.~2,
	no.~1, pp. 37--49, 2014.
	
	\bibitem{gao2016obfuscated}
	Y.~Gao, G.~Li, H.~Ma, S.~F. Al-Sarawi, O.~Kavehei, D.~Abbott, and D.~C.
	Ranasinghe, ``Obfuscated challenge-response: A secure lightweight
	authentication mechanism for {PUF}-based pervasive devices,'' in \emph{Percom
		Workshops}, 2016, pp. 1--6.
	
	\bibitem{maes2009soft}
	R.~Maes, P.~Tuyls, and I.~Verbauwhede, ``A soft decision helper data algorithm
	for {SRAM PUF}s,'' in \emph{IEEE Int. Symp. Information Theory}.\hskip 1em
	plus 0.5em minus 0.4em\relax IEEE, 2009, pp. 2101--2105.
	
	\bibitem{maes2013accurate}
	R.~Maes, ``An accurate probabilistic reliability model for silicon {PUFs},'' in
	\emph{CHES}, 2013, pp. 73--89.
	
	\bibitem{herder2017trapdoor}
	C.~Herder, L.~Ren, M.~van Dijk, M.-D.~M. Yu, and S.~Devadas, ``Trapdoor
	computational fuzzy extractors and stateless cryptographically-secure
	physical unclonable functions,'' \emph{IEEE Trans. Dependable Secure
		Comput.}, vol.~14, no.~1, pp. 65--82, 2017.
	
	\bibitem{gao2020physical}
	Y.~Gao, S.~F. Al-Sarawi, and D.~Abbott, ``Physical unclonable functions,''
	\emph{Nature Electronics}, vol.~3, no.~2, pp. 81--91, 2020.
	
	\bibitem{maiti2010large}
	A.~Maiti, J.~Casarona, L.~McHale, and P.~Schaumont, ``A large scale
	characterization of {RO-PUF},'' in \emph{HOST}, 2010, pp. 94--99.
	
	\bibitem{ruhrmair2013pufs}
	U.~Ruhrmair and M.~Van~Dijk, ``{PUFs} in security protocols: Attack models and
	security evaluations,'' in \emph{IEEE Symp. Security and Privacy}, 2013, pp.
	286--300.
	
	\bibitem{yulockdown}
	M.-D. Yu, M.~Hiller, J.~Delvaux, R.~Sowell, S.~Devadas, and I.~Verbauwhede, ``A
	lockdown technique to prevent machine learning on {PUFs} for lightweight
	authentication,'' \emph{IEEE Trans. Multi-Scale Comput. Syst.}, vol.~2,
	no.~3, pp. 146--159, 2016.
	
	\bibitem{ruhrmair2013puf}
	U.~Ruhrmair, J.~Solter, F.~Sehnke, X.~Xu, A.~Mahmoud, V.~Stoyanova, G.~Dror,
	J.~Schmidhuber, W.~Burleson, and S.~Devadas, ``{PUF} modeling attacks on
	simulated and silicon data,'' \emph{IEEE Trans. Inf. Forensics Security},
	vol.~8, no.~11, pp. 1876--1891, 2013.
	
	\bibitem{yu2011lightweight}
	M.-D. Yu, D.~M'~Raihi, R.~Sowell, and S.~Devadas, ``Lightweight and secure
	{PUF} key storage using limits of machine learning,'' in \emph{CHES}.\hskip
	1em plus 0.5em minus 0.4em\relax Springer, 2011, pp. 358--373.
	
	\bibitem{yu2012performance}
	M.-D. Yu, R.~Sowell, A.~Singh, D.~M'Raihi, and S.~Devadas, ``Performance
	metrics and empirical results of a {PUF} cryptographic key generation
	{ASIC},'' in \emph{HOST}.\hskip 1em plus 0.5em minus 0.4em\relax IEEE, 2012,
	pp. 108--115.
	
	\bibitem{lim2004extracting}
	D.~Lim, ``Extracting secret keys from integrated circuits,'' Master's thesis,
	Massachusetts Institute of Technology, 2004.
	
	\bibitem{ruhrmair2010modeling}
	U.~R{\"u}hrmair, F.~Sehnke, J.~S{\"o}lter, G.~Dror, S.~Devadas, and
	J.~Schmidhuber, ``Modeling attacks on physical unclonable functions,'' in
	\emph{CCS}, 2010, pp. 237--249.
	
	\bibitem{becker2015pitfalls}
	G.~T. {Becker}, ``On the pitfalls of using {Arbiter-PUFs} as building blocks,''
	\emph{IEEE Trans. Comput.-Aided Design Integr. Circuits Syst.}, vol.~34,
	no.~8, pp. 1295--1307, 2015.
	
	\bibitem{becker2015gap}
	G.~T. Becker, ``The gap between promise and reality: On the insecurity of {XOR}
	arbiter {PUFs},'' in \emph{CHES}, 2015, pp. 535--555.
	
	\bibitem{xu2016using}
	X.~Xu, W.~Burleson, and D.~E. Holcomb, ``Using statistical models to improve
	the reliability of delay-based {PUFs},'' in \emph{IEEE Computer Society
		Annual Symp. VLSI}.\hskip 1em plus 0.5em minus 0.4em\relax IEEE, 2016, pp.
	547--552.
	
	\bibitem{yu2010secure}
	M.-D. Yu and S.~Devadas, ``Secure and robust error correction for physical
	unclonable functions,'' \emph{IEEE Design \& Test of Computers}, vol.~27,
	no.~1, pp. 48--65, 2010.
	
	\bibitem{maes2009low}
	R.~{M}aes, P.~Tuyls, and I.~Verbauwhede, ``Low-overhead implementation of a
	soft decision helper data algorithm for {SRAM} {PUFs},'' in
	\emph{CHES}.\hskip 1em plus 0.5em minus 0.4em\relax Springer, 2009, pp.
	332--347.
	
	\bibitem{ranasinghe2005random}
	D.~C. Ranasinghe, D.~Lim, S.~Devadas, D.~Abbott, and P.~H. Cole, ``Random
	numbers from metastability and thermal noise,'' \emph{Electronics Letters},
	vol.~41, no.~16, p.~1, 2005.
	
	\bibitem{holcomb2009power}
	{Holcomb, Daniel E}, W.~P. Burleson, and K.~Fu, ``{Power-up SRAM} state as an
	identifying fingerprint and source of true random numbers,'' \emph{IEEE
		Trans. Comput.}, vol.~58, no.~9, pp. 1198--1210, 2009.
	
	\bibitem{van2012efficient}
	V.~van~der Leest, E.~van~der Sluis, G.-J. Schrijen, P.~Tuyls, and H.~Handschuh,
	``Efficient implementation of true random number generator based on sram
	pufs,'' in \emph{Cryptography and Security: From Theory to
		Applications}.\hskip 1em plus 0.5em minus 0.4em\relax Springer, 2012, pp.
	300--318.
	
	\bibitem{wang2012flash}
	Y.~Wang, W.-k. Yu, S.~Wu, G.~Malysa, G.~E. Suh, and E.~C. Kan, ``Flash memory
	for ubiquitous hardware security functions: True random number generation and
	device fingerprints,'' in \emph{IEEE Symp. Security and Privacy}.\hskip 1em
	plus 0.5em minus 0.4em\relax IEEE, 2012, pp. 33--47.
	
	\bibitem{zheng2017true}
	G.~Zheng, Y.~Lyu, and D.~Wang, ``True random number generator based on ring
	oscillator {PUF}s,'' in \emph{Proc. 2nd Int. Conf. Multimedia Systems and
		Signal Processing}.\hskip 1em plus 0.5em minus 0.4em\relax ACM, 2017, pp.
	1--5.
	
	\bibitem{maes2013physically}
	R.~Maes, ``Physically unclonable functions: Properties,'' in \emph{Physically
		Unclonable Functions}.\hskip 1em plus 0.5em minus 0.4em\relax Springer, 2013,
	pp. 49--80.
	
	\bibitem{sahoo2017multiplexer}
	D.~P. Sahoo, D.~Mukhopadhyay, R.~S. Chakraborty, and P.~H. Nguyen, ``A
	multiplexer-based arbiter {PUF} composition with enhanced reliability and
	security,'' \emph{IEEE Trans. Comput.}, vol.~67, no.~3, pp. 403--417, 2017.
	
	\bibitem{nguyen2018interpose}
	P.~H. Nguyen, D.~P. Sahoo, C.~Jin, K.~Mahmood, U.~R{\"u}hrmair, and M.~van
	Dijk, ``The interpose {PUF}: Secure {PUF} design against state-of-the-art
	machine learning attacks,'' \emph{IACR Transactions on Cryptographic Hardware
		and Embedded Systems}, pp. 243--290, 2019.
	
	\bibitem{gassend2002controlled}
	B.~{Ga}ssend, D.~Clarke, M.~Van~Dijk, and S.~Devadas, ``Controlled physical
	random functions,'' in \emph{ACSAC}.\hskip 1em plus 0.5em minus 0.4em\relax
	IEEE, 2002, pp. 149--160.
	
	\bibitem{gassend2008controlled}
	B.~Gassend, M.~V. Dijk, D.~Clarke, E.~Torlak, S.~Devadas, and P.~Tuyls,
	``Controlled physical random functions and applications,'' \emph{ACM
		Transactions on Information and System Security (TISSEC)}, vol.~10, no.~4,
	p.~3, 2008.
	
	\bibitem{tuyls2006read}
	P.~Tuyls, G.-J. Schrijen, B.~{\v{S}}kori{\'c}, J.~Van~Geloven, N.~Verhaegh, and
	R.~Wolters, ``Read-proof hardware from protective coatings,'' in
	\emph{CHES}.\hskip 1em plus 0.5em minus 0.4em\relax Springer, 2006, pp.
	369--383.
	
	\bibitem{immler2018b}
	V.~Immler, J.~Obermaier, M.~K{\"o}nig, M.~Hiller, and G.~Sig, ``{B-TREPID}:
	Batteryless tamper-resistant envelope with a {PUF} and integrity detection,''
	in \emph{HOST}.\hskip 1em plus 0.5em minus 0.4em\relax IEEE, 2018, pp.
	49--56.
	
	\bibitem{ruhrmair2014efficient}
	U.~R{\"u}hrmair, X.~Xu, J.~S{\"o}lter, A.~Mahmoud, M.~Majzoobi, F.~Koushanfar,
	and W.~Burleson, ``Efficient power and timing side channels for physical
	unclonable functions,'' in \emph{CHES}.\hskip 1em plus 0.5em minus
	0.4em\relax Springer, 2014, pp. 476--492.
	
	\bibitem{tajik2014physical}
	S.~Tajik, E.~Dietz, S.~Frohmann, J.-P. Seifert, D.~Nedospasov, C.~Helfmeier,
	C.~Boit, and H.~Dittrich, ``Physical characterization of arbiter {PUFs},'' in
	\emph{CHES}.\hskip 1em plus 0.5em minus 0.4em\relax Springer, 2014, pp.
	493--509.
	
	\bibitem{boit2016ic}
	C.~Boit, S.~Tajik, P.~Scholz, E.~Amini, A.~Beyreuther, H.~Lohrke, and
	J.~Seifert, ``From {IC} debug to hardware security risk: The power of
	backside access and optical interaction,'' in \emph{Proc. IEEE Int. Symp.
		Physical and Failure Analysis of Integrated Circuits}, 2016, pp. 365--369.
	
	\bibitem{wild2017fair}
	A.~Wild, G.~T. Becker, and T.~G{\"u}neysu, ``A fair and comprehensive
	large-scale analysis of oscillation-based {PUFs} for {FPGAs},'' in \emph{27th
		Int. Conf. FPL}, 2017, pp. 1--7.
	
	\bibitem{hesselbarth2018large}
	R.~Hesselbarth, F.~Wilde, C.~Gu, and N.~Hanley, ``Large scale {RO} {PUF}
	analysis over slice type, evaluation time and temperature on 28nm {Xilinx}
	{FPGAs},'' in \emph{HOST}, 2018, pp. 126--133.
	
	\bibitem{tobisch2015scaling}
	J.~Tobisch and G.~T. Becker, ``On the scaling of machine learning attacks on
	{PUFs} with application to noise bifurcation,'' in \emph{International
		Workshop on Radio Frequency Identification: Security and Privacy
		Issues}.\hskip 1em plus 0.5em minus 0.4em\relax Springer, 2015, pp. 17--31.
	
	\bibitem{delvaux2015helper}
	J.~Delvaux, D.~Gu, D.~Schellekens, and I.~Verbauwhede, ``Helper data algorithms
	for {PUF}-based key generation: Overview and analysis,'' \emph{IEEE Trans.
		Comput.-Aided Design Integr. Circuits Syst.}, vol.~34, no.~6, pp. 889--902,
	2015.
	
	\bibitem{robust2017becker}
	G.~T. Becker, ``Robust fuzzy extractors and helper data manipulation attacks
	revisited: Theory vs practice,'' \emph{IEEE Trans. Dependable Secure
		Comput.}, 2017, DOI:10.1109/TDSC.2017.2762675.
	
	\bibitem{konigsmark2016polypuf}
	S.~T.~C. Konigsmark, D.~Chen, and M.~D. Wong, ``{PolyPUF}: Physically secure
	self-divergence,'' \emph{IEEE Trans. Comput.-Aided Design Integr. Circuits
		Syst.}, vol.~35, no.~7, pp. 1053--1066, 2016.
	
	\bibitem{ye2016rpuf}
	J.~Ye, Y.~Hu, and X.~Li, ``{RPUF}: Physical unclonable function with randomized
	challenge to resist modeling attack,'' in \emph{AsianHOST}.\hskip 1em plus
	0.5em minus 0.4em\relax IEEE, 2016, pp. 1--6.
	
	\bibitem{idriss2017lightweight}
	T.~Idriss and M.~Bayoumi, ``Lightweight highly secure puf protocol for mutual
	authentication and secret message exchange,'' in \emph{Int. Conf. RFID
		Technology \& Application}.\hskip 1em plus 0.5em minus 0.4em\relax IEEE,
	2017, pp. 214--219.
	
	\bibitem{sadeghi2010enhancing}
	A.-R. Sadeghi, I.~Visconti, and C.~Wachsmann, ``Enhancing rfid security and
	privacy by physically unclonable functions,'' in \emph{Towards
		Hardware-Intrinsic Security}.\hskip 1em plus 0.5em minus 0.4em\relax
	Springer, 2010, pp. 281--305.
	
	\bibitem{gao2018puf}
	Y.~Gao, H.~Ma, S.~F. Al-Sarawi, D.~Abbott, and D.~C. Ranasinghe, ``{PUF-FSM}: A
	controlled strong {PUF},'' \emph{IEEE Trans. Comput.-Aided Design Integr.
		Circuits Syst.}, vol.~37, no.~5, pp. 1104--1108, 2018.
	
	\bibitem{van2012reverse}
	A.~Van~Herrewege, S.~Katzenbeisser, R.~Maes, R.~Peeters, A.-R. Sadeghi,
	I.~Verbauwhede, and C.~Wachsmann, ``Reverse fuzzy extractors: Enabling
	lightweight mutual authentication for {PUF-enabled RFIDs},'' in
	\emph{Financial Cryptography and Data Security}.\hskip 1em plus 0.5em minus
	0.4em\relax Springer, 2012, pp. 374--389.
	
	\bibitem{maes2014countering}
	R.~Maes and V.~van~der Leest, ``Countering the effects of silicon aging on
	{SRAM PUFs},'' in \emph{HOST}.\hskip 1em plus 0.5em minus 0.4em\relax IEEE,
	2014, pp. 148--153.
	
	\bibitem{liu2017acro}
	C.~Q. Liu, Y.~Cao, and C.~H. Chang, ``{ACRO-PUF}: A low-power, reliable and
	aging-resilient current starved inverter-based ring oscillator physical
	unclonable function,'' \emph{IEEE Trans. Circuits and Syst. I: Reg. Papers},
	vol.~64, no.~12, pp. 3138--3149, 2017.
	
	\bibitem{kirkpatrick2010software}
	M.~S. Kirkpatrick and E.~Bertino, ``Software techniques to combat drift in
	{PUF}-based authentication systems,'' in \emph{Workshop on Secure Component
		and System Identification}, 2010, p.~9.
	
	\bibitem{maes2012experimental}
	R.~Maes, V.~Ro{\v{z}}i{\'c}, I.~Verbauwhede, P.~Koeberl, E.~Van~der Sluis, and
	V.~Van~der Leest, ``Experimental evaluation of physically unclonable
	functions in 65 nm {CMOS},'' in \emph{Proc. ESSCIRC}.\hskip 1em plus 0.5em
	minus 0.4em\relax IEEE, 2012, pp. 486--489.
	
	\bibitem{chuang2017physically}
	K.-H. Chuang, E.~Bury, R.~Degraeve, B.~Kaczer, G.~Groeseneken, I.~Verbauwhede,
	and D.~Linten, ``Physically unclonable function using {CMOS} breakdown
	position,'' in \emph{Proc. IEEE Int. Reliab. Phys. Symp}, 2017,
	DOI:10.1109/IRPS.2017.7936312.
	
	\bibitem{che2014non}
	W.~Che, J.~Plusquellic, and S.~Bhunia, ``A non-volatile memory based physically
	unclonable function without helper data,'' in \emph{ICCAD}, 2014, pp.
	148--153.
	
	\bibitem{khan2017phase}
	R.~S. Khan, N.~Noor, C.~Jin, J.~Scoggin, Z.~Woods, S.~Muneer, A.~Ciardullo,
	P.~H. NGUYEN, A.~GOKIRMAK, M.~van Dijk \emph{et~al.}, ``Phase change memory
	and its applications in hardware security,'' in \emph{Security Opportunities
		in Nano Devices and Emerging Technologies}.\hskip 1em plus 0.5em minus
	0.4em\relax CRC Press, 2017, pp. 115--136.
	
	\bibitem{maes2012pufky}
	R.~Maes, A.~Van~Herrewege, and I.~Verbauwhede, ``{PUFKY}: A fully functional
	{PUF}-based cryptographic key generator,'' in \emph{CHES}, 2012, pp.
	302--319.
	
	\bibitem{becker2014active}
	G.~T. Becker, R.~Kumar \emph{et~al.}, ``Active and passive side-channel attacks
	on delay based {PUF} designs.'' \emph{IACR Cryptology ePrint Archive}, vol.
	2014, p. 287, 2014.
	
	\bibitem{yan2016puf}
	W.~Yan, F.~Tehranipoor, and J.~A. Chandy, ``{PUF}-based fuzzy authentication
	without error correcting codes,'' \emph{IEEE Trans. Comput.-Aided Design
		Integr. Circuits Syst.}, vol.~36, no.~9, pp. 1445--1457, 2016.
	
	\bibitem{ozturk2008towards}
	E.~{\"O}zt{\"u}rk, G.~Hammouri, and B.~Sunar, ``Towards robust low cost
	authentication for pervasive devices,'' in \emph{Percom}.\hskip 1em plus
	0.5em minus 0.4em\relax IEEE, 2008, pp. 170--178.
	
	\bibitem{gao2018lightweight}
	Y.~Gao, Y.~Su, L.~Xu, and D.~C. Ranasinghe, ``Lightweight (reverse) fuzzy
	extractor with multiple reference {PUF} responses,'' \emph{IEEE Trans. Inf.
		Forensics Security}, vol.~14, no.~7, pp. 1887--1901, 2019.
	
	\bibitem{su2019secucode}
	Y.~Su, Y.~Gao, M.~Chesser, O.~Kavehei, A.~Sample, and D.~Ranasinghe,
	``Secucode: Intrinsic {PUF} entangled secure wireless code dissemination for
	computational {RFID} devices,'' \emph{IEEE Trans. Dependable Secure Comput.},
	2019.
	
	\bibitem{su2019hash}
	Y.~Su, Y.~Gao, O.~Kavehei, and D.~C. Ranasinghe, ``Hash functions and
	benchmarks for resource constrained passive devices: A preliminary study,''
	in \emph{PerCom Workshops}.\hskip 1em plus 0.5em minus 0.4em\relax IEEE,
	2019, pp. 1020--1025.
	
\end{thebibliography}
\end{document}